\documentclass[aps,showpacs,preprintnumbers,amsmath,amssymb,twocolumn,superscriptaddress,prx,nofootinbib,floatfix,10
pt]{revtex4-2}

\usepackage{graphicx}  
\usepackage{epstopdf}  
\usepackage{amsfonts}
\usepackage{amsmath}
\usepackage{amssymb}
\usepackage{amsthm}
\usepackage{bm}
\usepackage{dcolumn}
\usepackage{tabularx}
\usepackage{array}
\usepackage[dvipsnames]{xcolor}
\usepackage{mathrsfs}
\usepackage{mathtools}
\usepackage{verbatim}
\usepackage{comment}
\usepackage{ragged2e}
\usepackage{physics}
\usepackage{hyperref}
\usepackage{url}
\usepackage{cleveref}
\usepackage{tikz}
\usepackage{caption}
\usepackage{subcaption}
\usepackage{booktabs}
\usepackage{nicefrac}
\usepackage{blindtext}
\usepackage{lipsum}
\usepackage[bottom]{footmisc}
\usepackage{centernot}
\usepackage{eurosym}

\usepackage{footmisc}
\usepackage{lipsum} % For generating dummy text
\usepackage{etoolbox} % For the \AtEndEnvironment command

\DeclareCaptionJustification{justified}{\justifying}
\hypersetup{colorlinks=true, citecolor=orange, urlcolor=blue, linkcolor=magenta}
\usepackage[linesnumbered,lined,boxed,commentsnumbered,ruled,vlined]{algorithm2e}
\usepackage{algpseudocode}

\theoremstyle{definition}

\begin{document}

\title{Spectral signatures of structural change in financial networks}

\author{Valentina Macchiati}
\affiliation{Scuola Normale Superiore, P.zza dei Cavalieri 7, 56126 Pisa (Italy)}
\author{Emiliano Marchese}
\affiliation{IMT School for Advanced Studies, P.zza San Francesco 19, 55100 Lucca (Italy)}
\author{Piero Mazzarisi}
\affiliation{Università degli Studi di Siena, P.zza S.Francesco 7-8, 53100 Siena (Italy)}
\author{Diego Garlaschelli}
\affiliation{IMT School for Advanced Studies, P.zza San Francesco 19, 55100 Lucca (Italy)}
\affiliation{Lorentz Institute for Theoretical Physics, University of Leiden, Niels Bohrweg, 2, Leiden, NL-2333 CA, The Netherlands}
\affiliation{INdAM-GNAMPA Istituto Nazionale di Alta Matematica `Francesco Severi', P.le Aldo Moro 5, 00185 Rome (Italy)}
\author{Tiziano Squartini}
\email{tiziano.squartini@imtlucca.it}
\affiliation{IMT School for Advanced Studies, P.zza San Francesco 19, 55100 Lucca (Italy)}
\affiliation{Scuola Normale Superiore, P.zza dei Cavalieri 7, 56126 Pisa (Italy)}
\affiliation{INdAM-GNAMPA Istituto Nazionale di Alta Matematica `Francesco Severi', P.le Aldo Moro 5, 00185 Rome (Italy)}

\date{\today}

\begin{abstract}
The level of systemic risk in economic and financial systems is strongly determined by the structure of the underlying networks of interdependent entities that can propagate shocks and stresses. Since changes in network structure imply changes in risk levels, it is important to identify structural transitions potentially leading to system-wide crises. Methods have been proposed to assess whether a real-world network is in a (quasi-)stationary state by checking the consistency of its structural evolution with appropriate maximum-entropy ensembles of graphs. While previous analyses of this kind have focused on dyadic and triadic motifs, hence disregarding higher-order structures, here we consider closed walks of any length. Specifically, we study the ensemble properties of the spectral radius of random graph models calibrated on real-world evolving networks. Our approach is shown to work remarkably well for directed networks, both binary and weighted. As illustrative examples, we consider the Electronic Market for Interbank Deposit (e-MID), the Dutch Interbank Network (DIN) and the International Trade Network (ITN) in their evolution across the 2008 crisis. By monitoring the deviation of the spectral radius from its ensemble expectation, we find that the ITN remains in a (quasi-)equilibrium state throughout the period considered, while both the DIN and e-MID exhibit a clear out-of-equilibrium behaviour. The spectral deviation therefore captures ongoing topological changes, extending over all length scales, to provide a compact proxy of the resilience of economic and financial networks.
\end{abstract}


\maketitle

\section{Introduction}

As witnessed by two major recent crises (i.e. the global financial one in 2008 and the Covid-19 pandemic in 2020), having a clear understanding of the intricate structure of economic and financial systems - be they interbank~\cite{bargigli2015multiplex}, interfirm~\cite{carvalho2019production,ialongo2022reconstructing} or trade networks~\cite{garlaschelli2005structure} - is crucial, especially under stress conditions. The interconnectedness of economic and financial agents is, in fact, known to play a major role both during the phase of distress accumulation and after a crisis outbreak in sustaining and reinforcing shock propagation~\cite{huser2015too}. Back in 2008, banks sought to minimise individual risk by diversifying their portfolios: the \emph{simultaneous} character of such diversification, however, led to an unexpected level of mutual dependency whose net consequence was that of amplifying the effects of individual defaults~\cite{squartini2013early,delpini2020portfolio}.

A particularly relevant question addresses the (quasi-)stationarity of the temporal evolution of a given, real-world, economic or financial network, i.e. \emph{does the system undergo smooth, structural changes controlled by few driving parameters?} Should this be the case, the behaviour of the network under analysis would be predictable solely in terms of the dynamics of those parameters; otherwise, the lack of stationarity may lead to abrupt - hence, uncontrollable - regime shifts.

The problem of the (non) stationarity of real-world, economic and financial networks has been addressed by studying whether they can be considered typical members of an evolving, (quasi-)equilibrium ensemble of graphs with given properties~\cite{squartini2015stationarity}: while such properties are treated as constraints - hence, assumed to be the `independent variables' undergoing an autonomous evolution - the other network properties are treated as `dependent vareiables' - hence, assumed to vary only as a consequence of the former ones. Broadly speaking, three different situations can occur:

\begin{itemize}
\item[$\bullet$] The observed network properties are systematically found to agree with what is expected from the evolution of the enforced constraints. In this case, one can conclude that the real-world network is (quasi-)stationary - and its evolution is driven by the dynamics of the constraints;
\item[$\bullet$] The observed network properties slightly deviate from equilibrium expectations, but the deviating patterns remain coherent. In this case, the network can still be considered (quasi-)stationary - even if its evolution cannot be claimed to be completely driven by the chosen constraints (very likely, with the addition of other appropriate constraints, one would go back to the first situation);
\item[$\bullet$] The observed network properties significantly deviate from the (quasi-)equilibrium expectations, showing different deviating patterns at different times. In this case, the network can be considered non-stationary.
\end{itemize}

Analyses of this kind have indeed led to individuate early-warning signals of upcoming, critical events, although the indicators considered so far have just involved dyadic and triadic `debt loops' with different levels of reciprocity~\cite{battiston2016complexity,macchiati2024interbank,squartini2013early}. The present paper aims to extend the study of early-warnings' emergence by considering closed walks of any length \emph{at once}. Such a request can be handled by exploiting the theorem stating that $a_{ij}^{(n)}$, i.e. the generic entry of the $n$-th power of the adjacency matrix $\mathbf{A}$, counts the total number of closed walks of length $n$ connecting node $i$ with node $j$: all `debt loops' can be, then, compactly accounted for by carrying out a double sum, over $n$ and over the diagonal entries of $\mathbf{A}$.

From a computational perspective, such a calculation can be greatly sped up by proxying the trace of the adjacency matrix with its principal eigenvalue $\lambda_1$, which, then, becomes the only relevant statistics whose $z$-score needs to be explicitly calculated. Such an appealing simplification, however, comes at a price: the expressions of $\langle\lambda_1\rangle$ and $\text{Var}[\lambda_1]$, i.e. of the expected value of $\lambda_1$ and its variance, are explicitly known in few cases only, i.e. \emph{i)} when the random network model is the binary, undirected version of the Erd\"os-R\'enyi (ER) model~\cite{furedi1981eigenvalues}; \emph{ii)} when the random network model is the Chung-Lu (CL) model, either in its binary, undirected version~\cite{chung2003eigenvalues,oliveira2009concentration,chung2011spectra,lu2012spectra,dionigi2023central} or in its binary, directed version~\cite{restrepo2007approximating,burstein2017asymptotics}; \emph{iii)} if the edges are treated as i.n.i.d. (independent, non-identically distributed) random variables, each one obeying a different Poisson distribution~\cite{zhang2014spectra}; \emph{iv)} if the considered graphs are infinitely large, locally tree-like and directed~\cite{neri2020linear}.

Let us remark that the existing estimations obtained under hypotheses are rarely satisfied by empirical configurations. For instance, the presence of cycles contradicts the assumption of observing locally tree-like structures, and the heterogeneity of the (in- and out-) degree distributions severely limits the applicability of the CL model. On a more general ground, the vast majority of the approaches above requires the knowledge of the (in- and out-) degree sequences, i.e. of a kind of information that data confidentiality issues make often (if not always) unavailable; moreover, none provides estimations of a network spectral properties taking its weighted marginals (i.e. in-strengths and out-strengths) as the sole input.

Motivated by the evidence that general results about the statistical properties of a network principal eigenvalue are currently missing, we propose an approach to their study that is applicable under \emph{any} random network model. The generality of our approach comes at a price: our results rest upon the validity of several approximations that need to be explicitly verified whenever a particular configuration is studied. Still, although our assumptions may appear quite drastic, our approach works remarkably well for directed networks, be they binary (BDN) or weighted (WDN).

A BDN is described by an adjacency matrix $\mathbf{A}$ whose generic entry satisfies the relationships $a_{ij}=1$ if a link points from node $i$ towards node $j$ and $a_{ij}=0$ otherwise. Moreover, $a_{ij}$ will, in general, differ from $a_{ji}$. A WDN is described by an adjacency matrix $\mathbf{W}$ whose generic entry satisfies the relationships $w_{ij}>0$ if a weighted link points from node $i$ towards node $j$ and $w_{ij}=0$ otherwise. Moreover, $w_{ij}$ will, in general, differ from $w_{ji}$.

\section{Detecting structural changes}

Structural changes can be spotted by comparing the empirical abundance of a quantity of interest with the corresponding expected value, calculated under a properly defined benchmark model\footnote{Hereby, the expressions `random network model', `benchmark model' and `null model' will be used interchangeably.}. To this aim, a very useful indicator is represented by the $z$-score

\begin{equation}
z[X]=\frac{X-\langle X\rangle}{\sigma[X]}
\end{equation}
where $X$ is the empirical abundance of the quantity $X$, $\langle X\rangle$ is its expected occurrence under the chosen null model and $\sigma[X]=\sqrt{\langle X^2\rangle-\langle X\rangle^2}$ is the standard deviation of $X$ under the same null model. In words, $z[X]$ quantifies the number of standard deviations by which the empirical abundance of $X$ differs from the expected one after checking for the Gaussianity of $X$ under the null model - often ensured by the fact that $X$ is the sum of several random variables - a result $|X|\leq2$ ($|X|\leq3$) indicates that the empirical abundance of $X$ is compatible with the expected one, at the $5\%$ ($1\%$) level of statistical significance; on the other hand, a value $|X|>2$ ($|X|>3$) indicates that the empirical abundance of $X$ is not compatible with the expected one, at the same significance level. In the latter case, a value $z[X]>0$ ($z[X]<0$) indicates the tendency of the pattern to be over-represented (under-represented) in the data with respect to the chosen benchmark.

\subsection{Dyadic signature of structural changes}

Moving from the observation that

\begin{equation}
\sum_{j=1}^Na_{ij}a_{jk}=[\mathbf{A}^2]_{ik}
\end{equation}
we will pose

\begin{align}
X&=\sum_{i=1}^N\sum_{j=1}^Na_{ij}a_{ji}=\sum_{i=1}^N[\mathbf{A}^2]_{ii}=\text{Tr}\left[\mathbf{A}^2\right],
\end{align}
noticing that the total number of links having a partner pointing in the opposite direction coincides with the trace of the second power of the adjacency matrix. The position above leads to

\begin{equation}
\langle X\rangle=\sum_{i=1}^N\sum_{j=1}^Np_{ij}p_{ji}=\sum_{i=1}^N[\mathbf{P}^2]_{ii}=\text{Tr}\left[\mathbf{P}^2\right]
\end{equation}
and to

\begin{align}
\sigma[X]&=\sqrt{\text{Var}\left[\sum_{i=1}^N\sum_{j=1}^Na_{ij}a_{ji}\right]}\nonumber\\
&=\sqrt{\text{Var}\left[2\cdot\sum_{i=1}^N\sum_{j(>i)}a_{ij}a_{ji}\right]}\nonumber\\
&=\sqrt{4\cdot\sum_{i=1}^N\sum_{j(>i)}\text{Var}[a_{ij}a_{ji}]}\nonumber\\
&=2\cdot\sqrt{\sum_{i=1}^N\sum_{j(>i)}p_{ij}p_{ji}(1-p_{ij}p_{ji})}
\end{align}
where $\mathbf{P}\equiv\{p_{ij}\}_{i,j=1}^N$ is the matrix of probability coefficients induced by the chosen null model, and the third passage follows from the evidence that the dyads induce independent random variables (see also Appendix~\hyperlink{AppA}{A}).

\subsection{Triadic signature of structural changes}

Analogously to the dyadic case, let us move from the observation that

\begin{equation}
\sum_{j=1}^N\sum_{k=1}^Na_{ij}a_{jk}a_{kl}=[\mathbf{A}^3]_{il}
\end{equation}
and pose

\begin{align}
X&=\sum_{i=1}^N\sum_{j=1}^N\sum_{k=1}^Na_{ij}a_{jk}a_{ki}=\sum_{i=1}^N[\mathbf{A}^3]_{ii}=\text{Tr}\left[\mathbf{A}^3\right],
\end{align}
noticing that the total number of triangles is proportional to the trace of the third power of the adjacency matrix. The position above leads to

\begin{equation}
\langle X\rangle=\sum_{i=1}^N\sum_{j=1}^N\sum_{k=1}^Np_{ij}p_{jk}p_{ki}=\sum_{i=1}^N[\mathbf{P}^3]_{ii}=\text{Tr}\left[\mathbf{P}^3\right]
\end{equation}
and to

\begin{align}
\sigma[X]&=\sqrt{\text{Var}\left[\sum_{i=1}^N\sum_{j=1}^N\sum_{k=1}^Na_{ij}a_{jk}a_{ki}\right]}\nonumber\\
&=\sqrt{\text{Var}\left[3\cdot\sum_{i=1}^N\sum_{j(>i)}\sum_{k(>j)}(a_{ij}a_{jk}a_{ki}+a_{ik}a_{kj}a_{ji})\right]}\nonumber\\
&=\sqrt{9\cdot\text{Var}\left[\sum_{i=1}^N\sum_{j(>i)}\sum_{k(>j)}(a_{ij}a_{jk}a_{ki}+a_{ik}a_{kj}a_{ji})\right]}\nonumber\\
&=3\cdot\sqrt{\text{Var}\left[\sum_{i=1}^N\sum_{j(>i)}\sum_{k(>j)}(a_{ij}a_{jk}a_{ki}+a_{ik}a_{kj}a_{ji})\right]}
\end{align}
where $\mathbf{P}\equiv\{p_{ij}\}_{i,j=1}^N$ is the matrix of probability coefficients induced by the chosen null model. Since triads do not induce independent random variables, the explicit expression of $\sigma[X]$ is derived in Appendix~\hyperlink{AppB}{B}. Let us notice that, in case the considered networks are sparse, one can simplify the expression above upon posing

\begin{align}
\sigma[X]&\simeq3\cdot\sqrt{\sum_{i=1}^N\sum_{j(>i)}\sum_{k(>j)}\text{Var}[a_{ij}a_{jk}a_{ki}+a_{ik}a_{kj}a_{ji}]}
\end{align}
with

\begin{align}
\text{Var}[a_{ij}a_{jk}a_{ki}+a_{ik}a_{kj}a_{ji}]&\simeq\text{Var}[a_{ij}a_{jk}a_{ki}]+\text{Var}[a_{ik}a_{kj}a_{ji}]
\end{align}
and

\begin{align}
\text{Var}[a_{ij}a_{jk}a_{ki}]&=p_{ij}p_{jk}p_{ki}(1-p_{ij}p_{jk}p_{ki}),\\
\text{Var}[a_{ik}a_{kj}a_{ji}]&=p_{ik}p_{kj}p_{ji}(1-p_{ik}p_{kj}p_{ji}).
\end{align}

\subsection{Spectral signature of structural changes}

Let us now enlarge the set of patterns to be considered for detecting structural changes by accounting for closed walks of any length.

\subsubsection{The trace of the matrix exponential}

Let us start by considering the $N\times N$ adjacency matrix $\mathbf{A}$ of a BDN, with $a_{ii}=0$, $\forall\:i$: the following relationship

\begin{equation}\label{eq.a1}
\mathbf{I}+\mathbf{A}+\frac{\mathbf{A}^2}{2!}+\frac{\mathbf{A}^3}{3!}+\dots+\frac{\mathbf{A}^n}{n!}+\dots=\sum_{k=0}^\infty\frac{\mathbf{A}^k}{k!}\equiv e^\mathbf{A},
\end{equation}
where $\mathbf{A}^0\equiv\mathbf{I}$, defines the exponential of $\mathbf{A}$~\cite{estrada2005subgraph,estrada2000characterization,de2007estimating,gutman2007graph,estrada2008communicability,estrada2009returnability}. Let us, now, calculate the trace of such a matrix exponential: since it is invariant under diagonalisation, one obtains that

\begin{equation}\label{eq.a2}
\text{Tr}\left[e^\mathbf{A}\right]=\sum_{k=0}^\infty\frac{\text{Tr}\left[\mathbf{A}^k\right]}{k!}=\sum_{k=0}^\infty\frac{\text{Tr}\left[\mathbf{\Lambda}^k\right]}{k!}=\text{Tr}\left[e^\mathbf{\Lambda}\right]
\end{equation}
where $\mathbf{\Lambda}$ is the matrix obtained upon diagonalising $\mathbf{A}$ (see also Appendix~\hyperlink{AppC}{C}). As the number of walks of length $k$ starting from and ending at the same vertex can be counted by computing the trace of the $k$-th power of the adjacency matrix, i.e. $\text{Tr}\left[\mathbf{A}^k\right]=\sum_{i=1}^N\left[\mathbf{A}^k\right]_{ii}$, eq.~(\ref{eq.a2}) relates the number of walks of any length characterising a binary network $\mathbf{A}$ with its spectral properties. Such a quantity, named Estrada index, represents a graph invariant quantifying the communicability of a given network, i.e. the `participation' of each node to the walks present in the network itself~\cite{estrada2000characterization}.\\

Analogously, given the $N\times N$ adjacency matrix $\mathbf{W}$ of a WDN with $w_{ii}=0$, $\forall\:i$, the relationships

\begin{equation}\label{eq.w3}
\mathbf{I}+\mathbf{W}+\frac{\mathbf{W}^2}{2!}+\frac{\mathbf{W}^3}{3!}+\dots+\frac{\mathbf{W}^n}{n!}+\dots=\sum_{k=0}^\infty\frac{\mathbf{W}^k}{k!}\equiv e^\mathbf{W}
\end{equation}
and

\begin{equation}\label{eq.w4}
\text{Tr}\left[e^\mathbf{W}\right]=\sum_{k=0}^\infty\frac{\text{Tr}\left[\mathbf{W}^k\right]}{k!}=\sum_{k=0}^\infty\frac{\text{Tr}\left[\mathbf{\Omega}^k\right]}{k!}=\text{Tr}\left[e^\mathbf{\Omega}\right],
\end{equation}
where $\mathbf{\Omega}$ is the matrix obtained upon diagonalising $\mathbf{W}$, hold true. As a result, concerning the number of walks of length $k$ starting and ending at the same vertex can be extended to weighted networks, eq.~(\ref{eq.w4}) generalises the Estrada index to weighted configurations.\\

Let us explicitly notice that

\begin{itemize}
\item[$\bullet$] the absence of self-loops, i.e. $\text{Tr}\left[\mathbf{A}\right]=\text{Tr}\left[\mathbf{W}\right]=0$, implies that, whenever present, complex eigenvalues must appear in conjugate pairs;
\item[$\bullet$] eq.~(\ref{eq.a1}) implies that $\text{Tr}\left[e^\mathbf{A}\right]\geq0$, i.e. that the trace of the exponential of $\mathbf{A}$ is a real, non-negative number. Analogously, eq.~(\ref{eq.w3}) implies that $\text{Tr}\left[e^\mathbf{W}\right]\geq0$, i.e. that the trace of the exponential of $\mathbf{W}$ is a real, non-negative number;
\item[$\bullet$] When computing the number of closed walks of a certain length, edges must be counted repeatedly. For example, the closed walks of length 4 in a binary, directed network are \emph{i)} the proper cycles like $i\rightarrow j\rightarrow k\rightarrow l\rightarrow i$; \emph{ii)} the pairs of dyads like $i\rightarrow j\rightarrow k\rightarrow j\rightarrow i$; \emph{iii)} the single dyads like $i\rightarrow j\rightarrow i$. Equations~(\ref{eq.a2}) and~(\ref{eq.w4}) compactly account for all of them.
\end{itemize}

The third observation has relevant implications for economic and financial applications: when studying the propagation of a shock, in fact, it is extremely important to account for all possible patterns along which distress can propagate, including the ones leading to multiple reverberations among the same nodes~\cite{bardoscia2017pathways}. As (combinations of) cycles are supposed to lower the resilience of financial networks by amplifying external shocks~\cite{battiston2016complexity}, eqs.~(\ref{eq.a2}) and~(\ref{eq.w4}) suggest the trace of the exponential matrix to represent a compact proxy of the stability of the network itself.

\subsubsection{Expected value of the trace of the matrix exponential}

Let us now move to analyse the expected value of the quantity $\text{Tr}[e^\mathbf{A}]=\text{Tr}[e^\mathbf{\Lambda}]$, under a properly-defined benchmark model. We will suppose the latter one to be described by an $N\times N$ matrix $\mathbf{P}$ whose generic entry $p_{ij}$, with $i\neq j$, indicates the probability that nodes $i$ and $j$ are connected via a directed link. Following the same steps as above, we find

\begin{equation}\label{eq.p1}
\text{Tr}\left[e^\mathbf{P}\right]=\sum_{k=0}^\infty\frac{\text{Tr}\left[\mathbf{P}^k\right]}{k!}=\sum_{k=0}^\infty\frac{\text{Tr}\left[\mathbf{\Pi}^k\right]}{k!}=\text{Tr}\left[e^\mathbf{\Pi}\right]
\end{equation}
where $\mathbf{\Pi}$ is the matrix obtained upon diagonalising $\mathbf{P}$. 

Let us now inspect the relationship between eq.~(\ref{eq.a2}) and eq.~(\ref{eq.p1}). Since we are considering binary, adjacency matrices, the matrix $\mathbf{P}$ satisfies the relationship $\langle\mathbf{A}\rangle=\mathbf{P}$, a compact notation stating for $\langle a_{ij}\rangle=p_{ij}$, $\forall\:i\neq j$. To extend this result to higher powers of the adjacency matrix, an explicit expression for the quantity $\langle\mathbf{A}^n\rangle=f(\mathbf{P})$, $\forall\:n$ is needed. Here, we adopt the recipe defining the so-called \emph{delta method}~\cite{oehlert1992note} and prescribing to identify $f(\mathbf{P})$ with $\mathbf{P}^n$. According to it, the expected value of the number of closed walks of any length satisfies the chain of inequalities

\begin{widetext}
\begin{equation}\label{eq.a5}
\langle\text{Tr}\left[e^\mathbf{A}\right]\rangle=\sum_{k=0}^\infty\frac{\langle\text{Tr}\left[\mathbf{A}^k\right]\rangle}{k!}=\sum_{k=0}^\infty\frac{\text{Tr}\left[\langle\mathbf{A}^k\rangle\right]}{k!}\geq\sum_{k=0}^\infty\frac{\text{Tr}\left[\langle\mathbf{A}\rangle^k\right]}{k!}=\sum_{k=0}^\infty\frac{\text{Tr}\left[\mathbf{P}^k\right]}{k!}=\text{Tr}\left[e^\mathbf{P}\right];
\end{equation}
\end{widetext}
a relationship leading to $\langle\text{Tr}\left[e^\mathbf{\Lambda}\right]\rangle\geq\text{Tr}\left[e^\mathbf{\Pi}\right]$.

The inequality can be understood upon considering a reciprocated dyad and noticing that relationships like $\langle[\mathbf{A}^4]_{ii}\rangle=\langle a_{ij}a_{ji}a_{ij}a_{ji}\rangle=\langle a_{ij}a_{ji}\rangle=\langle a_{ij}\rangle\langle a_{ji}\rangle=p_{ij}p_{ji}\geq p_{ij}^2p_{ji}^2=[\mathbf{P}^4]_{ii}$ hold true; in other words, estimating the number of closed walks of a certain length via the delta method implies overweighing the edges constituting them, whence the mismatch between the correct and the approximated expression. Such a mismatch is absent in case no link is reciprocated: given a square loop, in fact, $\langle[\mathbf{A}^4]_{ii}\rangle=\langle a_{ij}a_{jk}a_{kl}a_{li}\rangle=\langle a_{ij}\rangle\langle a_{jk}\rangle\langle a_{kl}\rangle\langle a_{li}\rangle=p_{ij}p_{jk}p_{kl}p_{li}=[\mathbf{P}^4]_{ii}$. In other words, the larger the number of reciprocal links\footnote{Let us remind that $L^\leftrightarrow=\sum_{i=1}^N\sum_{j(\neq i)}a_{ij}a_{ji}$.}, the less accurate the approximation provided by the delta method. Hereby, we will assume that the symbol $\gtrsim$ can replace the symbol $\geq$.\\

Analogously, upon posing $\langle\mathbf{W}\rangle=\mathbf{Q}$, the expected value of the quantity $\text{Tr}[e^\mathbf{W}]=\text{Tr}[e^\mathbf{\Omega}]$ can be approximated as follows

\begin{widetext}
\begin{equation}\label{eq.w5}
\langle\text{Tr}\left[e^\mathbf{W}\right]\rangle=\sum_{k=0}^\infty\frac{\langle\text{Tr}\left[\mathbf{W}^k\right]\rangle}{k!}=\sum_{k=0}^\infty\frac{\text{Tr}\left[\langle\mathbf{W}^k\rangle\right]}{k!}\geq\sum_{k=0}^\infty\frac{\text{Tr}\left[\langle\mathbf{W}\rangle^k\right]}{k!}=\sum_{k=0}^\infty\frac{\text{Tr}\left[\mathbf{Q}^k\right]}{k!}=\text{Tr}\left[e^\mathbf{Q}\right],
\end{equation}
\end{widetext}
a relationship leading to $\langle\text{Tr}\left[e^\mathbf{\Omega}\right]\rangle\geq\text{Tr}\left[e^\mathbf{\Psi}\right]$, where $\mathbf{\Psi}$ is the matrix obtained upon diagonalising $\mathbf{Q}$.

The inequality can be understood upon considering a weighted, reciprocated dyad and noticing that relationships like $\langle[\mathbf{W}^4]_{ii}\rangle=\langle w_{ij}w_{ji}w_{ij}w_{ji}\rangle=\langle (w_{ij}w_{ji})^2\rangle=\langle w_{ij}w_{ji}\rangle^2+\text{Var}[w_{ij}w_{ji}]=\langle w_{ij}\rangle^2\langle w_{ji}\rangle^2+\text{Var}[w_{ij}w_{ji}]\geq\langle w_{ij}\rangle^2\langle w_{ji}\rangle^2=[\mathbf{Q}]^4_{ii}$ hold true; as in the binary case, estimating the total weight of closed walks of a certain length via the delta method implies overweighing the edges constituting them. Such a mismatch is absent if no link is reciprocated, as evident upon considering a weighted, square loop. Hereby, we will assume that the symbol $\gtrsim$ can replace the symbol $\geq$.

\subsubsection{Expected value of the spectral radius}

Let us now recall the statement of the generalised Perron-Frobenius (GPF) theorem~\cite{perron1907theorie,frobenius1912matrizen}.\\

\textbf{GPF Theorem.} \emph{Whenever non-negative, irreducible matrices are considered, a unique, real, positive eigenvalue exists whose modulus is maximum and (only) the corresponding left and right eigenvectors have positive components}.\\

Requiring irreducibility, sometimes stated as regularity, implies requiring the existence of a natural number $n$ such that $[\mathbf{A}^n]_{ij}>0$, $\forall\:i,j$. In other words, when directed networks are considered, requiring irreducibility is equivalent to requiring strongly connectedness. In case such a requirement is not satisfied, the Perron-Frobenius theorem must be weakened as follows.\\

\textbf{WPF Theorem.} \emph{Whenever non-negative matrices are considered, a real, non-negative eigenvalue exists whose modulus is maximum and with associated, non-negative left and right eigenvectors}.\\

The eigenvalue mentioned in any variant of the Perron-Frobenius theorem will be referred to as the \emph{principal eigenvalue} or \emph{spectral radius}. The relationship between the matrices $\mathbf{A}$ and $\mathbf{\Lambda}$ encoded into eq. (\ref{eq.a2}) can be further simplified upon noticing that, in case the spectral radius exists, is unique\footnote{A reducible square matrix $\mathbf{M}$ can be written in a block triangular form~\cite{varga1962iterative}, each matrix $\mathbf{B}_{ii}$ on the diagonal being either irreducible or zero. As the spectrum of such a matrix is the union of the spectra of the $\mathbf{B}_{ii}$s, the GPF Theorem can be applied to each $\mathbf{B}_{ii}$: the Perron–Frobenius eigenvalue of $\mathbf{M}$ is, thus, the largest of the Perron–Frobenius eigenvalues of the $\mathbf{B}_{ii}$s, hence coinciding with the one of the maximal strongly-connected component of the network under study.} and the spectral gap is (much) larger than zero\footnote{Although the condition $\lambda_1-\lambda_2\gg0$ can be relaxed, the formulas provided in the present paper hold for this case.}, the sum $\text{Tr}\left[e^\mathbf{\Lambda}\right]=\sum_{i=1}^Ne^{\lambda_i}$ is exponentially dominated by the addendum $e^{\lambda_1}$, an observation allowing us to write

\begin{equation}\label{eq.a6}
\text{Tr}\left[e^\mathbf{A}\right]\gtrsim e^{\lambda_1};
\end{equation}
analogously, 

\begin{equation}\label{eq.w6}
\text{Tr}\left[e^\mathbf{W}\right]\gtrsim e^{\omega_1}.
\end{equation}

Let us now inspect the relationships between eqs.~(\ref{eq.a5}) and~(\ref{eq.a6}) and between eqs.~(\ref{eq.w5}) and~(\ref{eq.w6}). Putting everything together, we obtain

\begin{widetext}
\begin{align}\label{eq.ap}
\langle e^{\lambda_1}\rangle\lesssim\langle\text{Tr}\left[e^\mathbf{A}\right]\rangle=\sum_{k=0}^\infty\frac{\langle\text{Tr}\left[\mathbf{A}^k\right]\rangle}{k!}&=\sum_{k=0}^\infty\frac{\text{Tr}\left[\langle\mathbf{A}^k\rangle\right]}{k!}\gtrsim\sum_{k=0}^\infty\frac{\text{Tr}\left[\mathbf{P}^k\right]}{k!}=\text{Tr}\left[e^\mathbf{P}\right]\gtrsim e^{\pi_1},\\
\langle e^{\omega_1}\rangle\lesssim\langle\text{Tr}\left[e^\mathbf{W}\right]\rangle=\sum_{k=0}^\infty\frac{\langle\text{Tr}\left[\mathbf{W}^k\right]\rangle}{k!}&=\sum_{k=0}^\infty\frac{\text{Tr}\left[\langle\mathbf{W}^k\rangle\right]}{k!}\gtrsim\sum_{k=0}^\infty\frac{\text{Tr}\left[\mathbf{Q}^k\right]}{k!}=\text{Tr}\left[e^\mathbf{Q}\right]\gtrsim e^{\phi_1}.
\end{align}
\end{widetext}

The two chains of (in-)equalities above motivate us to explore the possibility of deriving an (approximated) expression for the expected value of the spectral radius. According to the delta method, the expected value of a function, $f$, of a random variable, $x$, can be computed by Taylor-expanding $f(x)$ around $\langle x\rangle=\mu$, taking the expected value of the resulting expression and retaining only the lowest order of the expansion. Such a prescription allows us to write $\langle e^{\lambda_1}\rangle\simeq e^{\langle\lambda_1\rangle}$ and $\langle e^{\omega_1}\rangle\simeq e^{\langle\omega_1\rangle}$, two positions further leading to the results

\begin{equation}\label{eq.ap2}
\langle\lambda_1\rangle\simeq\pi_1
\end{equation}
and

\begin{equation}\label{eq.wq2}
\langle\omega_1\rangle\simeq\phi_1.
\end{equation}

Equations~(\ref{eq.ap2}) and~(\ref{eq.wq2}) are the main result of our paper, as they establish a (fundamental, although approximated) relationship between the empirical value of the spectral radius of a directed network, be it binary or weighted, and its expected counterpart: in words, the delta method suggests us to identify the latter with the spectral radius of the matrix defining the chosen random network model. Since the calculation of the expected number, or of the expected weight, of walks boils down to calculate the spectral radius of a single matrix, i.e. $\mathbf{P}$ or $\mathbf{Q}$, eqs.~(\ref{eq.ap2}) and~(\ref{eq.wq2}) have deep implications from a purely computational point of view as well: in fact, they prevent the network ensemble induced by $\mathbf{P}$ or $\mathbf{Q}$ from being explicitly sampled.

\subsubsection{Variance of the spectral radius}

Now, let us focus on the variance of the spectral radius calculation. To this aim, we will move from the known expressions of $\langle\lambda_1\rangle$, treating them as subject to statistical variability. For instance, let us recall that 

\begin{equation}
\pi_1\simeq\frac{\bra{k}\ket{k}}{2L}=\sum_{i=1}^N\frac{k_i^2}{2L}
\end{equation}
for binary, undirected networks under the Chung-Lu model, according to which $p_{ij}=k_ik_j/2L$, $\forall\:i,j$; upon considering that all quantities defining such an expression are random variables themselves, one is led to write

\begin{equation}
\text{Var}[\lambda_1]=\text{Var}\left[\sum_{i=1}^N\frac{k_i^2}{2L}\right]
\end{equation}
and evaluate such an expression either analytically or numerically. In what follows, we will numerically evaluate the spectral radius variance of our random network models.

\subsubsection{Statistical significance of the spectral radius}

Let us now define the quantity to be inspected for spotting the presence of a spectral signature of structural changes: it reads

\begin{equation}
z[\lambda_1]=\frac{\lambda_1-\langle\lambda_1\rangle}{\sigma[\lambda_1]}\simeq\frac{\lambda_1-\pi_1}{\sigma[\lambda_1]}
\end{equation}
and is nothing but the $z$-score of the spectral radius $\lambda_1$. As already stressed, the statistical meaning of such a quantity is guaranteed by the Gaussianity of the quantity whose $z$-score is to be calculated. Such a property of the spectral radius is guaranteed by the analytical results obtained in~\cite{dionigi2023central} and by the numerical checks carried out in Appendix~\hyperlink{AppD}{D} and depicted in fig.~\ref{fig:11}.

\section{Random network models}

Let us now discuss a set of null models to be employed for the subsequent steps of our analysis. To this aim, we will consider some members of the family of Exponential Random Graph Models (ERGMs), i.e. the entropy-based benchmarks that preserve a given set of constraints, otherwise being maximally random. More specifically, we follow the approach introduced in~\cite{park2004statistical} and further developed in~\cite{squartini2011analytical}, which prescribes to carry out a constrained maximisation of Shannon entropy

\begin{align}
S=-\sum_\mathbf{G}P(\mathbf{G})\ln P(\mathbf{G}),
\end{align}
the sum running over the ensemble $\mathbb{G}$ of $N\times N$ directed networks, be they binary (in which case $\mathbf{G}\equiv\mathbf{A}$) or weighted (in which case $\mathbf{G}\equiv\mathbf{W}$).

\subsection{Erd\"os-R\'enyi Model}

The Erd\"os-R\'enyi Model~\cite{squartini2011analytical} is induced by the Hamiltonian 

\begin{align}
H(\mathbf{A})=\alpha L(\mathbf{A}),
\end{align}
where $L(\mathbf{A})=\sum_{i=1}^N\sum_{j(\neq i)}a_{ij}$ represents the total number of directed edges, and $\alpha$ is the Lagrange multiplier associated with such a global constraint. The probability of the generic configuration $\mathbf{A}$ reads

\begin{align}
P_\text{ER}(\mathbf{A})=p^{L(\mathbf{A})}(1-p)^{N(N-1)-L(\mathbf{A})}
\end{align}
where $p=e^{-\alpha}/(1+e^{-\alpha})$ is the probability that a link points from node $i$ towards node $j$.

In order to tune the unknown parameter defining the Erd\"os-R\'enyi Model to ensure that $\langle L\rangle_\text{ER}=L(\mathbf{A}^*)$, we maximise the likelihood function $\mathcal{L}_\text{ER}=\ln P_\text{ER}(\mathbf{A}^*)$ with respect to it. Such a recipe leads us to find

\begin{align}
p=\frac{L(\mathbf{A}^*)}{N(N-1)},\quad\forall\:i\neq j
\end{align}
with obvious meaning of the symbols.

\subsubsection{Expected value of the spectral radius}

Although eq.~(\ref{eq.ap2}) provides a general recipe for estimating the expected value of the spectral radius of any random network model, a more explicit expression can be derived for the Erd\"os-R\'enyi Model. Specifically, let us consider the following equation

\begin{equation}
\sum_{k=0}^\infty\frac{\text{Tr}\left[\mathbf{P}^k\right]}{k!}=N+\sum_{k=2}^\infty\frac{(Np)^k}{k!}
\end{equation}
where $\mathbf{P}\equiv\mathbf{P}_\text{ER}=\{p_{ij}\}_{i,j=1}^N$, $p_{ij}\equiv p$, $\forall\:i\neq j$ and each addendum encodes the information about the order of magnitude of the specific contribution to the total number of cycles - to see this explicitly, let us consider that $\text{Tr}\left[\mathbf{A}^2\right]=\sum_{i=1}^N\left[\mathbf{A}^2\right]_{ii}=\sum_{i=1}^N\sum_{j(\neq i)}a_{ij}a_{ji}$ whose expected value reads $\langle\text{Tr}\left[\mathbf{A}^2\right]\rangle=\sum_{i=1}^N\sum_{j(\neq i)}\langle a_{ij}a_{ji}\rangle=\sum_{i=1}^N\sum_{j(\neq i)}p^2\simeq(Np)^2$ and analogously for the higher orders of the expansion. As adding and subtracting $1$ and $Np$ leads to

\begin{align}
\sum_{k=0}^\infty\frac{\text{Tr}\left[\mathbf{P}^k\right]}{k!}&=\sum_{k=0}^\infty\frac{(Np)^k}{k!}+N(1-p)-1\nonumber\\
&=e^{Np}+N(1-p)-1\gtrsim e^{Np}\simeq e^{\langle k\rangle},
\end{align}
eq.~(\ref{eq.ap}) can be employed to derive the chain of relationships

\begin{equation}
\pi_1\simeq Np\simeq\langle k\rangle,
\end{equation}
stating that the spectral radius, $\pi_1$, of the $N\times N$ matrix of i.i.d. Bernoulli random variables $\mathbf{P}\equiv\mathbf{P}_\text{ER}$ can be accurately approximated by their sum along any row or any column; in network terms, this can be rephrased by saying that the expected value of the spectral radius under the Erd\"os-R\'enyi Model coincides with the expected value of the degree of each node.\\

A second way of identifying $\pi_1$ rests upon the following relationship:

\begin{equation}
\mathbf{P}\cdot\mathbf{1}=(N-1)p\cdot\mathbf{1}=\langle k\rangle\cdot\mathbf{1};
\end{equation}
since $\mathbf{P}$ obeys the GPF Theorem, the equation above allows us to identify the value of its spectral radius quite straightforwardly by posing

\begin{equation}
\pi_1=(N-1)p=\langle k\rangle\equiv\lambda_1^\text{ER}.
\end{equation}

Such a result is consistent with the one stating that the spectral radius of the deterministic matrix $a_{ii}\equiv\nu$, $\forall\:i=j$ and $a_{ij}\equiv\mu$, $\forall\:i\neq j$ is equal to $\lambda_1=(N-1)\mu+\nu$.\\

A third way of identifying the expected value of $\lambda_1$ rests upon the results from~\cite{furedi1981eigenvalues}, i.e.

\begin{equation}\label{eq.furkom}
\lambda_1=\sum_{i=1}^N\sum_j\frac{a_{ij}}{N}+\frac{\sigma^2}{\mu},
\end{equation}
where $a_{ii}\equiv\nu$, $\forall\:i=j$, $\langle a_{ij}\rangle=\mu$ and $\text{Var}[a_{ij}]=\sigma^2$, $\forall\:i\neq j$. Since, in our case, $\nu=0$, $\mu=p$ and $\sigma^2=p(1-p)$, $\forall\:i\neq j$, such an expression leads to

\begin{equation}
\overline{\lambda_1}=\sum_{i=1}^N\sum_j\frac{p}{N}+\frac{\sigma^2}{\mu}=(N-1)p+(1-p).
\end{equation}

\subsubsection{Variance of the spectral radius}

Equation~(\ref{eq.furkom}) offers a straightforward way to calculate the variance of the spectral radius. It is, in fact, enough to evaluate the expression

\begin{align}
\text{Var}[\lambda_1]=\sum_{i=1}^N\sum_{j(\neq i)}\frac{\text{Var}[a_{ij}]}{N^2}&\simeq p(1-p)\equiv\text{Var}[\lambda_1^\text{ER}]
\end{align}
with the symbol $\simeq$ replacing the more correct expression $\lim_{N\to\infty} N(N-1)p(1-p)/N^2=p(1-p)$, indicating that $\text{Var}[\lambda_1]$ tends to $p(1-p)$ in the (asymptotic) regime $N\to\infty$.

\subsection{Binary Configuration Model}

The Binary Configuration Model~\cite{squartini2011analytical} is induced by the Hamiltonian 

\begin{align}
H(\mathbf{A})=\sum_{i=1}^N[\alpha_ik_i(\mathbf{A})+\beta_ih_i(\mathbf{A})]
\end{align}
where $k_i(\mathbf{A})=\sum_{j(\neq i)}a_{ij}$ represents the out-degree of node $i$, i.e. the number of nodes pointed by it and $h_i(\mathbf{A})=\sum_{j(\neq i)}a_{ji}$ represents the in-degree of node $i$, i.e. the number of nodes it is pointed by; the vectors $\{\alpha_i\}_{i=1}^N$ and $\{\beta_i\}_{i=1}^N$ represent the Lagrange multipliers associated with those above, local constraints. The probability of the generic configuration $\mathbf{A}$ reads

\begin{align}
P_\text{BCM}(\mathbf{A})=\prod_{i=1}^N\prod_{j(\neq i)}p_{ij}^{a_{ij}}(1-p_{ij})^{1-a_{ij}}
\end{align}
where $p_{ij}=e^{-\alpha_i-\beta_j}/(1+e^{-\alpha_i-\beta_j})$ is the probability that a link points from node $i$ towards node $j$.

To tune the unknown parameters defining the Binary Configuration Model to ensure that $\langle k_i\rangle_\text{BCM}=k_i(\mathbf{A}^*)$, $\forall\:i$ and $\langle h_i\rangle_\text{BCM}=h_i(\mathbf{A}^*)$, $\forall\:i$, we maximise the likelihood function $\mathcal{L}_\text{BCM}=\ln P_\text{BCM}(\mathbf{A}^*)$ with respect to them. Such a recipe leads us to solve

\begin{align}
k_i(\mathbf{A}^*)&=\sum_{j(\neq i)}\frac{e^{-\alpha_i-\beta_j}}{1+e^{-\alpha_i-\beta_j}},\quad\:\forall\:i\\
h_i(\mathbf{A}^*)&=\sum_{j(\neq i)}\frac{e^{-\alpha_j-\beta_i}}{1+e^{-\alpha_j-\beta_i}},\quad\:\forall\:i
\end{align}
with obvious meaning of the symbols.

\subsubsection{Expected value of the spectral radius}

According to eq.~(\ref{eq.ap2}), $\pi_1$ is the spectral radius of the $N\times N$ matrix of i.n.i.d. random variables $\mathbf{P}\equiv\mathbf{P}_\text{BCM}=\{p_{ij}\}_{i,j=1}^N$, with $p_{ij}=e^{-\alpha_i-\beta_j}/(1+e^{-\alpha_i-\beta_j})$, $\forall\:i\neq j$.\\

As for the Erd\"os-R\'enyi Model, a more explicit expression can also be derived for the Binary Configuration Model. To this aim, let us consider that a way to identify $\pi_1$ in case $p_{ij}=k_ih_j/L$, $\forall\:i,j$ rests upon the relationship

\begin{equation}
\mathbf{P}=\frac{\mathbf{k}\otimes\textbf{h}}{L}=\frac{\ket{k}\bra{h}}{L},
\end{equation}
indicating that the matrix $\mathbf{P}$ characterising the Binary Configuration Model can be obtained as the direct product of the vector of out-degrees, $\mathbf{k}$, and the vector of in-degrees, $\mathbf{h}$. Employing the bra-ket formalism allows the calculations to be carried out quite easily, as

\begin{equation}
\mathbf{P}\ket{k}=\frac{\ket{k}\bra{h}}{L}\ket{k}=\frac{\bra{h}\ket{k}}{L}\ket{k}
\end{equation}
where $\bra{h}\ket{k}=\sum_{i=1}^Nk_ih_i$. Since $\mathbf{P}$ obeys the GPF Theorem, the equation above allows us to identify the value of its spectral radius\footnote{Notice that $\bra{h}\mathbf{P}=\bra{h}\frac{\ket{k}\bra{h}}{L}=\bra{h}\frac{\bra{h}\ket{k}}{L}$ as well.} quite straightforwardly as $\pi_1=\bra{h}\ket{k}/L=\sum_{i=1}^Nk_ih_i/L$. The sparse-case approximation of the Binary Configuration Model is, however, defined by the position $p_{ij}=k_ih_j/L$, $\forall\:i\neq j$, a piece of evidence leading us to write

\begin{equation}
\pi_1\simeq\frac{\bra{h}\ket{k}}{L}=\sum_{i=1}^N\frac{k_ih_i}{L}\equiv\lambda_1^\text{CL}.
\end{equation}

\subsubsection{Variance of the spectral radius}

The expression $\pi_1=\bra{h}\ket{k}/L=\sum_{i=1}^Nk_ih_i/L$ offers a straightforward way to calculate the variance of the spectral radius. Upon considering that all quantities defining such an expression are random variables themselves, one is led to write

\begin{equation}
\text{Var}[\lambda_1]=\text{Var}\left[\sum_{i=1}^N\frac{k_ih_i}{L}\right]\equiv\text{Var}[\lambda_1^\text{CL}]
\end{equation}
and evaluate such an expression either analytically or numerically. In what follows, we will proceed by evaluating it numerically.

\subsection{Reciprocal Configuration Model}

The Reciprocal Configuration Model~\cite{squartini2011analytical} is induced by the Hamiltonian 

\begin{align}
H(\mathbf{A})=\sum_{i=1}^N[\alpha_ik_i^\rightarrow(\mathbf{A})+\beta_ik_i^\leftarrow(\mathbf{A})+\gamma_ik_i^\leftrightarrow(\mathbf{A})]
\end{align}
where $k_i^\rightarrow(\mathbf{A})=\sum_{j(\neq i)}a_{ij}^\rightarrow$ represents the non-reciprocated out-degree of node $i$, $k_i^\leftarrow(\mathbf{A})=\sum_{j(\neq i)}a_{ij}^\leftarrow$ represents the non-reciprocated in-degree of node $i$ and $k_i^\leftrightarrow(\mathbf{A})=\sum_{j(\neq i)}a_{ij}^\leftrightarrow$ represents the reciprocated degree of node $i$; the vectors $\{\alpha_i\}_{i=1}^N$, $\{\beta_i\}_{i=1}^N$ and $\{\gamma_i\}_{i=1}^N$ represent the Lagrange multipliers associated with those above, local constraints. The probability of the generic configuration $\mathbf{A}$ reads

\begin{align}
P_\text{RCM}(\mathbf{A})=\prod_{i=1}^N\prod_{j(>i)}(p_{ij}^\rightarrow)^{a_{ij}^\rightarrow}(p_{ij}^\leftarrow)^{a_{ij}^\leftarrow}(p_{ij}^\leftrightarrow)^{a_{ij}^\leftrightarrow}(p_{ij}^\times)^{a_{ij}^\times}
\end{align}
where

\begin{equation}
p_{ij}^\rightarrow=\frac{e^{-\alpha_i-\beta_j}}{1+e^{-\alpha_i-\beta_j}+e^{-\alpha_j-\beta_i}+e^{-\gamma_i-\gamma_j}}
\end{equation}
is the probability that a non-reciprocated link points from node $i$ towards node $j$,

\begin{equation}
p_{ij}^\leftarrow=\frac{e^{-\alpha_j-\beta_i}}{1+e^{-\alpha_i-\beta_j}+e^{-\alpha_j-\beta_i}+e^{-\gamma_i-\gamma_j}}
\end{equation}
is the probability that a non-reciprocated link points from node $j$ towards node $i$,

\begin{equation}
p_{ij}^\leftrightarrow=\frac{e^{-\gamma_i-\gamma_i}}{1+e^{-\alpha_i-\beta_j}+e^{-\alpha_j-\beta_i}+e^{-\gamma_i-\gamma_j}}
\end{equation}
is the probability that nodes $i$ and $j$ are connected by a reciprocated link and $p_{ij}^\times=1-p_{ij}^\rightarrow-p_{ij}^\leftarrow-p_{ij}^\leftrightarrow$ is the probability that $i$ and $j$ are disconnected.

To tune the unknown parameters defining the Reciprocal Configuration Model to ensure that $\langle k_i^\rightarrow\rangle_\text{RCM}=k_i^\rightarrow(\mathbf{A}^*)$, $\forall\:i$, $\langle k_i^\leftarrow\rangle_\text{RCM}=k_i^\leftarrow(\mathbf{A}^*)$, $\forall\:i$ and $\langle k_i^\leftrightarrow\rangle_\text{RCM}=k_i^\leftrightarrow(\mathbf{A}^*)$, $\forall\:i$, we maximise the likelihood function $\mathcal{L}_\text{RCM}=\ln P_\text{RCM}(\mathbf{A}^*)$ with respect to them. Such a recipe leads us to solve

\begin{align}
k_i^\rightarrow(\mathbf{A}^*)&=\sum_{j(\neq i)}\frac{e^{-\alpha_i-\beta_j}}{1+e^{-\alpha_i-\beta_j}+e^{-\alpha_j-\beta_i}+e^{-\gamma_i-\gamma_j}},\quad\:\forall\:i\\
k_i^\leftarrow(\mathbf{A}^*)&=\sum_{j(\neq i)}\frac{e^{-\alpha_j-\beta_i}}{1+e^{-\alpha_i-\beta_j}+e^{-\alpha_j-\beta_i}+e^{-\gamma_i-\gamma_j}},\quad\:\forall\:i\\
k_i^\leftrightarrow(\mathbf{A}^*)&=\sum_{j(\neq i)}\frac{e^{-\gamma_i-\gamma_i}}{1+e^{-\alpha_i-\beta_j}+e^{-\alpha_j-\beta_i}+e^{-\gamma_i-\gamma_j}},\quad\:\forall\:i
\end{align}
with obvious meaning of the symbols.

\subsubsection{Expected value of the spectral radius}

According to eq.~(\ref{eq.ap2}), $\pi_1$ is the spectral radius of the $N\times N$ matrix of i.n.i.d. random variables $\mathbf{P}\equiv\mathbf{P}_\text{RCM}=\{p_{ij}\}_{i,j=1}^N$, $p_{ij}=p_{ij}^\rightarrow+p_{ij}^\leftrightarrow=(e^{-\alpha_i-\beta_j}+e^{-\gamma_i-\gamma_i})/(1+e^{-\alpha_i-\beta_j}+e^{-\alpha_j-\beta_i}+e^{-\gamma_i-\gamma_j})$, $\forall\:i\neq j$.\\

As for the Binary Configuration Model, more explicit expressions can also be derived for the Reciprocal Configuration Model. To this aim, let us consider that, in the sparse case, one can write

\begin{align}
\mathbf{P}^\rightarrow\ket{k^\rightarrow}&=\frac{\ket{k^\rightarrow}\bra{k^\leftarrow}}{L^\rightarrow}\ket{k^\rightarrow}=\frac{\bra{k^\leftarrow}\ket{k^\rightarrow}}{L^\rightarrow}\ket{k^\rightarrow},\\
\mathbf{P}^\leftrightarrow\ket{k^\leftrightarrow}&=\frac{\ket{k^\leftrightarrow}\bra{k^\leftrightarrow}}{2L^\leftrightarrow}\ket{k^\leftrightarrow}=\frac{\bra{k^\leftrightarrow}\ket{k^\leftrightarrow}}{2L^\leftrightarrow}\ket{k^\leftrightarrow}
\end{align}
where $\bra{k^\leftarrow}\ket{k^\rightarrow}=\sum_{i=1}^Nk^\leftarrow_ik^\rightarrow_i$
and $\bra{k^\leftrightarrow}\ket{k^\leftrightarrow}=\sum_{i=1}^Nk^\leftrightarrow_ik^\leftrightarrow_i$. Since $\mathbf{P}^\rightarrow$
and $\mathbf{P}^\leftrightarrow$ obey the GPF Theorem, the equations above allow us to identify the values of their spectral radius\footnote{An analogous observation to the one in the previous footnote can be made.} quite straightforwardly as

\begin{align}
\pi_1^\rightarrow&\simeq\frac{\bra{k^\leftarrow}\ket{k^\rightarrow}}{L^\rightarrow}=\sum_{i=1}^N\frac{k^\leftarrow_ik^\rightarrow_i}{L^\rightarrow}\equiv\lambda_1^{\text{CL}^\rightarrow},\\
\pi_1^\leftrightarrow&\simeq\frac{\bra{k^\leftrightarrow}\ket{k^\leftrightarrow}}{2L^\leftrightarrow}=\sum_{i=1}^N\frac{k^\leftrightarrow_ik^\leftrightarrow_i}{2L^\leftrightarrow}\equiv\lambda_1^{\text{CL}^\leftrightarrow}
\end{align}
(because of the definition of the sparse-case approximation of the Reciprocal Configuration Model, valid $\forall\:i\neq j$).

\subsubsection{Variance of the spectral radius}

The expressions above offer a straightforward way to calculate the corresponding variances. In fact, one is led to write

\begin{align}
\text{Var}[\lambda_1^\rightarrow]&=\text{Var}\left[\sum_{i=1}^N\frac{k^\leftarrow_ik^\rightarrow_i}{L^\rightarrow}\right]\equiv\text{Var}[\lambda_1^{\text{CL}^\rightarrow}],\\
\text{Var}[\lambda_1^\leftrightarrow]&=\text{Var}\left[\sum_{i=1}^N\frac{k^\leftrightarrow_ik^\leftrightarrow_i}{2L^\leftrightarrow}\right]\equiv\text{Var}[\lambda_1^{\text{CL}^\leftrightarrow}]
\end{align}
and evaluate such expressions either analytically or numerically. In what follows, we will proceed by evaluating them numerically.

\subsection{Global Reciprocity Model}

The Global Reciprocity Model~\cite{picciolo2012role} is a special case of the Reciprocal Configuration Model, induced by the Hamiltonian

\begin{align}
H(\mathbf{A})=\sum_{i=1}^N[\alpha_ik_i(\mathbf{A})+\beta_ih_i(\mathbf{A})]+\gamma L^\leftrightarrow(\mathbf{A})
\end{align}
where $L^\leftrightarrow(\mathbf{A})=\sum_{i=1}^N\sum_{j(\neq i)}a_{ij}^\leftrightarrow$ represents the total number of reciprocated links; the parameters $\{\alpha_i\}_{i=1}^N$, $\{\beta_i\}_{i=1}^N$ and $\gamma$ represent the Lagrange multipliers associated with the aforementioned constraints. The probability of the generic configuration $\mathbf{A}$ reads

\begin{align}
P_\text{GRM}(\mathbf{A})=\prod_{i=1}^N\prod_{j(>i)}(p_{ij}^\rightarrow)^{a_{ij}^\rightarrow}(p_{ij}^\leftarrow)^{a_{ij}^\leftarrow}(p_{ij}^\leftrightarrow)^{a_{ij}^\leftrightarrow}(p_{ij}^\times)^{a_{ij}^\times}
\end{align}
where

\begin{equation}
p_{ij}^\rightarrow=\frac{e^{-\alpha_i-\beta_j}}{1+e^{-\alpha_i-\beta_j}+e^{-\alpha_j-\beta_i}+e^{-\alpha_i-\beta_j-\beta_i-\alpha_j-\gamma}}
\end{equation}
is the probability that a non-reciprocated link points from node $i$ towards $j$,

\begin{equation}
p_{ij}^\leftarrow=\frac{e^{-\alpha_j-\beta_i}}{1+e^{-\alpha_i-\beta_j}+e^{-\alpha_j-\beta_i}+e^{-\alpha_i-\beta_j-\beta_i-\alpha_j-\gamma}}
\end{equation}
is the probability that a non-reciprocated link points from node $j$ towards node $i$,

\begin{equation}
p_{ij}^\leftrightarrow=\frac{e^{-\alpha_i-\beta_j-\beta_i-\alpha_j-\gamma}}{1+e^{-\alpha_i-\beta_j}+e^{-\alpha_j-\beta_i}+e^{-\alpha_i-\beta_j-\beta_i-\alpha_j-\gamma}}
\end{equation}
is the probability that nodes $i$ and $j$ are connected by a reciprocated link and $p_{ij}^\times=1-p_{ij}^\rightarrow-p_{ij}^\leftarrow-p_{ij}^\leftrightarrow$ is the probability that $i$ and $j$ are disconnected.

To tune the unknown parameters defining the Global Reciprocity Model to ensure that $\langle k_i\rangle_\text{GRM}=k_i(\mathbf{A}^*)$, $\forall\:i$, $\langle h_i\rangle_\text{GRM}=h_i(\mathbf{A}^*)$, $\forall\:i$ and $\langle L^\leftrightarrow\rangle_\text{GRM}=L^\leftrightarrow(\mathbf{A}^*)$, $\forall\:i$, we maximise the likelihood function $\mathcal{L}_\text{GRM}=\ln P_\text{GRM}(\mathbf{A}^*)$ with respect to them. Such a recipe leads us to solve

\begin{widetext}
\begin{align}
k_i(\mathbf{A}^*)&=\sum_{j(\neq i)}\frac{e^{-\alpha_i-\beta_j}+e^{-\alpha_i-\beta_j-\beta_i-\alpha_j-\gamma}}{1+e^{-\alpha_i-\beta_j}+e^{-\alpha_j-\beta_i}+e^{-\alpha_i-\beta_j-\beta_i-\alpha_j-\gamma}},\quad\:\forall\:i\\
h_i(\mathbf{A}^*)&=\sum_{j(\neq i)}\frac{e^{-\alpha_j-\beta_i}+e^{-\alpha_i-\beta_j-\beta_i-\alpha_j-\gamma}}{1+e^{-\alpha_i-\beta_j}+e^{-\alpha_j-\beta_i}+e^{-\alpha_i-\beta_j-\beta_i-\alpha_j-\gamma}},\quad\:\forall\:i\\
L^\leftrightarrow(\mathbf{A}^*)&=\sum_{i=1}^N\sum_{j(\neq i)}\frac{e^{-\alpha_i-\beta_j-\beta_i-\alpha_j-\gamma}}{1+e^{-\alpha_i-\beta_j}+e^{-\alpha_j-\beta_i}+e^{-\alpha_i-\beta_j-\beta_i-\alpha_j-\gamma}}
\end{align}
\end{widetext}
with obvious meaning of the symbols.

In the case of the Global Reciprocity Model, $\pi_1$ is the spectral radius of the $N\times N$ matrix of i.n.i.d. random variables $\mathbf{P}\equiv\mathbf{P}_\text{GRM}=\{p_{ij}\}_{i,j=1}^N$, $p_{ij}=p_{ij}^\rightarrow+p_{ij}^\leftrightarrow=(e^{-\alpha_i-\beta_j}+e^{-\alpha_i-\beta_j-\beta_i-\alpha_j-\gamma})/(1+e^{-\alpha_i-\beta_j}+e^{-\alpha_j-\beta_i}+e^{-\alpha_i-\beta_j-\beta_i-\alpha_j-\gamma})$, $\forall\:i\neq j$.

\subsection{Density-Corrected Gravity Model}

The density-corrected Gravity Model~\cite{cimini2015systemic} is a two-step model inducing a probability for the generic configuration $\mathbf{A}$ reading

\begin{align}
P_\text{dcGM}(\mathbf{A})=\prod_{i=1}^N\prod_{j(\neq i)}p_{ij}^{a_{ij}}(1-p_{ij})^{1-a_{ij}}
\end{align}
where

\begin{figure*}[t!]
\centering
\includegraphics[width=\linewidth]{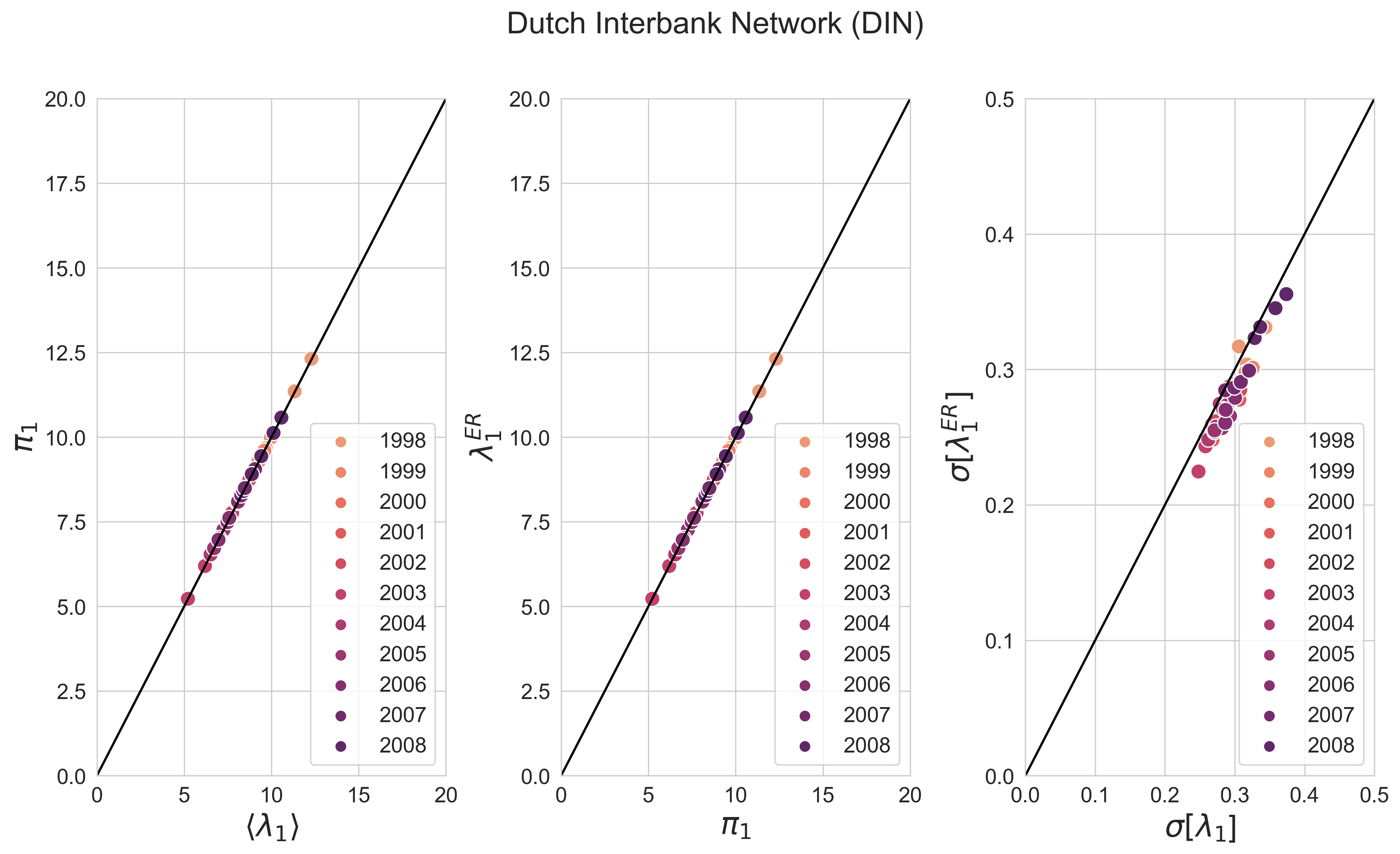}
\includegraphics[width=\linewidth]{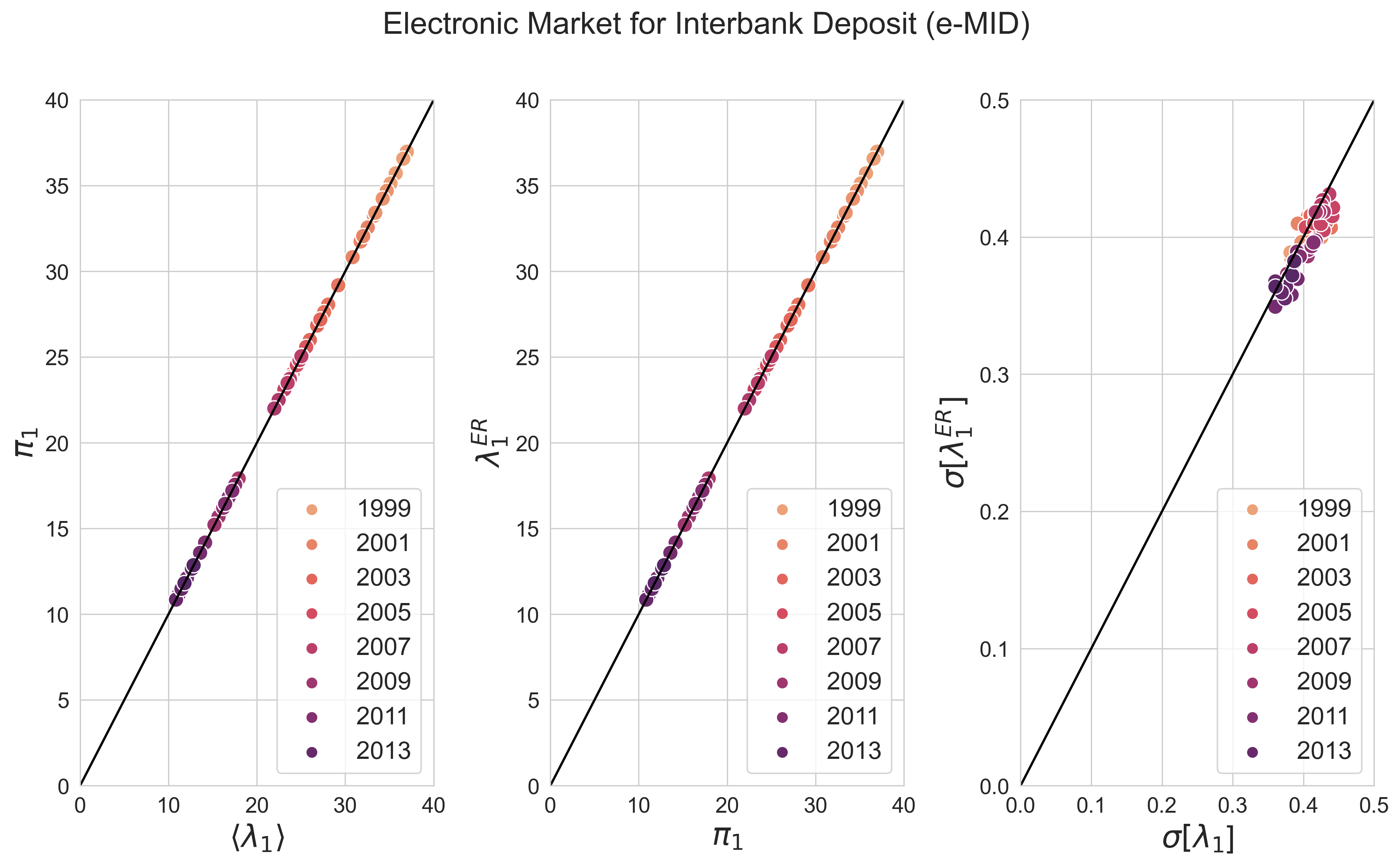}
\caption{Expected value and variance of the spectral radius for each of the quarters of the Dutch Interbank Network (DIN) and of the Electronic Market for Interbank Deposit (e-MID) according to the Erd\"os-R\'enyi Model. Left panels: the expected value of the spectral radius is very well approximated by the spectral radius of the matrix $\mathbf{P}=\{p\}_{i,j=1}^N$ characterising the Erd\"os-R\'enyi Model. Central panels: the spectral radius of the matrix $\mathbf{P}=\{p\}_{i,j=1}^N$ characterising the Erd\"os-R\'enyi Model, in turn, coincides with $\lambda_1^\text{ER}=(N-1)p=L/N$. Right panels: the variance of the spectral radius is slightly underestimated by the value $\text{Var}[\lambda_1^\text{ER}]=p(1-p)$.}
\label{fig:1}
\end{figure*}

\begin{equation}
p_{ij}=\frac{za_il_j}{1+za_il_j}
\end{equation}
is the probability that a link points from node $i$ towards node $j$ and $a_i=\sum_{j(\neq i)}w_{ij}$ is the out-strength of node $i$, $l_i=\sum_{j(\neq i)}w_{ji}$ is the in-strength of node $i$ and $z$ is a free parameter, determined by fixing the value of the total number of links\footnote{Analogously, one could have fixed the connectance, or link density, defined as $c=\frac{L}{N(N-1)}$.}, i.e. by solving the equation

\begin{equation}
L(\mathbf{A}^*)=\sum_{i=1}^N\sum_{j(\neq i)} \frac{za_il_j}{1+za_il_j}.
\end{equation}

The second step of the density-corrected Gravity Model, instead, is a conditional one, prescribing loading the link $a_{ij}=1$ with the value

\begin{equation}
w_{ij}=\frac{a_il_j}{Wp_{ij}},
\end{equation}
where $W=\sum_{i=1}^N\sum_{j(\neq i)}w_{ij}=\sum_{i=1}^Na_i=\sum_{i=1}^Nl_i$ is the total network volume. As a consequence of such a prescription, one recovers the result

\begin{equation}
\langle w_{ij}\rangle=\frac{a_il_j}{W};
\end{equation}
in other words, the dcGM ensures that the (financial equivalent of the) Gravity Model prescription is recovered on average.

\subsubsection{Expected value of the spectral radius}

According to eq.~(\ref{eq.ap2}), $\phi_1$ is the spectral radius of the $N\times N$ matrix of i.n.i.d. random variables $\mathbf{Q}\equiv\mathbf{Q}_\text{dcGM}=\{\langle w_{ij}\rangle\}_{i,j=1}^N$, $\langle w_{ij}\rangle=a_il_j/W$, $\forall\:i\neq j$.\\

As for the Binary Configuration Model, a more explicit expression can also be derived for the density-corrected Gravity Model. To this aim, let us consider that a way to identify $\phi_1$ in case $\langle w_{ij}\rangle=a_il_j/W$, $\forall\:i,j$ rests upon the relationship

\begin{equation}
\mathbf{Q}=\frac{\mathbf{a}\otimes\textbf{l}}{W}=\frac{\ket{a}\bra{l}}{W},
\end{equation}
indicating that the matrix $\mathbf{Q}$ characterising the dcGM can be obtained as the direct product of the vector of out-strengths, $\mathbf{a}$, and the vector of in-strengths, $\mathbf{l}$. Employing the bra-ket formalism allows the calculations to be carried out quite easily, as

\begin{equation}
\mathbf{Q}\ket{a}=\frac{\ket{a}\bra{l}}{W}\ket{a}=\frac{\bra{a}\ket{l}}{W}\ket{a}
\end{equation}
where $\bra{a}\ket{l}=\sum_{i=1}^Na_il_i$. Since $\mathbf{Q}$ obeys the GPF Theorem, the equation above allows us to identify the value of its spectral radius\footnote{Notice that $\bra{l}\mathbf{Q}=\bra{l}\frac{\ket{a}\bra{l}}{W}=\bra{l}\frac{\bra{a}\ket{l}}{W}$ as well.} quite straightforwardly as $\phi_1=\bra{a}\ket{l}/W=\sum_{i=1}^Na_il_i/W$. The density-corrected Gravity Model is, however, defined by the position $\langle w_{ij}\rangle=a_il_j/W$, $\forall\:i\neq j$, a piece of evidence leading us to write

\begin{equation}
\phi_1\simeq\frac{\bra{a}\ket{l}}{W}=\sum_{i=1}^N\frac{a_il_i}{W}\equiv\omega_1^\text{CL}.
\end{equation}

As the considered matrix is deterministic, the variance of its spectral radius is, by definition, zero.

\section{Data description}

\subsection{Dutch Interbank Network}

The Dutch Interbank Network (DIN) is represented as a binary, directed network whose nodes are anonymised, Dutch banks and links represent exposures (from contractual obligations to swaps) up to one year and larger than $1.5$ millions of euros. Data are reported quarterly from 1998Q1 to 2008Q4, hence consisting of 44 snapshots. Notice that the last four ends of quarters correspond to 2008, i.e. the first year of the global financial crisis~\cite{van2014finding}. Given the nature of the available data, a link pointing from bank $i$ to bank $j$ at time $t$ indicates the existence of a total exposure of more than 1.5 million euros, directed from $i$ to $j$, registered at the end of the particular quarter $t$.

\begin{figure*}[t!]
\centering
\includegraphics[width=\linewidth]{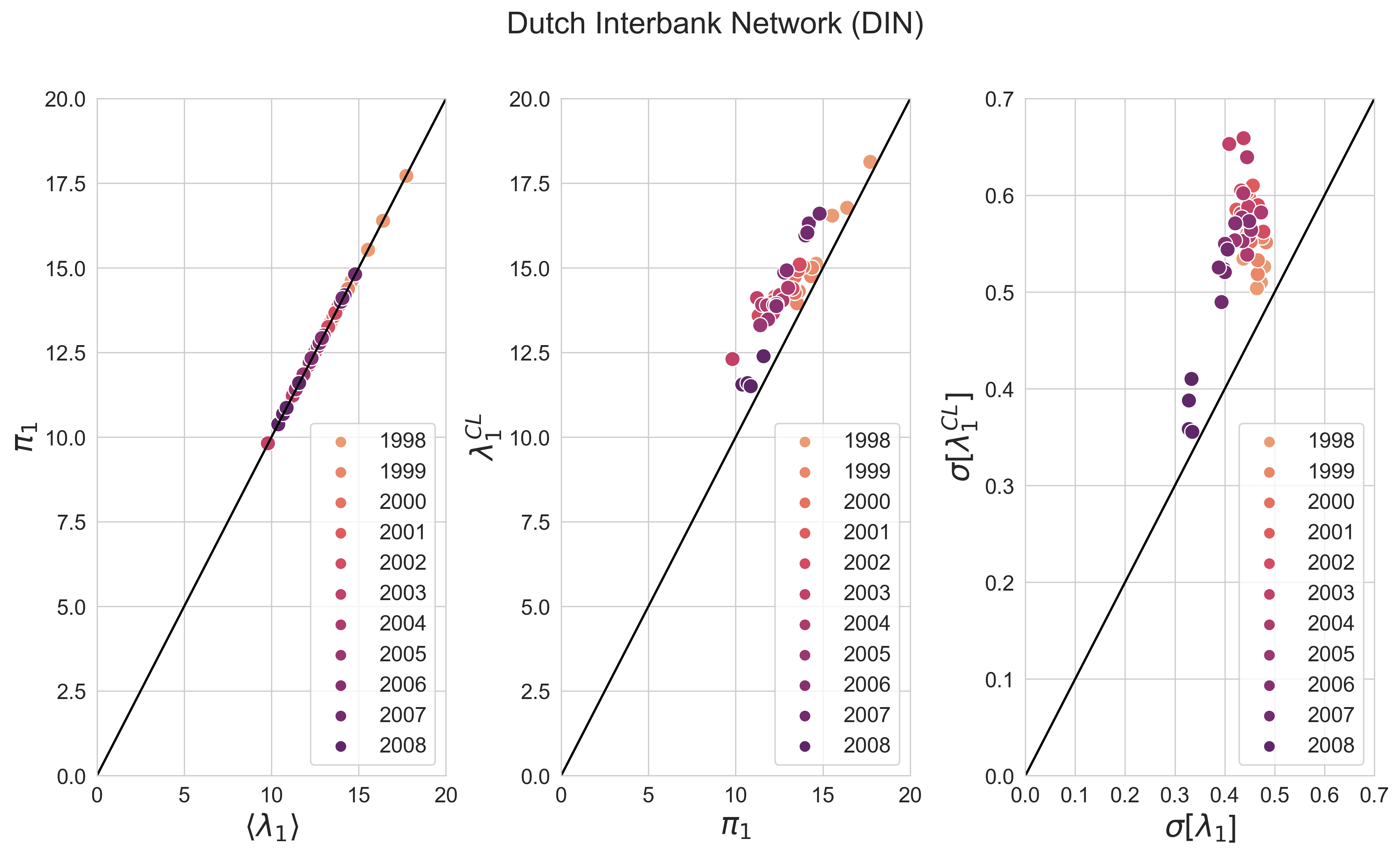}
\includegraphics[width=\linewidth]{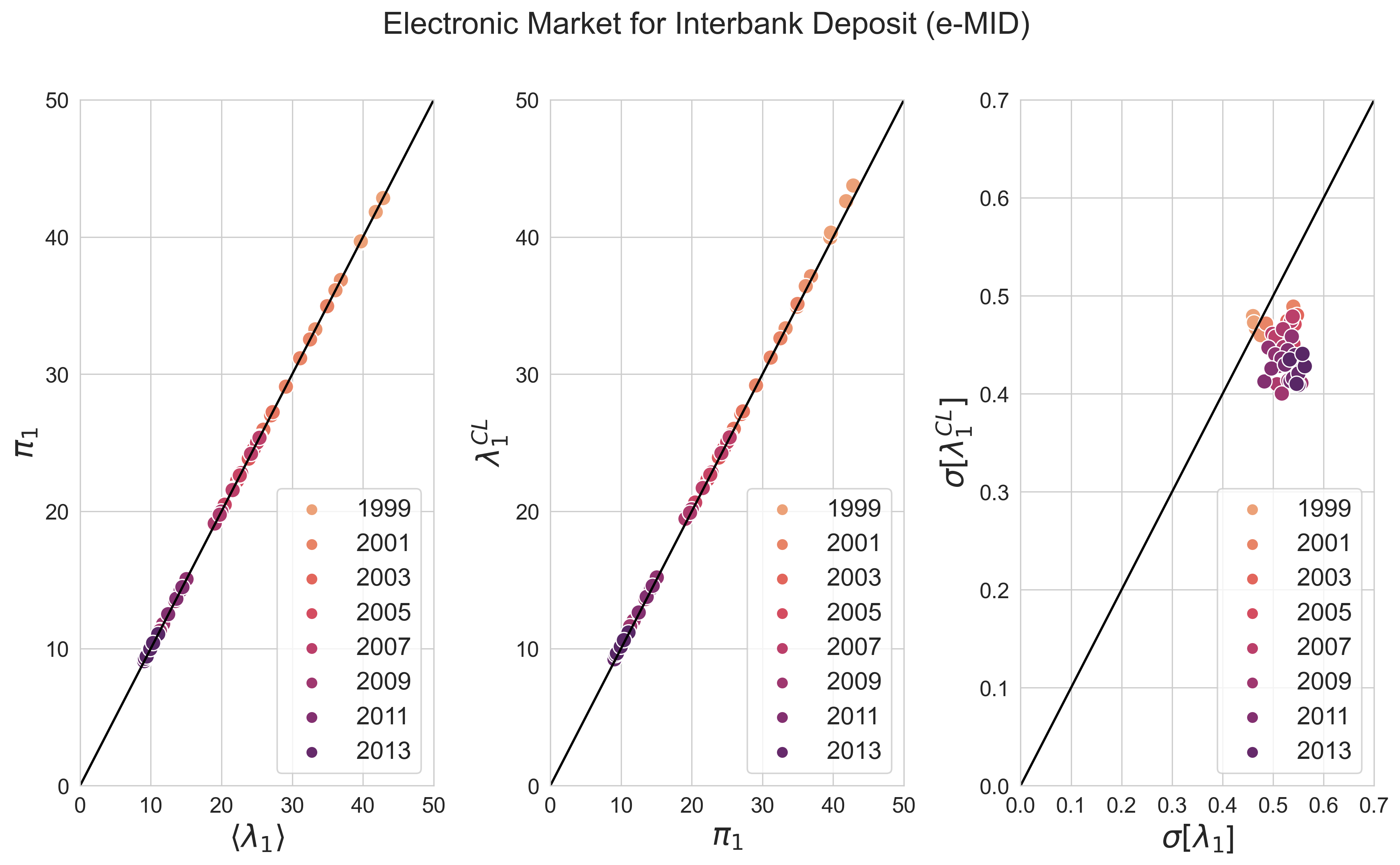}
\caption{Expected value and variance of the spectral radius for each of the quarters of the Dutch Interbank Network (DIN) and of the Electronic Market for Interbank Deposit (e-MID) according to the Binary Configuration Model. Left panels: the expected value of the spectral radius is very well approximated by the spectral radius of the matrix $\mathbf{P}=\{p_{ij}\}_{i,j=1}^N$ characterising the Binary Configuration Model. Central panels: the spectral radius of the matrix $\mathbf{P}=\{p_{ij}\}_{i,j=1}^N$ characterising the Binary Configuration Model is, overall, well approximated by $\lambda_1^\text{CL}=\sum_{i=1}^Nk_ih_i/L$. Right panels: the variance of the spectral radius is either overestimated or underestimated by the value $\text{Var}[\lambda_1^\text{CL}]=\text{Var}\left[\sum_{i=1}^Nk_ih_i/L\right]$.}
\label{fig:2}
\end{figure*}

\subsection{Electronic Market for Interbank Deposit}

The Electronic Market for Interbank Deposit (e-MID) is represented as a weighted, directed network whose nodes are anonymised, Italian banks and weights represent exposures in million euros\footnote{e-MID is a centralised interbank market for trading unsecured deposits, working as follows: a bank quotes an offer to lend or borrow money (minimum quote: $1.5$ million euros) at a certain maturity and interest rate; a second bank chooses (at least a part of) the quoted order (minimum quote: $50.000$ euros), and the trade is registered if and only if both counterparties have agreed on it. The following information is available for each active bank during the period: an anonymous ID identifying the bank and the country where it is legally settled. In~\cite{fricke2015core}, Fricke and Lux have highlighted \emph{i)} how the number of active, foreign banks largely varies over the considered period, experiencing a dramatic drop in correspondence of the Lehman-Brothers bailout; \emph{ii)} how the number of active Italian banks is quite stable over the period, although it decreases after the global financial crisis.}. Reported data cover the period January 1999-December 2014, on a daily frequency: a link with weight $w_{ij}$, pointing from bank $i$ to bank $j$ at time $t$ indicates the existence of the total exposure $w_{ij}\geq 50.000$ euros, directed from $i$ to $j$, registered at the end of the particular period $t$. Considering that $\simeq98\%$ of banks are Italian and that the volume of their transactions covers $\simeq85\%$ of the total volume (as of 2011), our analysis solely focuses on the subgraph induced by such a subset of nodes. We also examine all aggregation periods ranging from daily to yearly - although the figures will depict e-MID on a quarterly basis.

\subsection{International Trade Network}

The International Trade Network (ITN) is represented as a weighted, directed network whose nodes are countries and weights represent imports/exports in million euros. Data on yearly trade flows during the period 2000-2020 have been downloaded from the UN-COMTRADE website\footnote{\href{https://comtradeplus.un.org/}{https://comtradeplus.un.org/}}. To consistently compare data, a panel of 112 countries for which trade information was available for the entire period has been selected~\cite{guidi2024tracing}. Given the nature of the available data, a link whose weight is $w_{ij}$, pointing from country $i$ to country $j$ during the year $y$ indicates the existence of an exported amount of commodities whose value matches $w_{ij}$, directed from $i$ to $j$, during that year.

\section{Results}\label{sec:Results}

\subsection{Inspecting the accuracy of our approximations}

The derivation of our results rests upon several approximations whose accuracy must be explicitly checked case by case.\\

The first one concerns the expected value of the trace of the exponential of $\mathbf{A}$ - which has been proven to satisfy the relationship $\langle\text{Tr}\left[e^\mathbf{A}\right]\rangle\geq\text{Tr}\left[e^\mathbf{P}\right]$, hence being strictly larger than the trace of the exponential of $\mathbf{P}$ for any network with positive reciprocity, i.e. having $r=L^\leftrightarrow/L>0$. In order to check how close the two terms above are, we have explicitly computed the ratio $\text{Tr}\left[e^{\mathbf{P}}\right]/\langle\text{Tr}\left[e^{\mathbf{A}}\right]\rangle$ for all the snapshots of our systems. The results are reported in the seventh column of tables~\ref{tab:DIN} and~\ref{tab:eMID} in Appendix~\hyperlink{AppE}{E}. As evident, $\text{Tr}\left[e^{\mathbf{P}}\right]/\langle\text{Tr}\left[e^{\mathbf{A}}\right]\rangle\lesssim1$ irrespectively from the structural details of our configurations - in particular, even for configurations with a non-negligible level of reciprocity such as those constituting the DIN, for which $r\simeq 0.3$. In other words, the trace of the matrix $\mathbf{P}$ describing a random network model provides a quite accurate approximation of the expected value of the trace of the adjacency matrix $\mathbf{A}$ under the same model. As the 2008Q1, 2008Q2, 2008Q3 and 2008Q4 snapshots of the DIN confirm, the accuracy of the approximation above increases as $r$ decreases.

Analogously, $\text{Tr}\left[e^{\mathbf{Q}}\right]/\langle\text{Tr}\left[e^{\mathbf{W}}\right]\rangle\lesssim1$, as the seventh column of table~\ref{tab:ITN} in Appendix~\hyperlink{AppE}{E} shows.\\

The second one concerns the hypothesis that the trace of the exponential of $\mathbf{A}$ and the trace of the exponential of $\mathbf{P}$ are both dominated by their largest addendum, i.e. $\text{Tr}[e^\mathbf{A}]\gtrsim e^{\lambda_1}$ and $\text{Tr}[e^\mathbf{P}]\gtrsim e^{\pi_1}$. In order to check how close the two pairs above of terms are, we have explicitly computed the ratios $e^{\lambda_1}/\text{Tr}\left[e^{\mathbf{A}}\right]$ and $e^{\pi_1}/\text{Tr}\left[e^{\mathbf{P}}\right]$ for all the snapshots of our systems. The results are reported in the fifth and sixth columns of tables~\ref{tab:DIN} and~\ref{tab:eMID} in Appendix~\hyperlink{AppE}{E}. As evident, $e^{\lambda_1}/\text{Tr}\left[e^{\mathbf{A}}\right]\lesssim1$ and $e^{\pi_1}/\text{Tr}\left[e^{\mathbf{P}}\right]\lesssim1$ irrespectively from the structural details of our configurations. In words, the trace of the matrix $\mathbf{A}$ is exponentially dominated by the addendum $e^{\lambda_1}$ and the trace of the matrix $\mathbf{P}$ is exponentially dominated by the addendum $e^{\pi_1}$. The accuracy of the approximation remains steadily high.

Analogously, $e^{\omega_1}/\text{Tr}\left[e^{\mathbf{W}}\right]\lesssim1$ and $e^{\phi_1}/\text{Tr}\left[e^{\mathbf{Q}}\right]\lesssim1$, as the fifth and sixth column of table~\ref{tab:ITN} in Appendix~\hyperlink{AppE}{E} show.

\subsection{Expected value and variance of the\\spectral radius}

After having checked the goodness of our approximations, let us investigate the accuracy of the estimations of the expected value and variance of the spectral radius of our random network models.\\

\paragraph*{Erd\"os-R\'enyi Model.} As the last column of tables~\ref{tab:DIN} and~\ref{tab:eMID} shows, the expected value of the spectral radius of $\mathbf{A}$, evaluated numerically as the average over $|\mathbb{A}|=10^3$ configurations reading

\begin{equation}
\langle\lambda_1\rangle=\sum_{\mathbf{A}\in\mathbb{A}}\frac{\lambda_1(\mathbf{A})}{|\mathbb{A}|},
\end{equation}
is always very well approximated by the spectral radius of $\mathbf{P}$, i.e. $\pi_1$. The accuracy of such an estimation is pictorially confirmed by the left panels of fig.~\ref{fig:1}, showing the related scatter plot for each of the 44 snapshots constituting the DIN and for each of the 64 snapshots constituting the quarterly e-MID.

\begin{figure*}[t!]
\centering
\includegraphics[width=\linewidth]{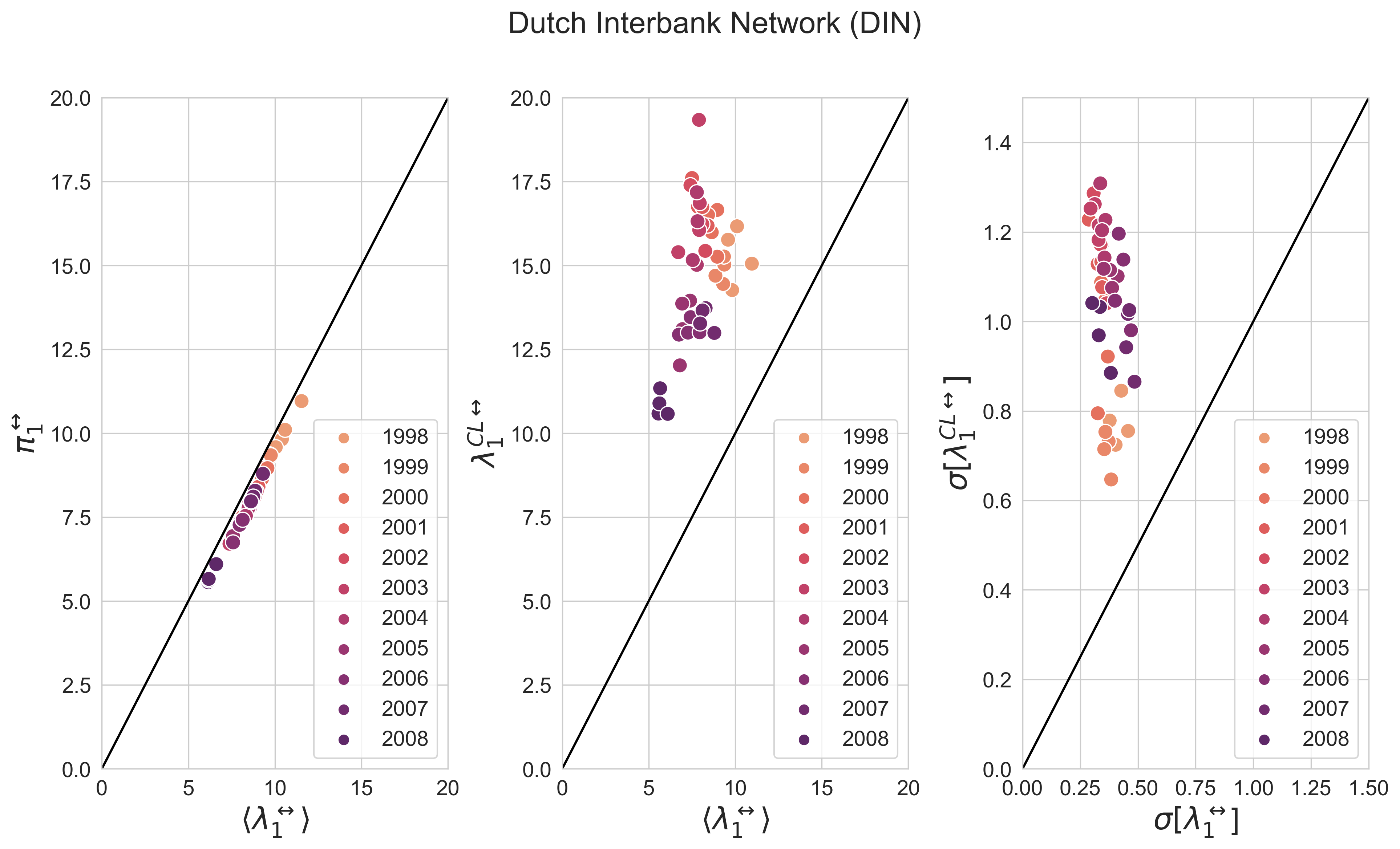}
\includegraphics[width=\linewidth]{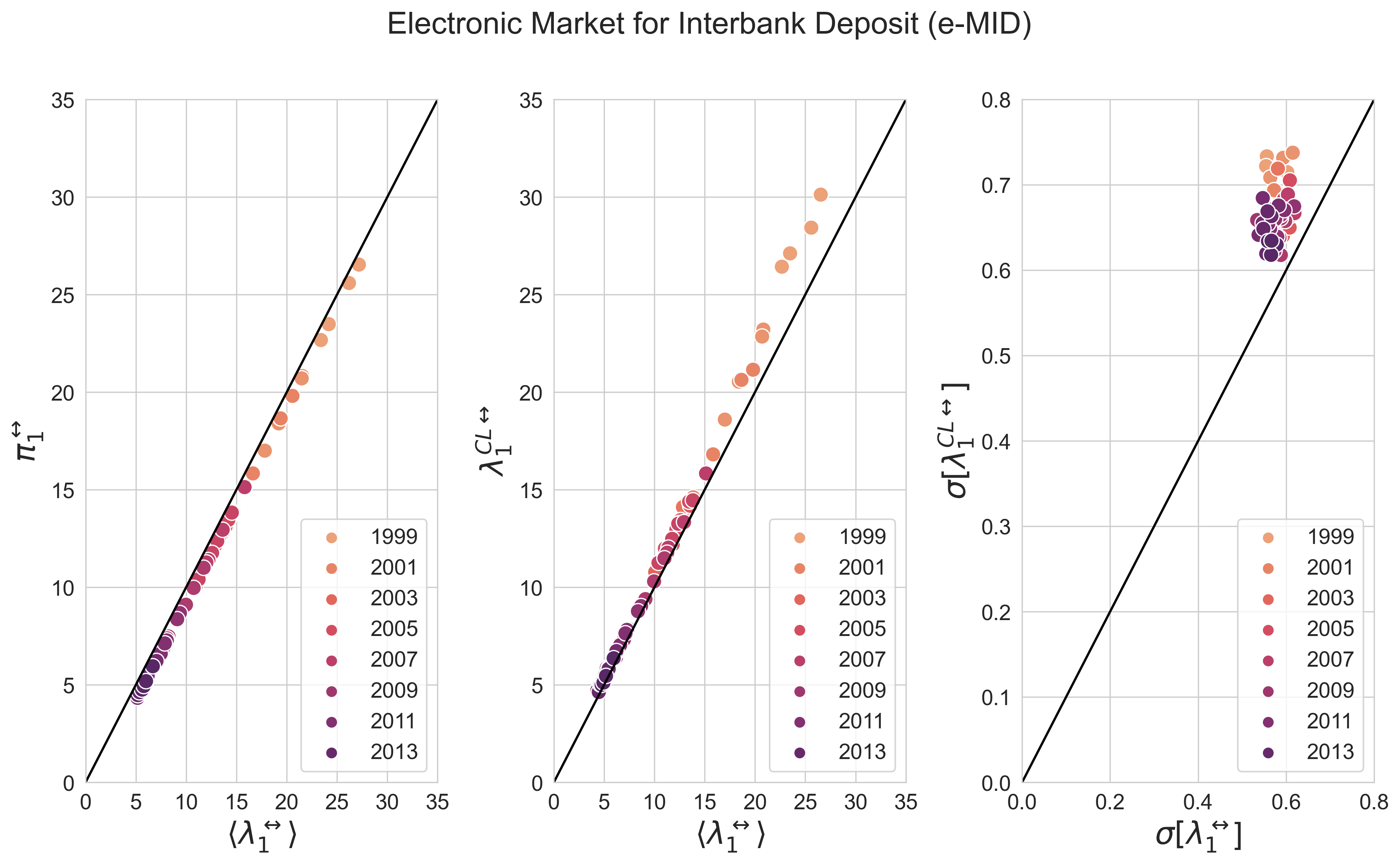}
\caption{Expected value and variance of the spectral radius for each of the quarters of the Dutch Interbank Network (DIN) and of the Electronic Market for Interbank Deposit (e-MID) according to the Reciprocal Configuration Model. Left panels: the expected value of the spectral radius is very well approximated by the spectral radius of the matrix $\mathbf{P}^\leftrightarrow=\{p_{ij}^\leftrightarrow\}_{i,j=1}^N$ characterising the Reciprocal Configuration Model. Central panels: the spectral radius of the matrix $\mathbf{P}^\leftrightarrow=\{p_{ij}^\leftrightarrow\}_{i,j=1}^N$ characterising the Reciprocal Configuration Model is, overall, well approximated by $\lambda_1^{\text{CL}^\leftrightarrow}=\sum_{i=1}^Nk_i^\leftrightarrow k_i^\leftrightarrow/2L$. Right panels: the variance of the spectral radius is underestimated by the value $\text{Var}[\lambda_1^{\text{CL}^\leftrightarrow}]=\text{Var}\left[\sum_{i=1}^Nk_i^\leftrightarrow k_i^\leftrightarrow/2L\right]$.}
\label{fig:3}
\end{figure*}

The central panels of fig.~\ref{fig:1}, instead, provide information about the explicit functional form of $\pi_1$, that matches the estimation reading $\lambda_1^\text{ER}=(N-1)p=L/N$.

The right panels of fig.~\ref{fig:1} provide information about the explicit functional form of the variance of the spectral radius by comparing

\begin{equation}
\text{Var}[\lambda_1]=\sum_{\mathbf{A}\in\mathbb{A}}\frac{[\lambda_1(\mathbf{A})-\langle\lambda_1\rangle]^2}{|\mathbb{A}|}
\end{equation}
with $\text{Var}[\lambda_1^\text{ER}]=p(1-p)$: as it can be appreciated, such an expression slightly underestimates the ensemble variance of the spectral radius.\\

\paragraph*{Binary Configuration Model.} As the last column of tables~\ref{tab:DIN} and~\ref{tab:eMID} shows, the expected value of the spectral radius of $\mathbf{A}$, evaluated numerically as the average over $|\mathbb{A}|=10^3$ configurations reading $\langle\lambda_1\rangle=\sum_{\mathbf{A}\in\mathbb{A}}\lambda_1(\mathbf{A})/|\mathbb{A}|$, is always very well approximated by the spectral radius of $\mathbf{P}$, i.e. $\pi_1$. The accuracy of such an estimation is pictorially confirmed by the left panels of fig.~\ref{fig:2}, showing the related scatter plot for each of the 44 snapshots constituting the DIN and for each of the 64 snapshots constituting the quarterly e-MID.

The central panels of fig.~\ref{fig:2}, instead, provide information about the explicit functional form of $\pi_1$ which is (overall) well approximated by the Chung-Lu estimation reading $\lambda_1^\text{CL}=\sum_{i=1}^Nk_ih_i/L$ for what concerns the e-MID and overestimated by the same expression for what concerns the DIN.

The right panels of fig.~\ref{fig:2} provide information about the explicit functional form of the variance of the spectral radius, by comparing $\text{Var}[\lambda_1]=\sum_{\mathbf{A}\in\mathbb{A}}[\lambda_1(\mathbf{A})-\langle\lambda_1\rangle]^2/|\mathbb{A}|$ with

\begin{equation}
\text{Var}[\lambda_1^\text{CL}]=\sum_{\mathbf{A}\in\mathbb{A}}\frac{[\lambda_1^\text{CL}(\mathbf{A})-\langle\lambda_1^\text{CL}\rangle]^2}{|\mathbb{A}|};
\end{equation}
as it can be appreciated, such an expression either underestimates (for what concerns the e-MID) or overestimates (for what concerns the DIN) the ensemble variance of the spectral radius. Notice also that such an expression calculates the variance of the spectral radius by evaluating $\lambda_1^\text{CL}(\mathbf{A})$, i.e. the numerical value of the Chung-Lu approximation, \emph{for each matrix in the sampled ensemble}. As fig.~\ref{fig:2} shows, these discrepancies seem to be due to a systematic mismatch caused by the configuration-specific values of the spectral radius - the DIN, for instance, obeys the relationship $\lambda_1^\text{CL}(\mathbf{A})>\lambda_1(\mathbf{A})$, $\forall\:\mathbf{A}$, a result potentially explaining the differences between $\lambda_1^\text{CL}$ and $\pi_1$ and between $\text{Var}[\lambda_1^\text{CL}]$ and $\text{Var}[\lambda_1]$ - in words, the numbers $\lambda_1^\text{CL}$s are not only larger than their ensemble counterparts but are also more dispersed (see also fig.~\ref{fig:13} in Appendix~\hyperlink{AppF}{F}).\\

\paragraph*{Reciprocal Configuration Model.} The Reciprocal Configuration Model performs similarly to the Binary Configuration Model. While the last column of tables~\ref{tab:DIN} and~\ref{tab:eMID} shows that the expected value of the spectral radius of $\mathbf{A}$, evaluated numerically as the average over $|\mathbb{A}|=10^3$ configurations reading $\langle\lambda_1\rangle=\sum_{\mathbf{A}\in\mathbb{A}}\lambda_1(\mathbf{A})/|\mathbb{A}|$, is always very well approximated by the spectral radius of $\mathbf{P}$, i.e. $\pi_1$, the left panels of fig.~\ref{fig:3}, show the scatter plot concerning the two sets of quantities $\langle\lambda_1^\leftrightarrow\rangle$ and $\pi_1^\leftrightarrow$ for each of the 44 snapshots constituting the DIN and for each of the 64 snapshots constituting the quarterly e-MID.

The central panels of fig.~\ref{fig:3}, instead, provide information about the explicit functional form of $\pi_1^\leftrightarrow$ which is (overall) well approximated by the Chung-Lu estimation reading $\lambda_1^\text{CL}=\sum_{i=1}^Nk_i^\leftrightarrow k_i^\leftrightarrow/2L$ for what concerns the e-MID and overestimated by the same expression for what concerns the DIN.

\begin{figure*}[t!]
\centering
\includegraphics[width=\linewidth]{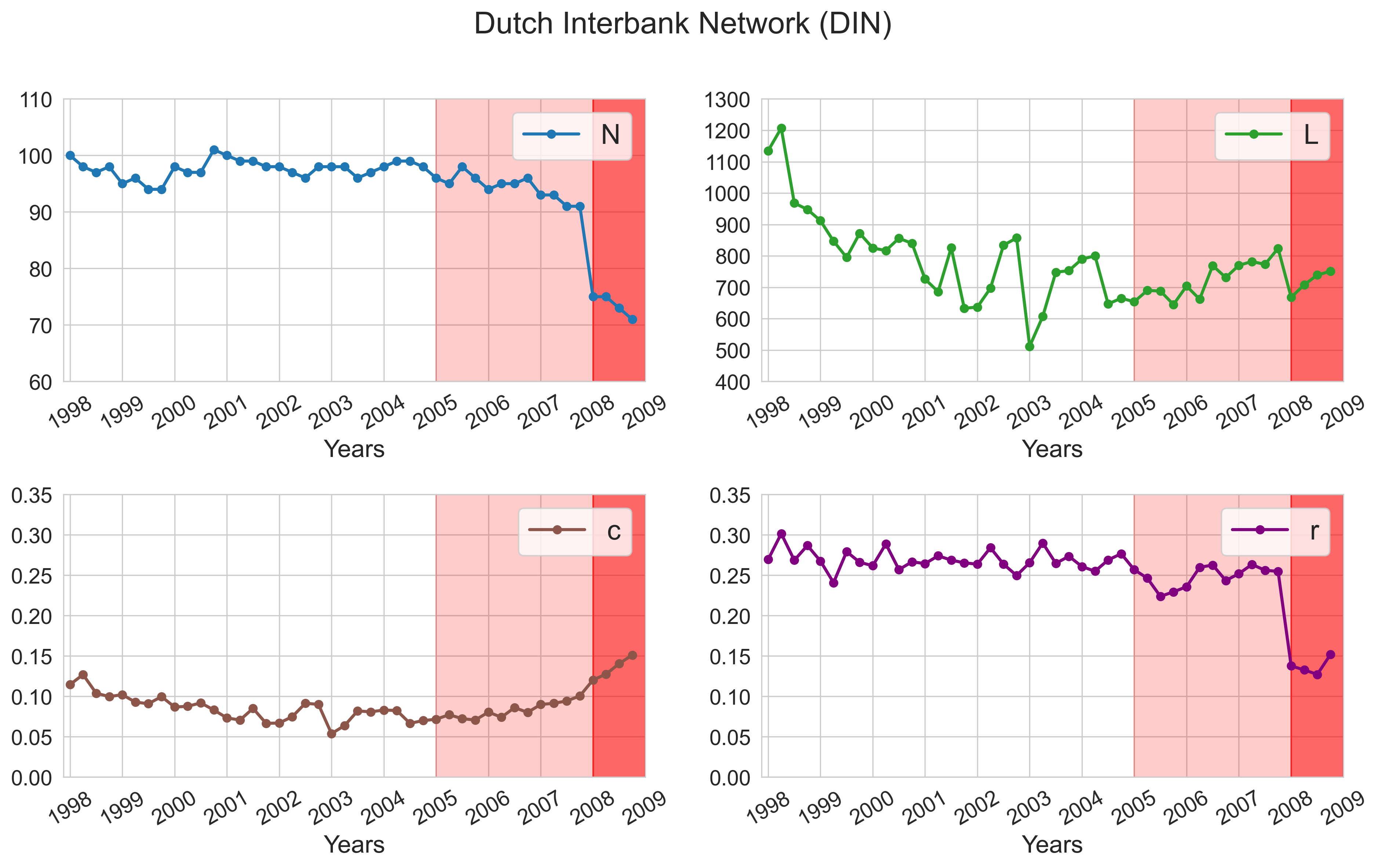}
\caption{Evolution of the number of nodes, links, connectance and reciprocity for all quarters of the DIN. The pre-crisis period (i.e. the years 2005, 2006 and 2007) is highlighted in light red while the global financial crisis (i.e. the year 2008) is highlighted in red.}
\label{fig:4}
\end{figure*}

The right panels of fig.~\ref{fig:3} provide information about the explicit functional form of the variance of the spectral radius by comparing $\text{Var}[\lambda_1^\leftrightarrow]=\sum_{\mathbf{A}\in\mathbb{A}}[\lambda_1^\leftrightarrow(\mathbf{A})-\langle\lambda_1^\leftrightarrow\rangle]^2/|\mathbb{A}|$ with

\begin{equation}
\text{Var}[\lambda_1^{\text{CL}^\leftrightarrow}]=\sum_{\mathbf{A}\in\mathbb{A}}\frac{[\lambda_1^{\text{CL}^\leftrightarrow}(\mathbf{A})-\langle\lambda_1^{\text{CL}^\leftrightarrow}\rangle]^2}{|\mathbb{A}|};
\end{equation}
as it can be appreciated, such an expression overestimates the ensemble variance of the spectral radius. As for the Binary Configuration Model, such an expression calculates the variance of the spectral radius by evaluating $\lambda_1^{\text{CL}^\leftrightarrow}(\mathbf{A})$, i.e. the numerical value of the Chung-Lu approximation, \emph{for each matrix in the sampled ensemble}. These discrepancies may, thus, be imputable to a systematic mismatch caused by the configuration-specific values of the spectral radius.\\

\begin{figure*}[t!]
\centering
\includegraphics[width=\linewidth]{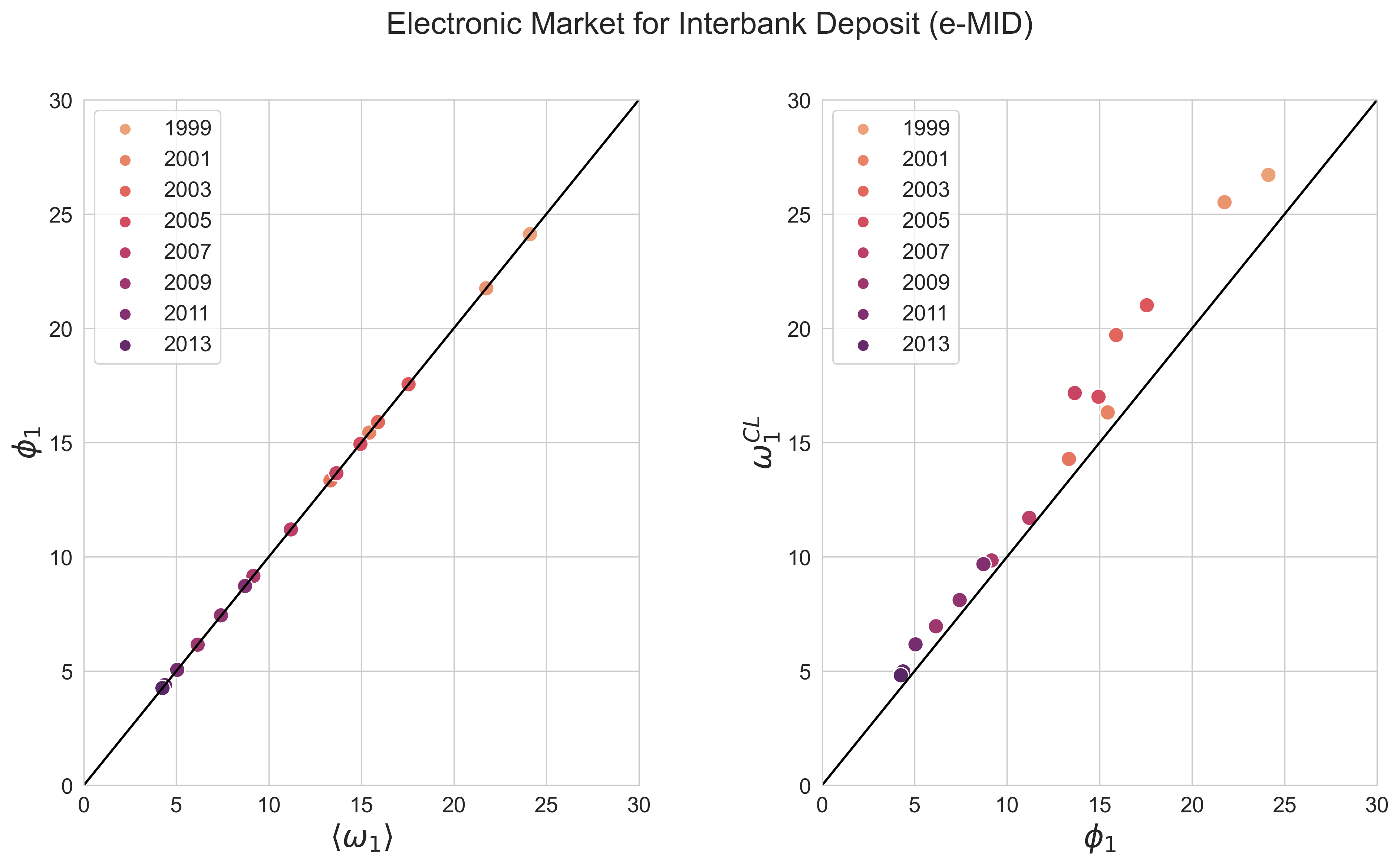}
\includegraphics[width=\linewidth]{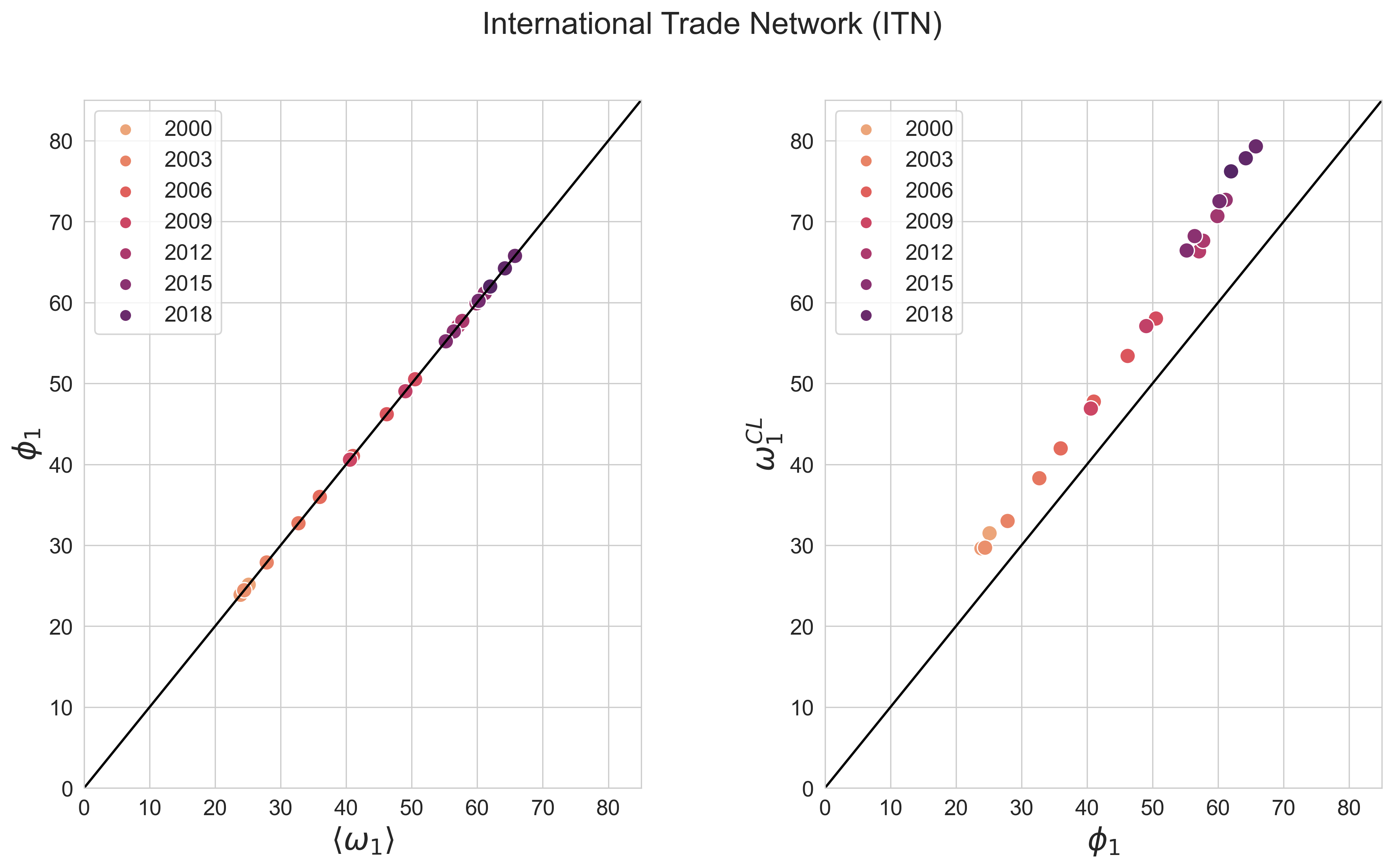}
\caption{Expected value of the spectral radius for each of the years of the Electronic Market for Interbank Deposit (e-MID) and the International Trade Network (ITN) according to the density-corrected Gravity Model. Left panels: the expected value of the spectral radius is very well approximated by the spectral radius of the matrix $\mathbf{Q}=\{\langle w_{ij}\rangle\}_{i,j=1}^N$ characterising the density-corrected Gravity Model. Right panels: the spectral radius of the matrix $\mathbf{Q}=\{\langle w_{ij}\rangle\}_{i,j=1}^N$ characterising the density-corrected Gravity Model is, overall, well approximated by $\phi_1^\text{CL}=\sum_{i=1}^Na_il_i/W$.}
\label{fig:12}
\end{figure*}

\paragraph*{Density-Corrected Gravity Model.} The last column of tables~\ref{tab:eMID} and~\ref{tab:ITN} shows that the expected value of the spectral radius of $\mathbf{W}$, evaluated numerically as the average over $|\mathbb{W}|=10^3$ configurations reading $\langle\omega_1\rangle=\sum_{\mathbf{W}\in\mathbb{W}}\omega_1(\mathbf{W})/|\mathbb{W}|$, is always very well approximated by the spectral radius of $\mathbf{Q}$, i.e. $\phi_1$, as the left panels of fig.~\ref{fig:12} pictorially confirm. Besides, the right panels of the same figure provide information about the explicit functional form of $\phi_1$ which is (overall) well approximated by the Chung-Lu estimation reading $\omega_1^\text{CL}=\sum_{i=1}^Na_il_i/W$ for each of the 16 snapshots constituting the yearly e-MID and for each of the 21 snapshots constituting the yearly ITN.

\begin{figure*}[t!]
\centering
\includegraphics[width=\linewidth]{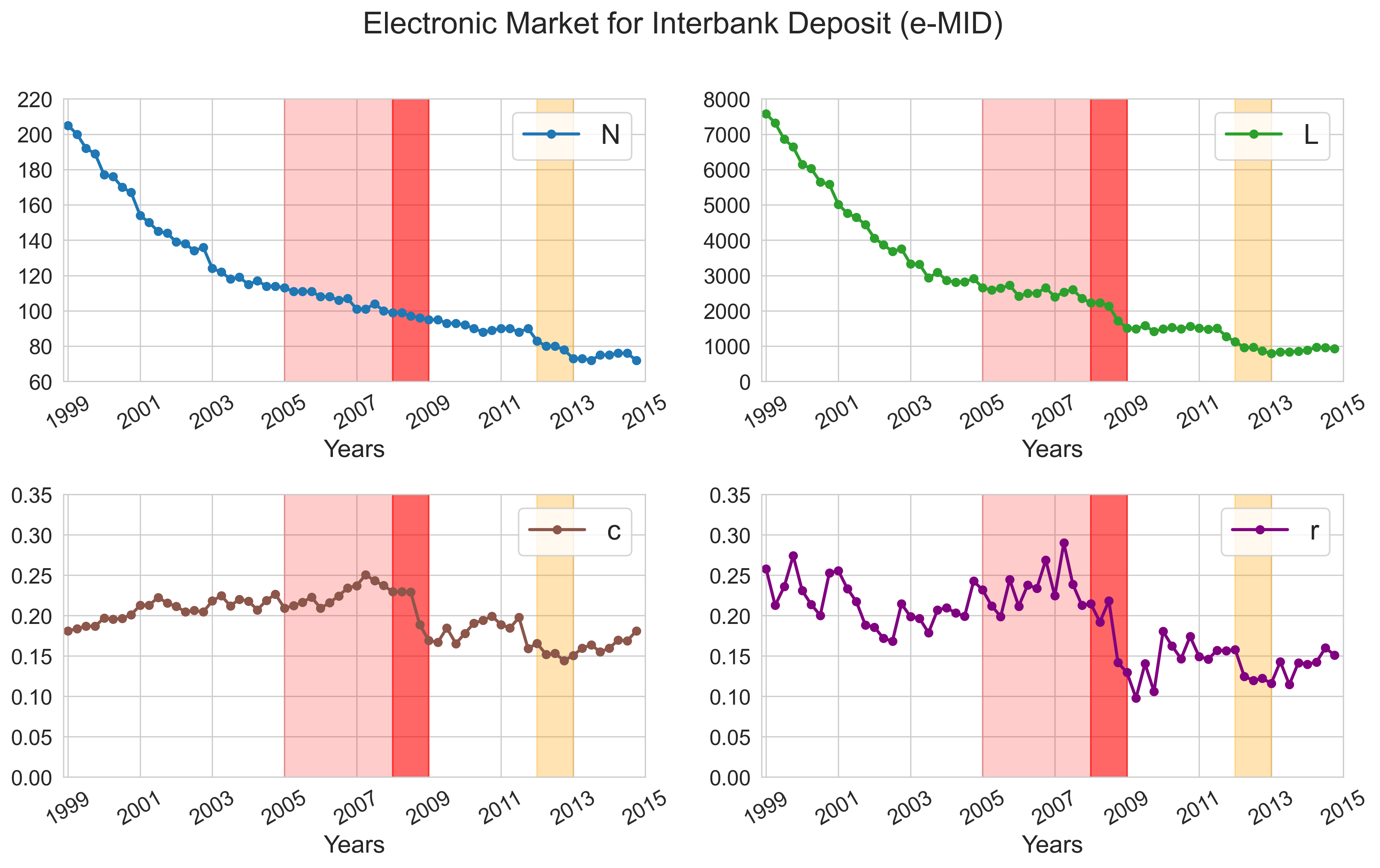}
\caption{Evolution of the number of nodes, links, connectance and reciprocity for all quarters of the e-MID. The pre-crisis period (i.e. the years 2005, 2006 and 2007) is highlighted in light red, while the global financial crisis (i.e. the year 2008) is highlighted in red. The period covered by the long-term refinancing operation (LTRO), promoted by the European Central Bank, is highlighted in yellow.}
\label{fig:5}
\end{figure*}

\subsection{Spectral signature of structural changes in financial networks}

Now, let us inspect the presence of structural changes affecting our networked configurations. To this aim, we will plot the evolution of $z[\lambda_1]$ across the periods covered by our datasets; we will proceed numerically by explicitly sampling the network ensemble induced by each of the benchmarks considered here per snapshot.

\subsubsection{Dutch Interbank Network}

As fig.~\ref{fig:4} clearly shows, the structural change undergone by the system in 2008 is signalled by several quantities: the total number of active Dutch banks sharply decreases as well as the total number of links, whose number diminishes in corresponding of the last year covered by our dataset; this, in turn, causes the connectance to rise. As already discussed in~\cite{squartini2013early}, one of the most evident signals of the global financial crisis is provided by reciprocity: for most of the period, it is characterised by an essentially constant trend, with small fluctuations around an average value of $\simeq0.26$; the last, four snapshots are, then, characterised by a drop of $\simeq40\%$, causing the empirical values to lie almost three sigmas away from the sample average - a trend indicating that the reciprocity of the DIN is anomalously low during the critical period and imputable to a decrease of the level of trust characterising the Dutch system.

An additional signal of the global financial crisis is provided by the empirical value of the spectral radius itself, which decreases in correspondence with 2008Q1 and remains constant across the last four snapshots of our dataset. As it is related to the number of closed walks in a network, its decrease may be related to the decline of reciprocity. However, the latter's trend appears as (much) less affected by the statistical fluctuations characterising the evolution of the DIN throughout its entire history.

Let us now comment on the signal provided by $z[\lambda_1]$. Even if the Erd\"os-R\'enyi Model is, from a merely financial perspective, an unlikely benchmark (its homogeneous nature forces the banks to be similar in size), employing it still allows us to conclude that the DIN is characterised by two structural changes - the first one taking place across 2005 and the second one taking place across 2008. More specifically, after a (more or less) stationary trend characterising the evolution of the DIN from 1998 to 2005 - in correspondence of which the number of closed walks is significantly large - a smooth trend characterising the pre-crisis phase is recovered; afterwards, an abrupt drop connecting the last quarter of 2007 with the first quarter of 2008 emerges. Such a result complements the ones presented in~\cite{squartini2013early} where such behaviour could have been revealed only by employing a heterogeneous benchmark (specifically, the Binary Configuration Model).

Employing the heterogeneous benchmarks - preserving the heterogeneity of banks by constraining the observed (reciprocal) degrees - leads to the same qualitative result. More quantitatively, instead, all such null models reveal that the number of closed walks is perfectly compatible with their predictions during the stationary phase of the system. Such a consistency confirms that, in the absence of distress, the topology of the DIN can be reconstructed quite accurately, solely employing the information provided by the number of (inward, outward and reciprocated) partners of each bank. It is noticed that the explanatory power of the Reciprocal Configuration Model is larger than that of the Global Reciprocity Model, which, in turn, is (only slightly) larger than that of the Binary Configuration Model.

As the build-up phase of the crisis began, a decreasing trend led to 2008, indicating that the local connectivity of banks became less and less informative about the network as a whole - emerges. Under the same benchmarks, the second regime shift is preceded by a short, rising trend. As already noticed in~\cite{squartini2013early}, maximum-entropy techniques yield a realistic guess of the real network only in tranquil times: when the network is under stress, instead, these models provide a sort of distorted picture of it, whose differences from the empirical situation constitute the structural changes we are looking for.

Apart from model-specific differences, however, the degree of informativeness about the changes affecting the DIN carried by the spectral radius seems quite independent of the model employed to spot the differences above.

\subsubsection{Electronic Market for Interbank Deposit}

For what concerns the e-MID, instead, the evolution of the total number of active Italian banks steadily decreases, hence not providing any clear indication about the presence of structural changes. On the contrary, the evolution of the total number of links provides a quite clear indication of the presence of two regime shifts as $L$ drops in correspondence of 2008 and 2012. Overall, the connectance and the reciprocity provide a very similar indication - the global financial crisis being characterised by a stronger signal than the one characterising the long-term refinancing operation (LTRO) promoted by the European Central Bank at the end of 2011\footnote{The two LTRO measures date December the 22nd, 2011 and February the 29th, 2012.}.

The evolution of the empirical value of the spectral radius is characterised by a drop in correspondence of the first crisis, originating a slightly fluctuating trend that lasts until 2012, the year in correspondence of which a second, decreasing trend can also be observed.

\begin{figure*}[t!]
\centering
\includegraphics[width=0.49\linewidth]{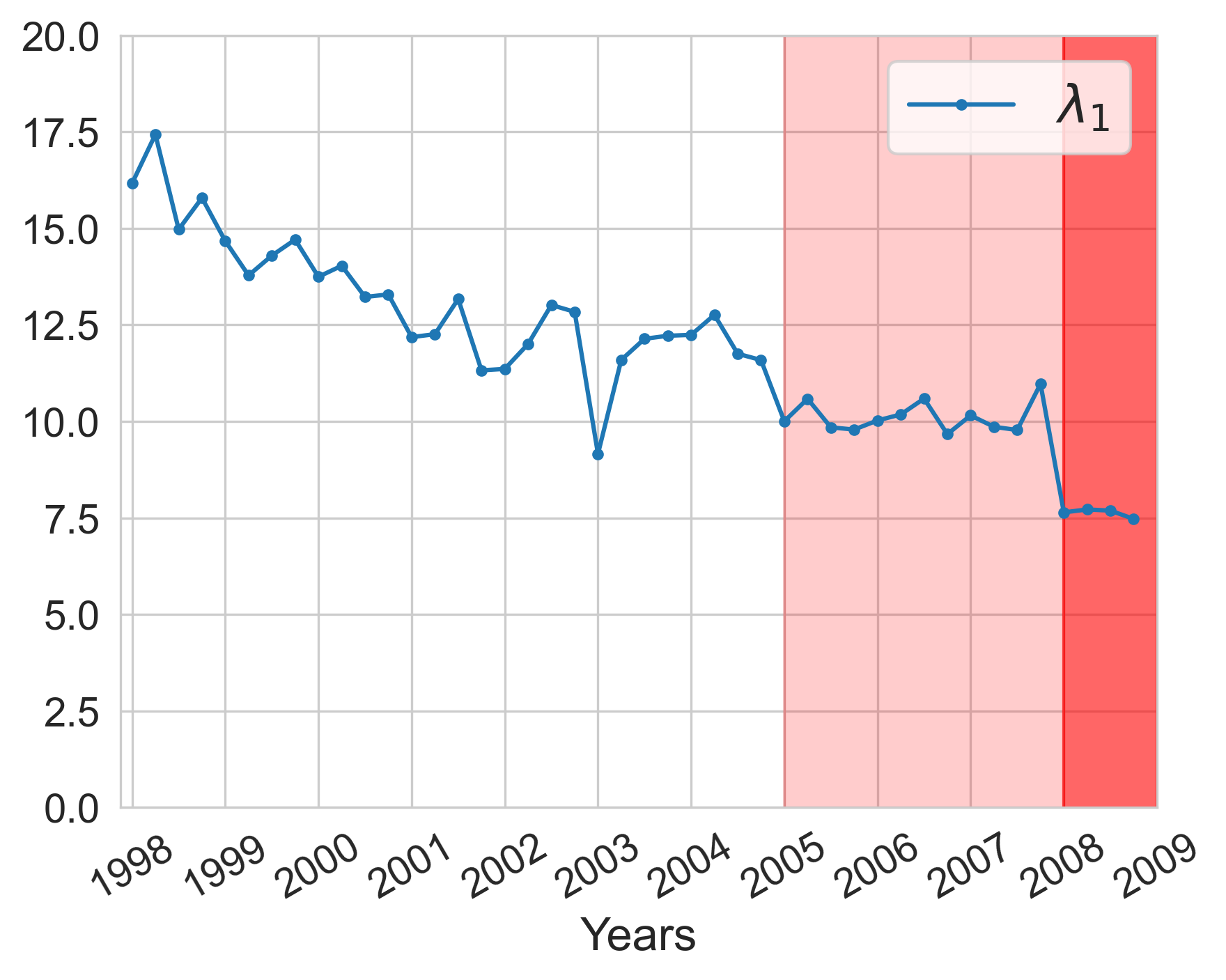}
\includegraphics[width=0.49\linewidth]{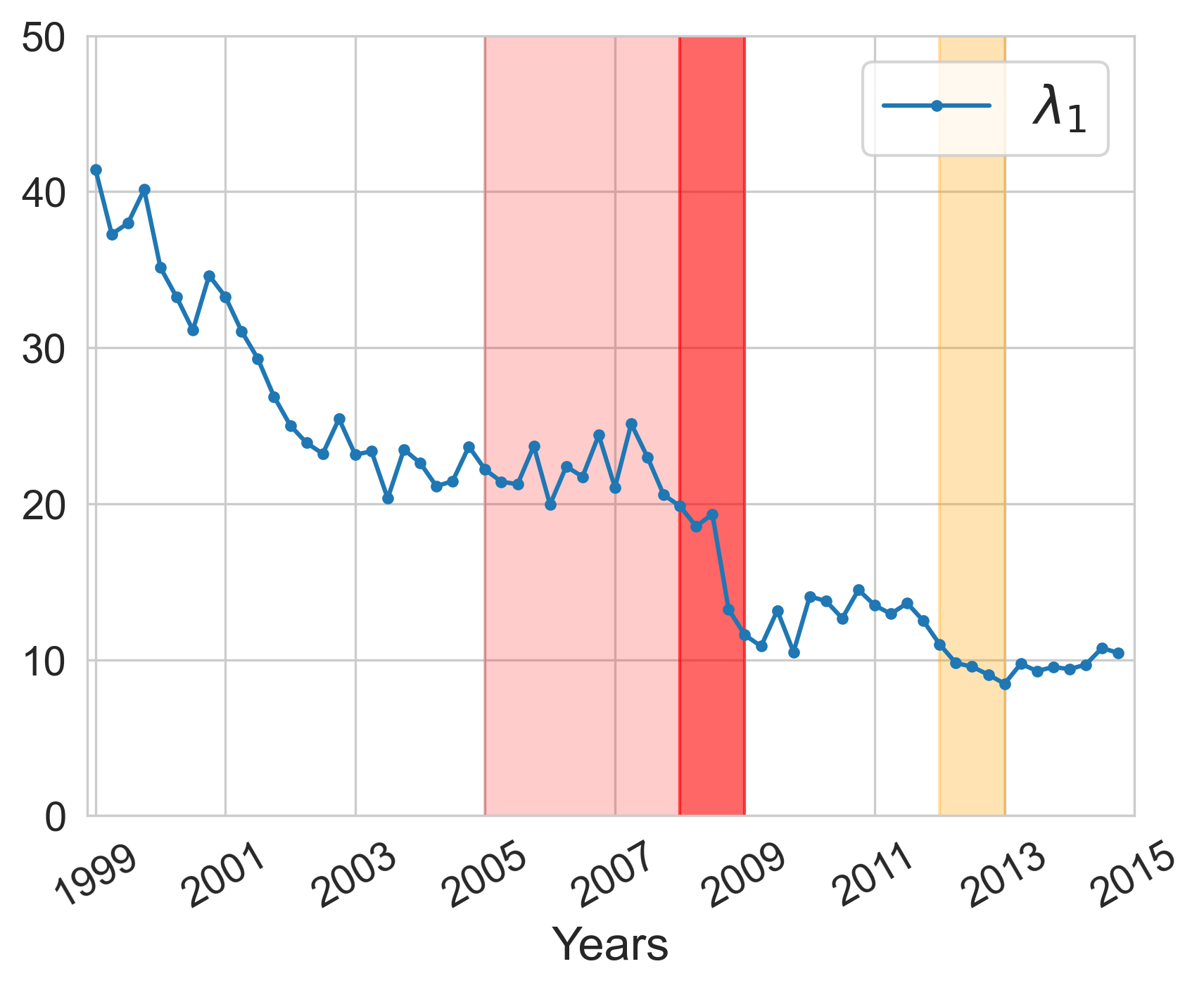}
\includegraphics[width=0.49\linewidth]{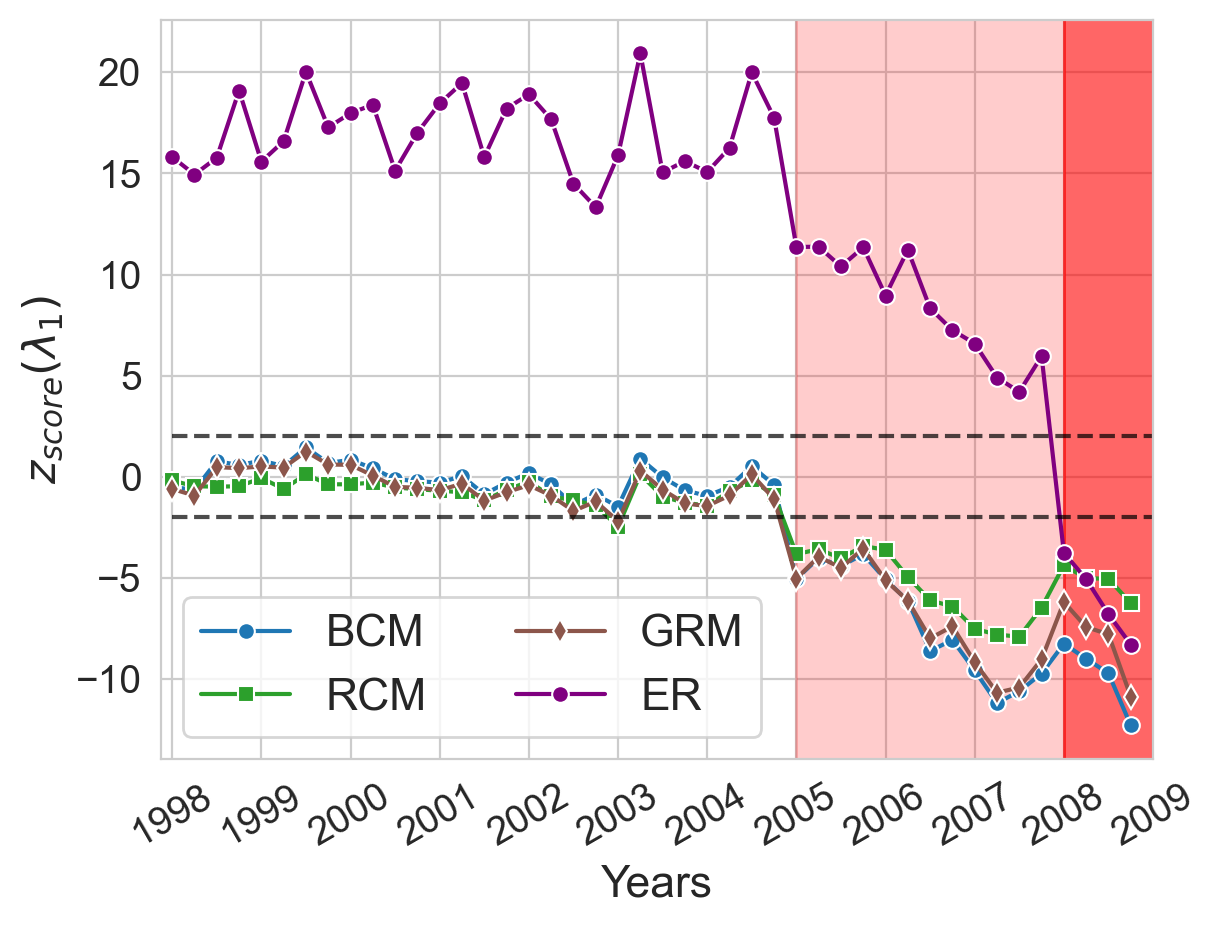}
\includegraphics[width=0.49\linewidth]{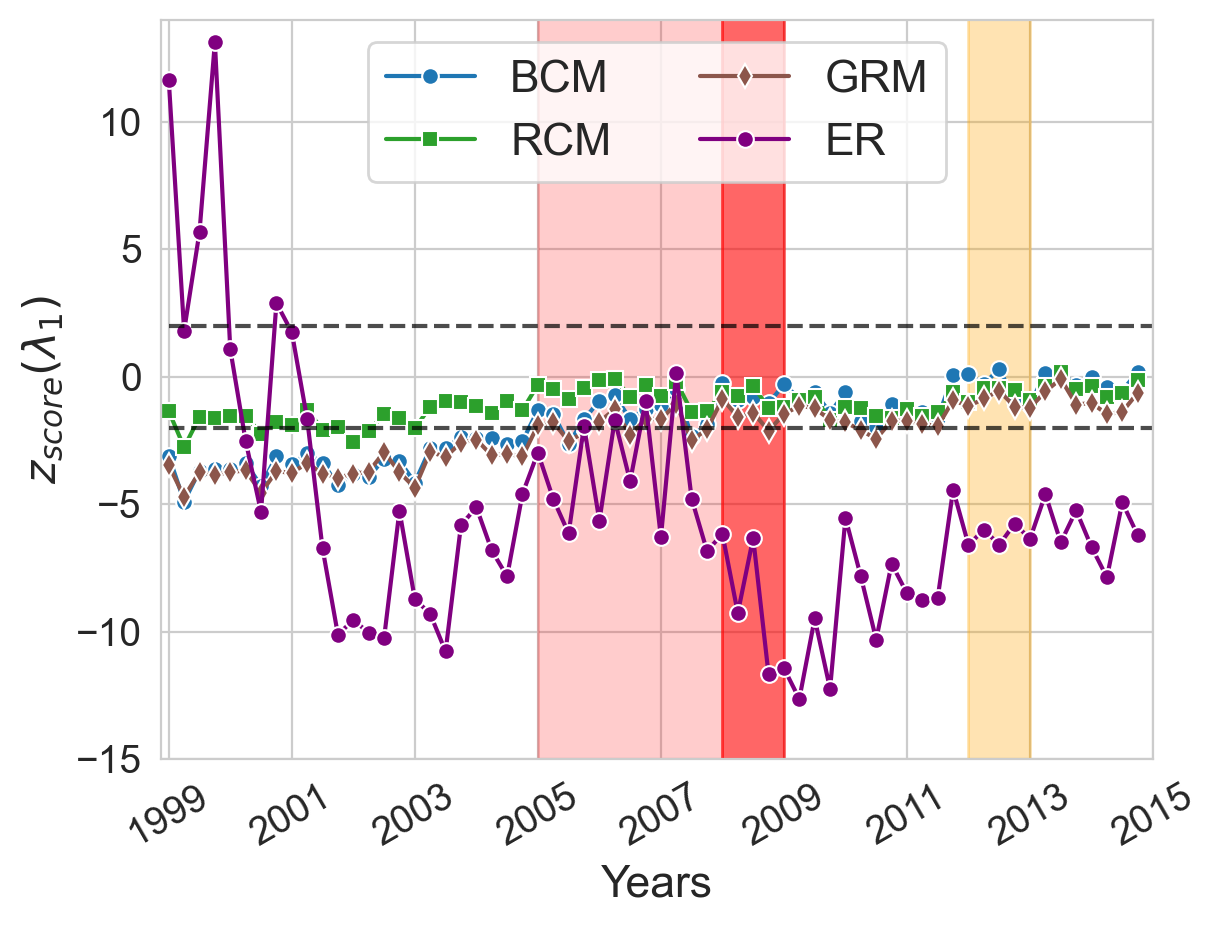}
\includegraphics[width=0.49\linewidth]{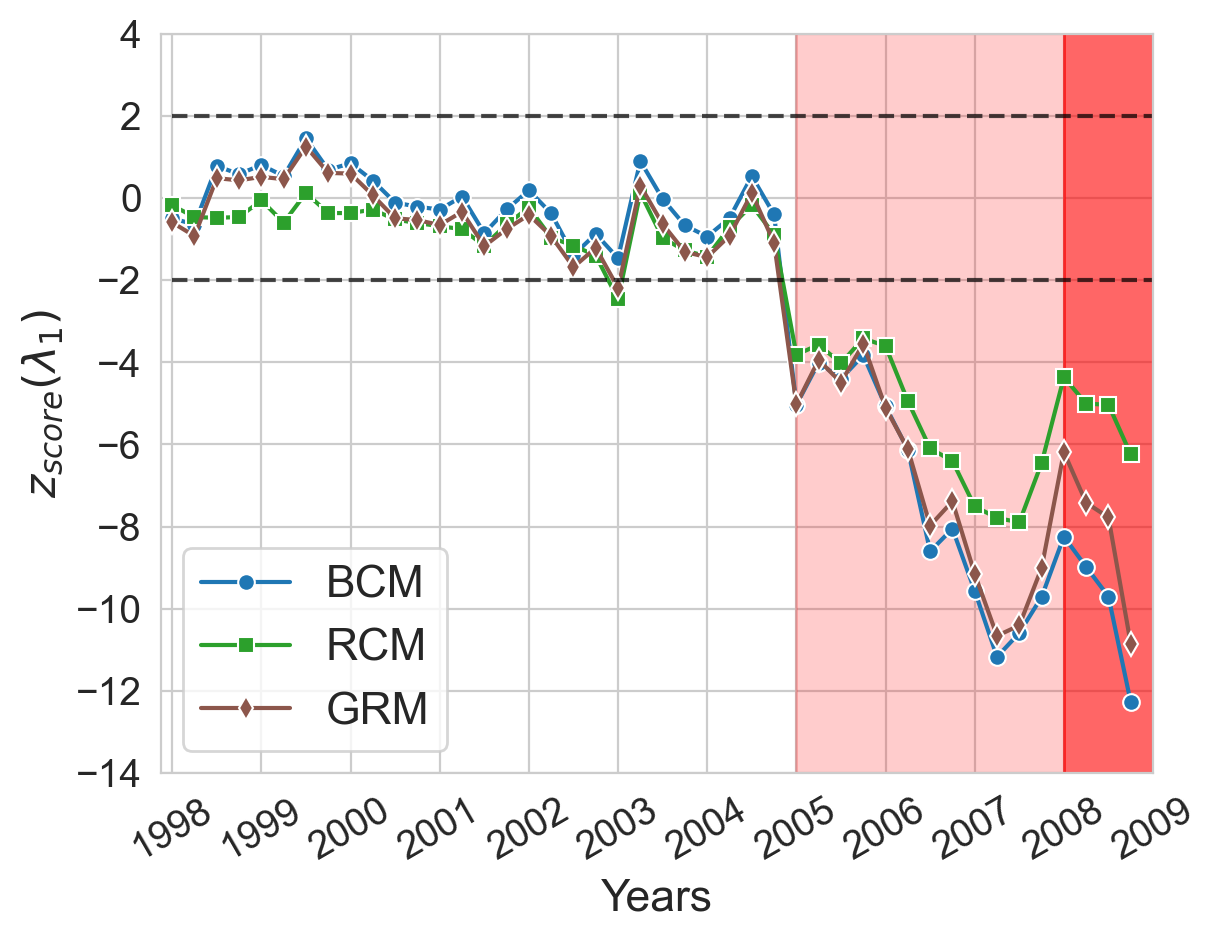}
\includegraphics[width=0.49\linewidth]{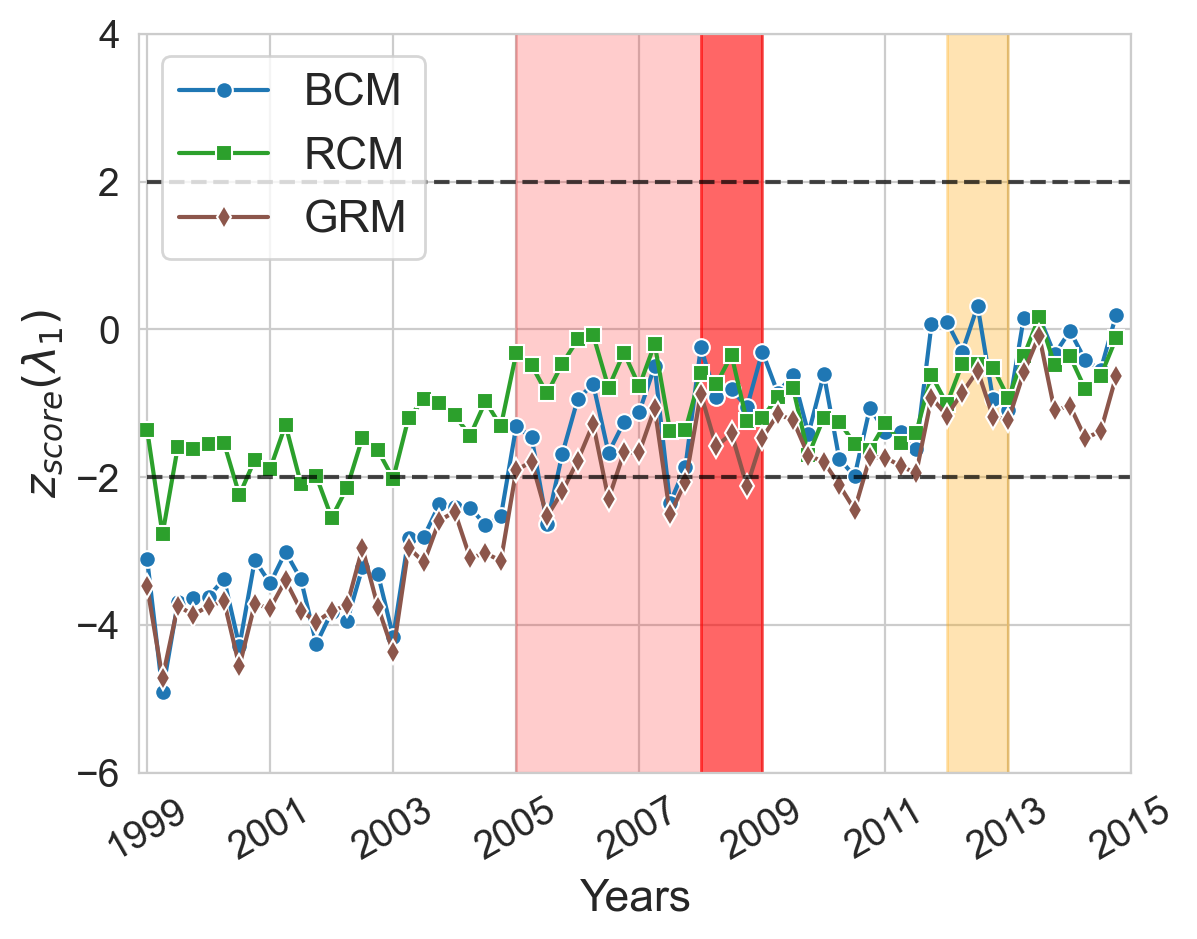}
\caption{Evolution of the spectral radius (top panels) and of its $z$-score $z[\lambda_1]=(\lambda_1-\langle\lambda_1\rangle)/\sigma[\lambda_1]$ (central and bottom panels) across the quarters of the Dutch Interbank Network (DIN) and of the Electronic Market for Interbank Deposit (e-MID). While the evolution of the (empirical value of the) spectral radius returns a signal for each of the events captured by our datasets - typically dropping in correspondence with a crisis - the evolution of its $z$-score returns early-signals for the same events. Interestingly, while each benchmark provides information about the evolution of the two systems considered here, they seem to behave oppositely: for instance, while the global financial crisis induces a statistically significant signal in the case of the DIN - which evolves from a regime of compatibility towards a regime of incompatibility with our heterogeneous benchmarks - it does not in the case of the e-MID - which evolves from a regime of incompatibility towards a regime of compatibility with the same null models.}
\label{fig:6}
\end{figure*}

Let us now comment on the signal provided by $z[\lambda_1]$. Employing a homogeneous benchmark such as the Erd\"os-R\'enyi Model allows us to conclude that the e-MID is characterised by three structural changes, the first one taking place across 2000, the second one taking place between 2007 and 2008 and the third one taking place across 2012.

More specifically, the evolution of the e-MID starts with a drop of the $z$-score of the spectral radius, indicating that the number of closed walks has become significantly smaller than expected during 2001. Afterwards, an increasing trend leading to a phase characterised by several closed walks compatible with the output of the prediction by the Erd\"os-R\'enyi Model becomes visible. Such a period is interrupted by the so-called pre-crisis phase, during which the trend of $\lambda_1$ reverts and becomes again significantly smaller than expected. From 2009 on, a second, increasing trend lasting until 2012 becomes visible: afterwards, the system stabilises.

Employing the heterogeneous benchmarks leads to quite different results: more quantitatively, the first regime shift disappears, replaced by a stationary trend lasting until 2003; afterwards, a rising trend leading the system to its (pre-)critical phase appears. Since 2009 on, a decreasing trend lasting a couple of years emerges to be followed, once more, by an increasing one. From this perspective, the DIN and the e-MID behave, somehow, oppositely: while the global financial crisis induces a statistically significant signal in the case of the DIN, it does not in the case of the e-MID. In a sense, maximum-entropy techniques can be used to reconstruct the e-MID when the system is under stress, while this should be avoided in tranquil times - e.g. the first years of the dataset - when the picture of it inferred from local constraints departs the most from the empirical one.

Differently from the DIN, the explanatory power of the Reciprocal Configuration Model (still larger than the one of the Global Reciprocity Model, which, however, performs similarly to the Binary Configuration Model) is so large that the measurements carried out on the e-MID (practically) always compatible with the predictions. Although such a piece of evidence speaks against the use of the Reciprocal Configuration Model to detect deviations from the average behaviour, statistical \emph{tendencies} can still be revealed, confirming once more that a dichotomous yes/no answer to the question \emph{is this pattern statistically significant?} may be quite unsatisfactory to gain a sufficiently deep insight into system behaviour.

\section{Discussion}

The so-called stability analysis represents an application of particular interest in the study of financial networks, a topic whose popularity has steadily increased since the turmoil due to the mortgage crisis~\cite{caccioli2018network}. The objective of this kind of analysis is to understand the relationship(s) between the topological structure of financial networks and their resilience to events like shocks, cascading failures, etc., by employing real data~\cite{battiston2016price}, reconstructed configurations~\cite{di2018assessing} or (simple) toy models~\cite{bardoscia2017pathways}. A direct way to explore this connection is by running stress tests on several different topological structures by measuring the effects of a simulated shock and the subsequent propagation of losses \emph{ex post}~\cite{ram2020net}: later works have related these results to the magnitude of the spectral radius of the so-called leverage matrix~\cite{bardoscia2017pathways} although no algorithm has been devised to estimate its magnitude from the (partial) information that is usually available in financial contexts.

With the present contribution, we have tackled a more general challenge, i.e. that of estimating the spectral radius of random network models calibrated on real-world evolving networks. To this aim, we have adopted several approximations that have led to the surprisingly simple recipe $\langle\lambda_1\rangle\simeq\pi_1$ for estimating the expected value of $\lambda_1$, with $\pi_1$ representing the spectral radius of the probabilistic matrix describing the chosen model. Despite our result is based on an approximation\footnote{We have explicitly verified that the properties of \emph{existence}, \emph{reality}, \emph{positivity}, \emph{maximality} and \emph{uniqueness} of the spectral radius hold for each, considered configuration.}, it turns out to be extremely accurate for any directed (binary or weighted) random network model considered.

Besides the theoretical relevance of such a result, its usefulness lies in spotting the structural changes separating a (financial) regime from another by exploiting the interplay between distress and topological changes. As the case studies of the DIN and the e-MID illustrate, deviations from the average behaviour can happen in both directions, either \emph{moving away from a less structured configuration} (hence becoming a less typical member of an equilibrium ensemble of graphs) or \emph{moving towards a less structured configuration} (hence becoming a more typical member of an equilibrium ensemble of graphs): from this perspective, each quantity characterising the original network can be straightforwardly assigned a level of significance - which is sensitive to the direction - by computing the related $z$-score, i.e. an index comparing the measured value with the one expected under a null model preserving some properties of the observed network but, otherwise, being maximally random.

Although our results become exact in case a perfectly non-reciprocal network is observed, future research calls for a more accurate evaluation of our approximations - hopefully, in terms of the reciprocity itself. Besides, extending the results of the present analysis to undirected, binary or weighted networks would enlarge their applicability beyond the economic and financial domains.

\section{Acknowledgments}

VM acknowledges support from the project NetRes - `Network analysis of economic and financial resilience', Italian DM n. 289, 25-03-2021 (PRO3 2021-2023 University joint program `Le Scuole Superiori ad Ordinamento Speciale: istituzioni a servizio del Paese'), CUP D67G22000130001 (\url{https://netres.imtlucca.it}) funded by the Italian Ministry of University and Research (MUR). This work is also supported by the European Union - NextGenerationEU - National Recovery and Resilience Plan (PNRR Research Infrastructures), project `SoBigData.it - Strengthening the Italian RI for Social Mining and Big Data Analytics' - Grant IR0000013 (Avviso MUR D.D. n. 3264, 28/12/2021)  (\url{https://pnrr.sobigdata.it/}) and the project ``Reconstruction, Resilience and Recovery of Socio-Economic Networks'' RECON-NET EP\_FAIR\_005 - PE0000013 ``FAIR'' - PNRR M4C2 Investment 1.3, financed by the European Union – NextGenerationEU.

We thank Anna Gallo for useful discussions.

\bibliography{biblio}

\clearpage

\onecolumngrid

\hypertarget{AppA}{}
\section*{Appendix A.\\Dyadic early-warning signals}

Upon defining 

\begin{align}
X&=\sum_{i=1}^N\sum_{j=1}^Na_{ij}a_{ji}=\sum_{i=1}^N[\mathbf{A}^2]_{ii}=\text{Tr}\left[\mathbf{A}^2\right],
\end{align}
we are left with the task of calculating its expected value and variance. The evidence that the expected value is a linear operator (i.e. $\langle aX+bY\rangle=a\langle X\rangle+b\langle Y\rangle$) and that the entries of a binary, directed network are treated as independent random variables under any of the random network models considered here, makes such a calculation straightforward. In fact,

\begin{equation}
\langle X\rangle=\left\langle\sum_{i=1}^N\sum_{j=1}^Na_{ij}a_{ji}\right\rangle=\sum_{i=1}^N\sum_{j=1}^N\langle a_{ij}a_{ji}\rangle=\sum_{i=1}^N\sum_{j=1}^N\langle a_{ij}\rangle\langle a_{ji}\rangle=\sum_{i=1}^N\sum_{j=1}^Np_{ij}p_{ji}.
\end{equation}

In order to calculate the variance of $X$, let us consider that $X$ can be re-written as

\begin{equation}
X=\sum_{i=1}^N\sum_{j=1}^Na_{ij}a_{ji}=2\sum_{i=1}^N\sum_{j(>i)}a_{ij}a_{ji}
\end{equation}
i.e. as a sum over dyads, treated as independent random variables under any random network models considered here. Since the variance of a sum of independent random variables coincides with the sum of their variances, one can write

\begin{align}
\text{Var}[X]&=\text{Var}\left[2\sum_{i=1}^N\sum_{j(>i)}a_{ij}a_{ji}\right]=4\cdot\sum_{i=1}^N\sum_{j(>i)}\text{Var}[a_{ij}a_{ji}];
\end{align}
then, since $a_{ij}a_{ji}\sim\text{Ber}[p_{ij}p_{ji}]$, one finds that

\begin{align}
\text{Var}[X]&=4\cdot\sum_{i=1}^N\sum_{j(>i)}p_{ij}p_{ji}(1-p_{ij}p_{ji}).
\end{align}

It is nevertheless instructive to follow an alternative road and consider that

\begin{align}
\text{Var}[X]=\text{Var}\left[\sum_{i=1}^N\sum_{j=1}^Na_{ij}a_{ji}\right]&=\sum_{i=1}^N\sum_{j=1}^N\text{Var}[a_{ij}a_{ji}]+2\cdot\sum_{i=1}^N\sum_{j(>i)}\text{Cov}[a_{ij}a_{ji},a_{ij}a_{ji}]\nonumber\\
&=\sum_{i=1}^N\sum_{j=1}^Np_{ij}p_{ji}(1-p_{ij}p_{ji})+2\cdot\sum_{i=1}^N\sum_{j(>i)}p_{ij}p_{ji}(1-p_{ij}p_{ji})\nonumber\\
&=2\cdot\sum_{i=1}^N\sum_{j(>i)}p_{ij}p_{ji}(1-p_{ij}p_{ji})+2\cdot\sum_{i=1}^N\sum_{j(>i)}p_{ij}p_{ji}(1-p_{ij}p_{ji})\nonumber\\
&=4\cdot\sum_{i=1}^N\sum_{j(>i)}p_{ij}p_{ji}(1-p_{ij}p_{ji}).
\end{align}

\begin{figure}[t!]
\centering
\includegraphics[width=\linewidth]{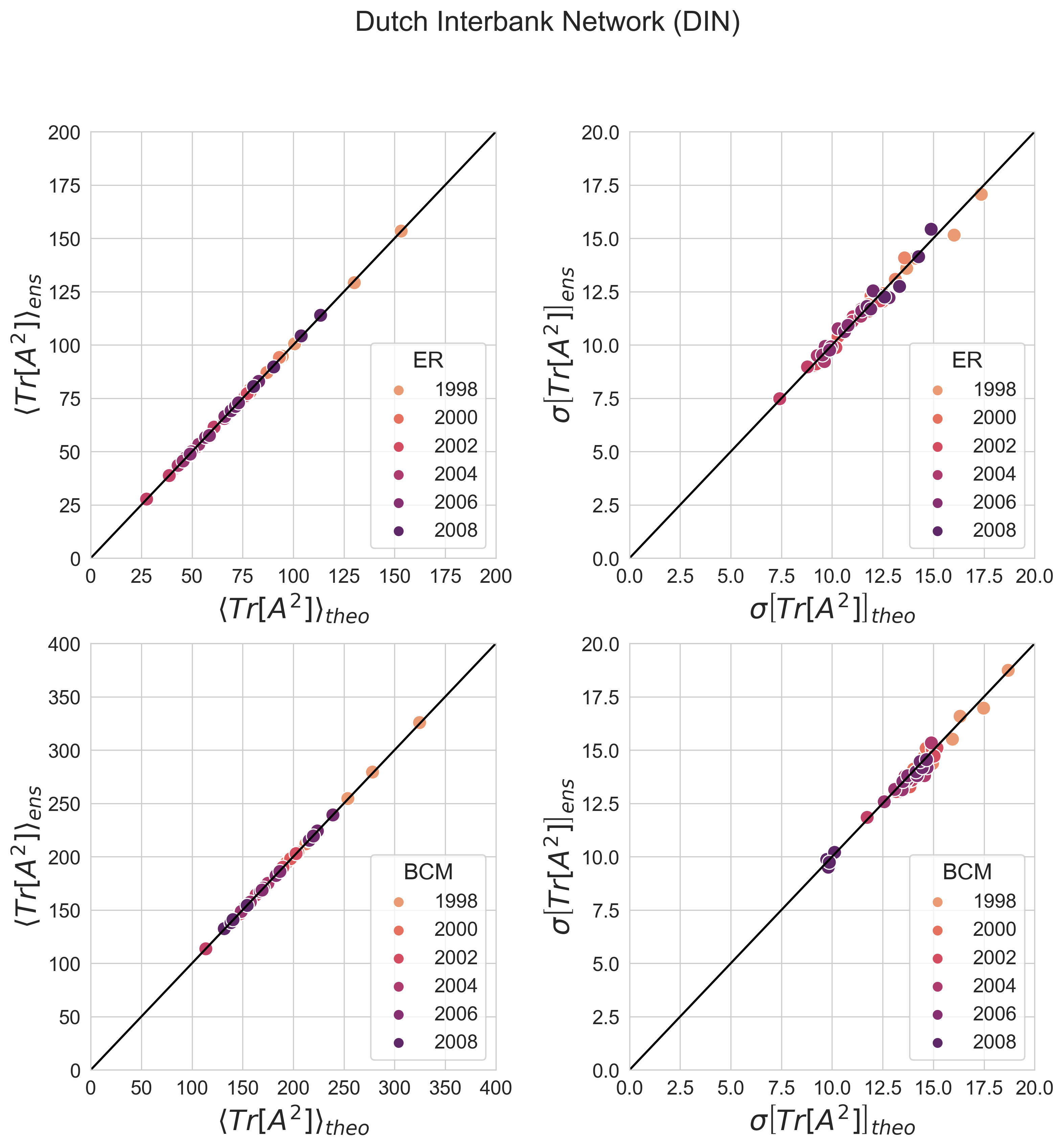}
\caption{Comparison between the analytical estimations of the expected value and variance of the number of dyads and the numerical counterparts, obtained by explicitly sampling the ensembles induced by the Erd\"os-R\'enyi Model (top panels) and the Binary Configuration Model (bottom panels). The numerical simulations have been carried out on the quarters of the Dutch Interbank Network (DIN); the number of sampled matrices per snapshot is $10^3$.}
\label{fig:7}
\end{figure}

\begin{figure}[t!]
\centering
\includegraphics[width=\linewidth]{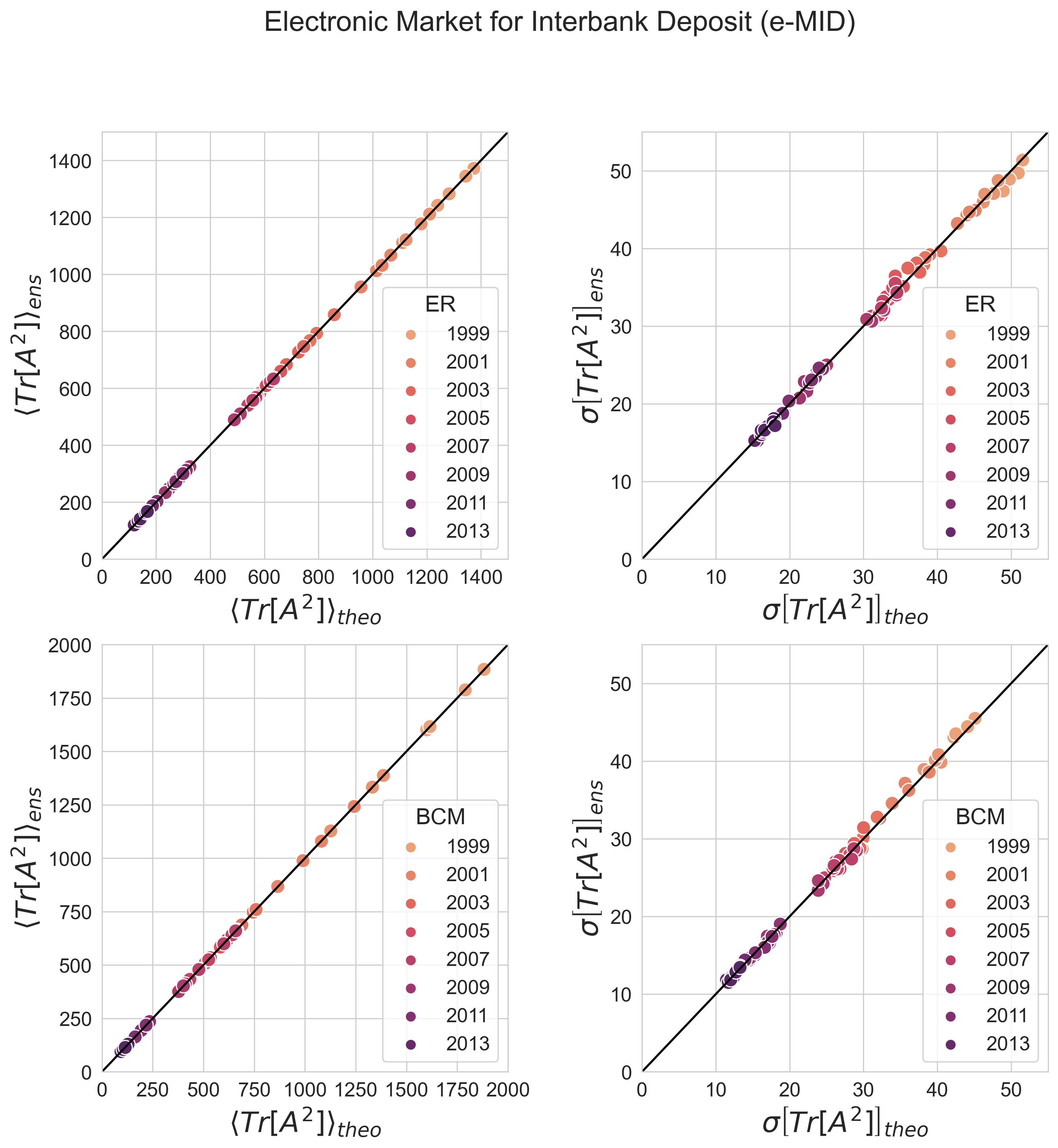}
\caption{Comparison between the analytical estimations of the expected value and variance of the number of dyads and the numerical counterparts, obtained by explicitly sampling the ensembles induced by the Erd\"os-R\'enyi Model (top panels) and the Binary Configuration Model (bottom panels). The numerical simulations have been carried out on the quarters of the Electronic Market for Interbank Deposit (e-MID); the number of sampled matrices per snapshot is $10^3$.}
\label{fig:8}
\end{figure}

The comparison between the analytical estimations of the expected value and the variance of the number of dyads and the numerical counterparts, obtained by explicitly sampling the ensembles induced by the Erd\"os-R\'enyi Model and the Binary Configuration Model is illustrated in figs.~\ref{fig:7} and~\ref{fig:8}.

\clearpage

\clearpage

\clearpage

\hypertarget{AppB}{}
\section*{Appendix B.\\Triadic early-warning signals}

Upon defining 

\begin{align}
X&=\sum_{i=1}^N\sum_{j=1}^N\sum_{k=1}^Na_{ij}a_{jk}a_{ki}=\sum_{i=1}^N[\mathbf{A}^3]_{ii}=\text{Tr}\left[\mathbf{A}^3\right],
\end{align}
we are left with the task of calculating its expected value and variance. Analogously to the dyadic case, calculating the expected value is straightforward. In fact,

\begin{equation}
\langle X\rangle=\left\langle\sum_{i=1}^N\sum_{j=1}^N\sum_{k=1}^Na_{ij}a_{jk}a_{ki}\right\rangle=\sum_{i=1}^N\sum_{j=1}^N\sum_{k=1}^N\langle a_{ij}a_{jk}a_{ki}\rangle=\sum_{i=1}^N\sum_{j=1}^N\sum_{k=1}^N\langle a_{ij}\rangle\langle a_{jk}\rangle\langle a_{ki}\rangle=\sum_{i=1}^N\sum_{j=1}^N\sum_{k=1}^Np_{ij}p_{jk}p_{ki}.
\end{equation}

In order to calculate the variance of $X$, let us, first, consider that $X$ can be re-written as

\begin{align}
X&=\sum_{i=1}^N\sum_{j=1}^N\sum_{k=1}^Na_{ij}a_{jk}a_{ki}=3\cdot\sum_{i=1}^N\sum_{j(>i)}\sum_{k(>j)}(a_{ij}a_{jk}a_{ki}+a_{ik}a_{kj}a_{ji})\equiv3\cdot\sum_{i<j<k}(a_{ij}a_{jk}a_{ki}+a_{ik}a_{kj}a_{ji})
\end{align}
i.e. as a sum over triads. Then, let us notice that

\begin{align}
\text{Var}[X]=3^2\cdot\left[\sum_\mathbf{I}\text{Var}[a_\mathbf{I}]+2\cdot\sum_{\mathbf{I<\mathbf{J}}}\text{Cov}[a_\mathbf{I},a_\mathbf{J}]\right]
\end{align}
where we have employed the multi-index notation, i.e. $\mathbf{I}\equiv(i,j,k)$ and $\mathbf{J}\equiv(l,m,n)$. More explicitly,

\begin{align}
\text{Var}[a_\mathbf{I}]&=\text{Var}[a_{ij}a_{jk}a_{ki}]+\text{Var}[a_{ik}a_{kj}a_{ji}]+\text{Cov}[a_{ij}a_{jk}a_{ki},a_{ik}a_{kj}a_{ji}]\nonumber\\
&=p_{ij}p_{jk}p_{ki}(1-p_{ij}p_{jk}p_{ki})+p_{ik}p_{kj}p_{ji}(1-p_{ik}p_{kj}p_{ji})+\text{Cov}[a_{ij}a_{jk}a_{ki},a_{ik}a_{kj}a_{ji}]
\end{align}
with $\text{Cov}[a_{ij}a_{jk}a_{ki},a_{ik}a_{kj}a_{ji}]$ depending on the adopted benchmark: under both the Erd\"os-R\'enyi Model and the Binary Configuration Model, it amounts at zero. Overall, thus,

\begin{equation}
\sum_\mathbf{I}\text{Var}[a_\mathbf{I}]=\sum_{i<j<k}[p_{ij}p_{jk}p_{ki}(1-p_{ij}p_{jk}p_{ki})+p_{ik}p_{kj}p_{ji}(1-p_{ik}p_{kj}p_{ji})].
\end{equation}

Moreover,

\begin{align}
\text{Cov}[a_\mathbf{I},a_\mathbf{J}]&=\langle(a_{ij}a_{jk}a_{ki}+a_{ik}a_{kj}a_{ji})\cdot(a_{lm}a_{mn}a_{nl}+a_{ln}a_{nm}a_{ml})\rangle-\langle a_{ij}a_{jk}a_{ki}+a_{ik}a_{kj}a_{ji}\rangle\cdot\langle a_{lm}a_{mn}a_{nl}+a_{ln}a_{nm}a_{ml}\rangle\nonumber\\
&=\langle(a_{ij}a_{jk}a_{ki}+a_{ik}a_{kj}a_{ji})\cdot(a_{lm}a_{mn}a_{nl}+a_{ln}a_{nm}a_{ml})\rangle-(p_{ij}p_{jk}p_{ki}+p_{ik}p_{kj}p_{ji})\cdot(p_{lm}p_{mn}p_{nl}+p_{ln}p_{nm}p_{ml})
\end{align}
is different from zero, i.e. any two triads co-variate as long as they share an edge. In this case, they form a diamond whose vertices can be labelled either as $i\equiv l$, $j\equiv m$, $k$, $n$ or as $i\equiv m$, $j\equiv l$, $k$, $n$ and induce the expression

\begin{align}
\text{Cov}[a_\mathbf{I},a_\mathbf{J}]&=p_{ij}p_{jk}p_{ki}p_{jn}p_{ni}-(p_{ij})^2p_{jk}p_{ki}p_{jn}p_{ni}+p_{ji}p_{ik}p_{kj}p_{in}p_{nj}-(p_{ji})^2p_{ik}p_{kj}p_{in}p_{nj}\nonumber\\
&=p_{ij}(1-p_{ij})p_{jk}p_{ki}p_{jn}p_{ni}+p_{ji}(1-p_{ji})p_{ik}p_{kj}p_{in}p_{nj}.
\end{align}

Let us now, calculate the number of times such an expression appears, i.e. the number of triples sharing an edge: since we need to first, choose the pair of nodes individuating the common edge and, then the pair of nodes individuating the `free' vertices of the two triads, such a number amounts at $\binom{N}{2}\binom{N-2}{2}=N(N-1)(N-2)(N-3)/4$; in case $N=4$, it amounts at $3!=6$ - indeed, let us concretely focus on the triads $(1,2,3)$, $(1,2,4)$, $(1,3,4)$, $(2,3,4)$: $(1,2,3)$ co-variates with $(1,2,4)$, $(1,3,4)$, $(2,3,4)$; $(1,2,4)$ co-variates with $(1,3,4)$, $(2,3,4)$; $(1,3,4)$ co-variates with $(2,3,4)$. Overall, then,

\begin{align}
\sum_{\mathbf{I<\mathbf{J}}}\text{Cov}[a_\mathbf{I},a_\mathbf{J}]=3!\cdot\sum_{i<j<k<n}[p_{ij}(1-p_{ij})p_{jk}p_{ki}p_{jn}p_{ni}+p_{ji}(1-p_{ji})p_{ik}p_{kj}p_{in}p_{nj}].
\end{align}

\begin{figure}[t!]
\centering
\includegraphics[width=\linewidth]{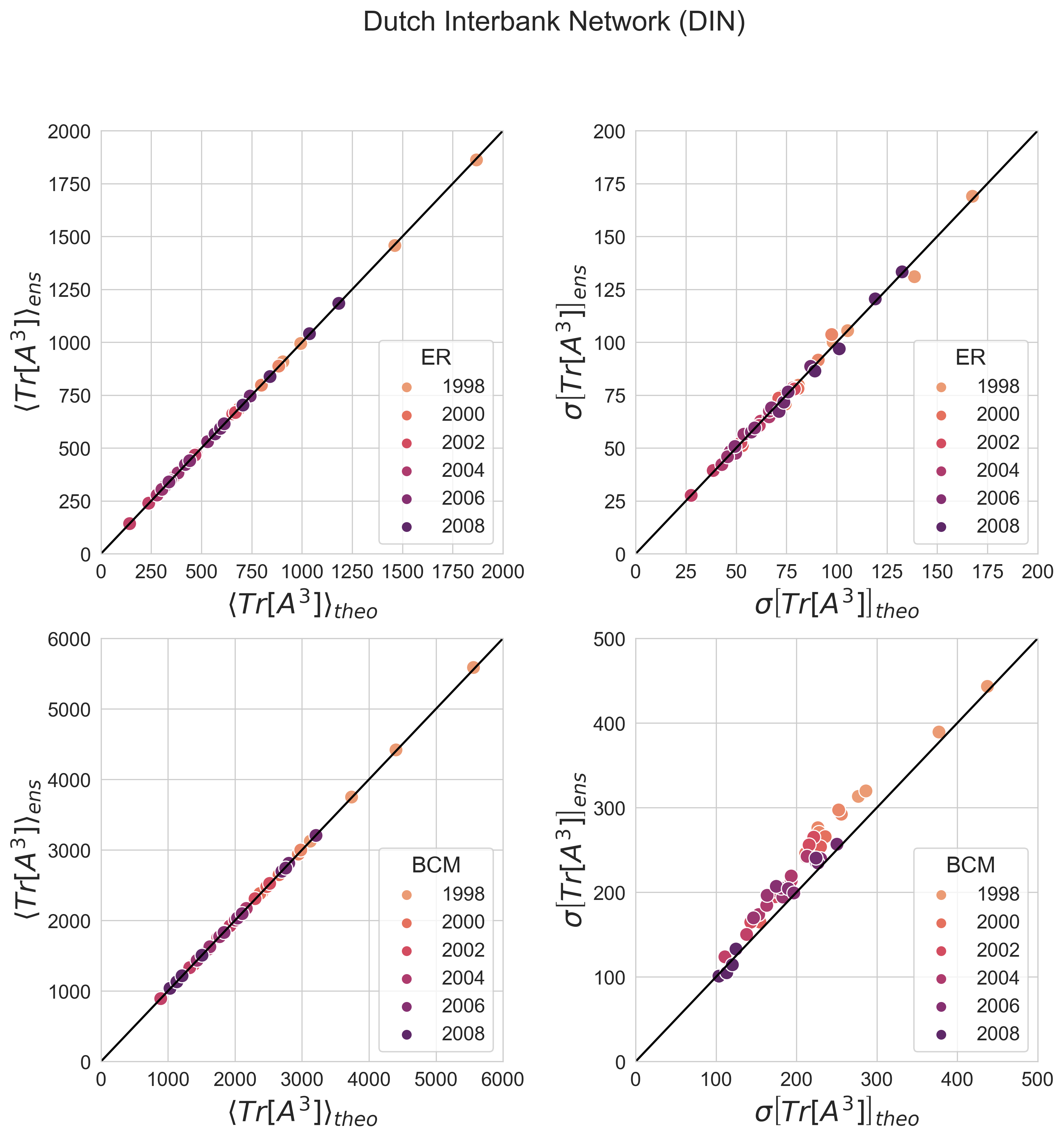}
\caption{Comparison between the analytical estimations of the expected value and variance of the number of triads and the numerical counterparts, obtained by explicitly sampling the ensembles induced by the Erd\"os-R\'enyi Model (top panels) and the Binary Configuration Model (bottom panels). The numerical simulations have been carried out on the quarters of the Dutch Interbank Network (DIN); the number of sampled matrices per snapshot is $10^3$.}
\label{fig:9}
\end{figure}

\begin{figure}[t!]
\centering
\includegraphics[width=\linewidth]{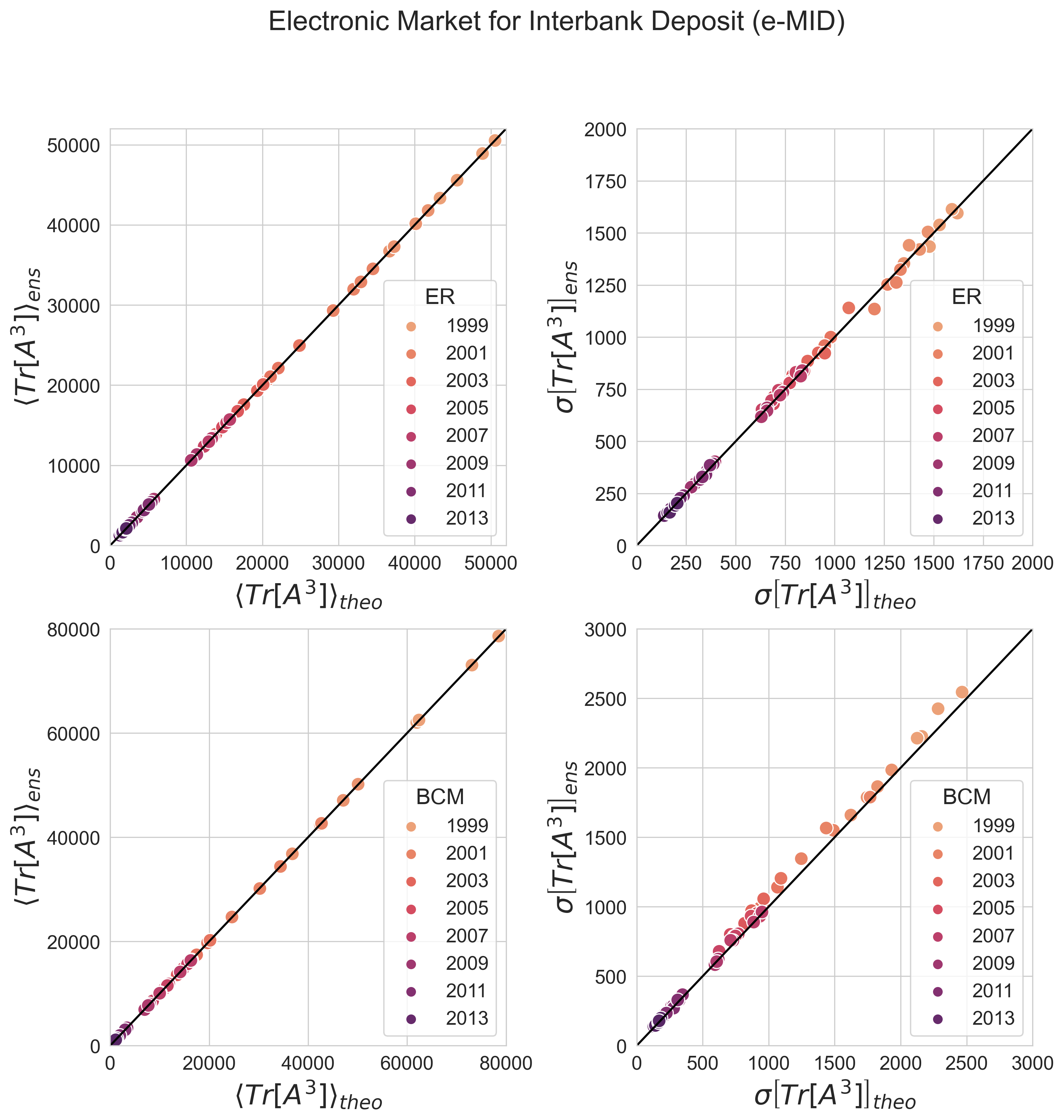}
\caption{Comparison between the analytical estimations of the expected value and variance of the number of triads and the numerical counterparts, obtained by explicitly sampling the ensembles induced by the Erd\"os-R\'enyi Model (top panels) and the Binary Configuration Model (bottom panels). The numerical simulations have been carried out on the quarters of the Electronic Market for Interbank Deposit (e-MID); the number of sampled matrices per snapshot is $10^3$.}
\label{fig:10}
\end{figure}

The comparison between the analytical estimations of the expected value and the variance of the number of triads and the numerical counterparts, obtained by explicitly sampling the ensembles induced by the Erd\"os-R\'enyi Model and the Binary Configuration Model is illustrated in figs.~\ref{fig:9} and~\ref{fig:10}.

\clearpage

\clearpage

\clearpage

\hypertarget{AppC}{}
\section*{Appendix C.\\Diagonalisation and trace of the matrix exponential}

In this Appendix, we will provide a sketch of the proof that

\begin{equation}
f(\mathbf{A})=\mathbf{F}f(\mathbf{\Lambda})\mathbf{F}^{-1}
\end{equation}
and that

\begin{equation}
\text{Tr}\left[f(\mathbf{A})\right]=\text{Tr}\left[\mathbf{F}f(\mathbf{\Lambda})\mathbf{F}^{-1}\right]=\text{Tr}\left[f(\mathbf{\Lambda})\mathbf{F}^{-1}\mathbf{F}\right]=\text{Tr}\left[f(\mathbf{\Lambda})\right],
\end{equation}
i.e. that the trace is invariant under a cyclic permutation of matrices, in the special case $f(\cdot)\equiv e^{(\cdot)}$ and where $\mathbf{F}$ is the matrix that diagonalises $\mathbf{A}$, i.e. the one ensuring that $\mathbf{F}^{-1}\mathbf{A}\mathbf{F}=\mathbf{\Lambda}$.\\

Since the function of a matrix is formally identical to its series expansion, one can write that

\begin{equation}
e^\mathbf{A}\equiv\mathbf{I}+\mathbf{A}+\frac{\mathbf{A}^2}{2!}+\frac{\mathbf{A}^3}{3!}+\dots+\frac{\mathbf{A}^n}{n!}+\dots;
\end{equation}
let us now diagonalise it:

\begin{eqnarray}
\mathbf{F}^{-1}e^\mathbf{A}\mathbf{F}&\equiv&\mathbf{F}^{-1}\mathbf{I}\mathbf{F}+\mathbf{F}^{-1}\mathbf{A}\mathbf{F}+\frac{\mathbf{F}^{-1}\mathbf{A}^2\mathbf{F}}{2!}+\frac{\mathbf{F}^{-1}\mathbf{A}^3\mathbf{F}}{3!}+\dots+\frac{\mathbf{F}^{-1}\mathbf{A}^n\mathbf{F}}{n!}+\dots\nonumber\\
&=&\mathbf{I}+\mathbf{\Lambda}+\frac{\left(\mathbf{F}^{-1}\mathbf{A}\mathbf{F}\right)\left(\mathbf{F}^{-1}\mathbf{A}\mathbf{F}\right)}{2!}+\frac{\left(\mathbf{F}^{-1}\mathbf{A}\mathbf{F}\right)\left(\mathbf{F}^{-1}\mathbf{A}\mathbf{F}\right)\left(\mathbf{F}^{-1}\mathbf{A}\mathbf{F}\right)}{3!}+\dots\nonumber\\
&=&\mathbf{I}+\mathbf{\Lambda}+\frac{\left(\mathbf{F}^{-1}\mathbf{A}\mathbf{F}\right)^2}{2!}+\frac{\left(\mathbf{F}^{-1}\mathbf{A}\mathbf{F}\right)^3}{3!}+\dots+\frac{\left(\mathbf{F}^{-1}\mathbf{A}\mathbf{F}\right)^n}{n!}+\dots\nonumber\\
&=&\mathbf{I}+\mathbf{\Lambda}+\frac{\mathbf{\Lambda}^2}{2!}+\frac{\mathbf{\Lambda}^3}{3!}+\dots+\frac{\mathbf{\Lambda}^n}{n!}+\dots\equiv e^\mathbf{\Lambda}.
\end{eqnarray}

Since all matrices appearing in the last row are diagonal, $e^\mathbf{\Lambda}$ also has diagonal entries. As a consequence,

\begin{eqnarray}
\text{Tr}\left[e^\mathbf{\Lambda}\right]&=&\sum_{i=1}^N\left(e^\mathbf{\Lambda}\right)_{ii}=\sum_{i=1}^Ne^{\lambda_i}=\sum_{i=1}^N1+\sum_{i=1}^N\lambda_i+\sum_{i=1}^N\frac{\lambda_i^2}{2!}+\sum_{i=1}^N\frac{\lambda_i^3}{3!}+\dots+\sum_{i=1}^N\frac{\lambda_i^n}{n!}+\dots\nonumber\\
&=&\text{Tr}\left[\mathbf{I}\right]+\text{Tr}\left[\mathbf{\Lambda}\right]+\frac{\text{Tr}\left[\mathbf{\Lambda}^2\right]}{2!}+\frac{\text{Tr}\left[\mathbf{\Lambda}^3\right]}{3!}+\dots+\frac{\text{Tr}\left[\mathbf{\Lambda}^n\right]}{n!}+\dots\nonumber\\
&=&\text{Tr}\left[\mathbf{I}\right]+\text{Tr}\left[\mathbf{F}^{-1}\mathbf{A}\mathbf{F}\right]+\frac{\text{Tr}\left[(\mathbf{F}^{-1}\mathbf{A}\mathbf{F})(\mathbf{F}^{-1}\mathbf{A}\mathbf{F})\right]}{2!}+\frac{\text{Tr}\left[(\mathbf{F}^{-1}\mathbf{A}\mathbf{F})(\mathbf{F}^{-1}\mathbf{A}\mathbf{F})(\mathbf{F}^{-1}\mathbf{A}\mathbf{F})\right]}{3!}+\dots\nonumber\\
&=&\text{Tr}\left[\mathbf{I}\right]+\text{Tr}\left[\mathbf{F}^{-1}\mathbf{A}\mathbf{F}\right]+\frac{\text{Tr}\left[\mathbf{F}^{-1}\mathbf{A}^2\mathbf{F}\right]}{2!}+\frac{\text{Tr}\left[\mathbf{F}^{-1}\mathbf{A}^3\mathbf{F}\right]}{3!}+\dots+\frac{\text{Tr}\left[\mathbf{F}^{-1}\mathbf{A}^n\mathbf{F}\right]}{n!}+\dots\nonumber\\
&=&\text{Tr}\left[\mathbf{I}\right]+\text{Tr}\left[\mathbf{A}\mathbf{F}\mathbf{F}^{-1}\right]+\frac{\text{Tr}\left[\mathbf{A}^2\mathbf{F}\mathbf{F}^{-1}\right]}{2!}+\frac{\text{Tr}\left[\mathbf{A}^3\mathbf{F}\mathbf{F}^{-1}\right]}{3!}+\dots+\frac{\text{Tr}\left[\mathbf{A}^n\mathbf{F}\mathbf{F}^{-1}\right]}{n!}+\dots\nonumber\\
&=&\text{Tr}\left[\mathbf{I}\right]+\text{Tr}\left[\mathbf{A}\right]+\frac{\text{Tr}\left[\mathbf{A}^2\right]}{2!}+\frac{\text{Tr}\left[\mathbf{A}^3\right]}{3!}+\dots+\frac{\text{Tr}\left[\mathbf{A}^n\right]}{n!}+\dots=\text{Tr}\left[e^\mathbf{A}\right],
\end{eqnarray}
where we have exploited the property of the trace of being invariant under circular shifts.

\clearpage

\hypertarget{AppD}{}
\section*{Appendix D.\\Ensemble distribution of the spectral radius}

\begin{figure*}[h!]
\centering
\includegraphics[width=\linewidth]{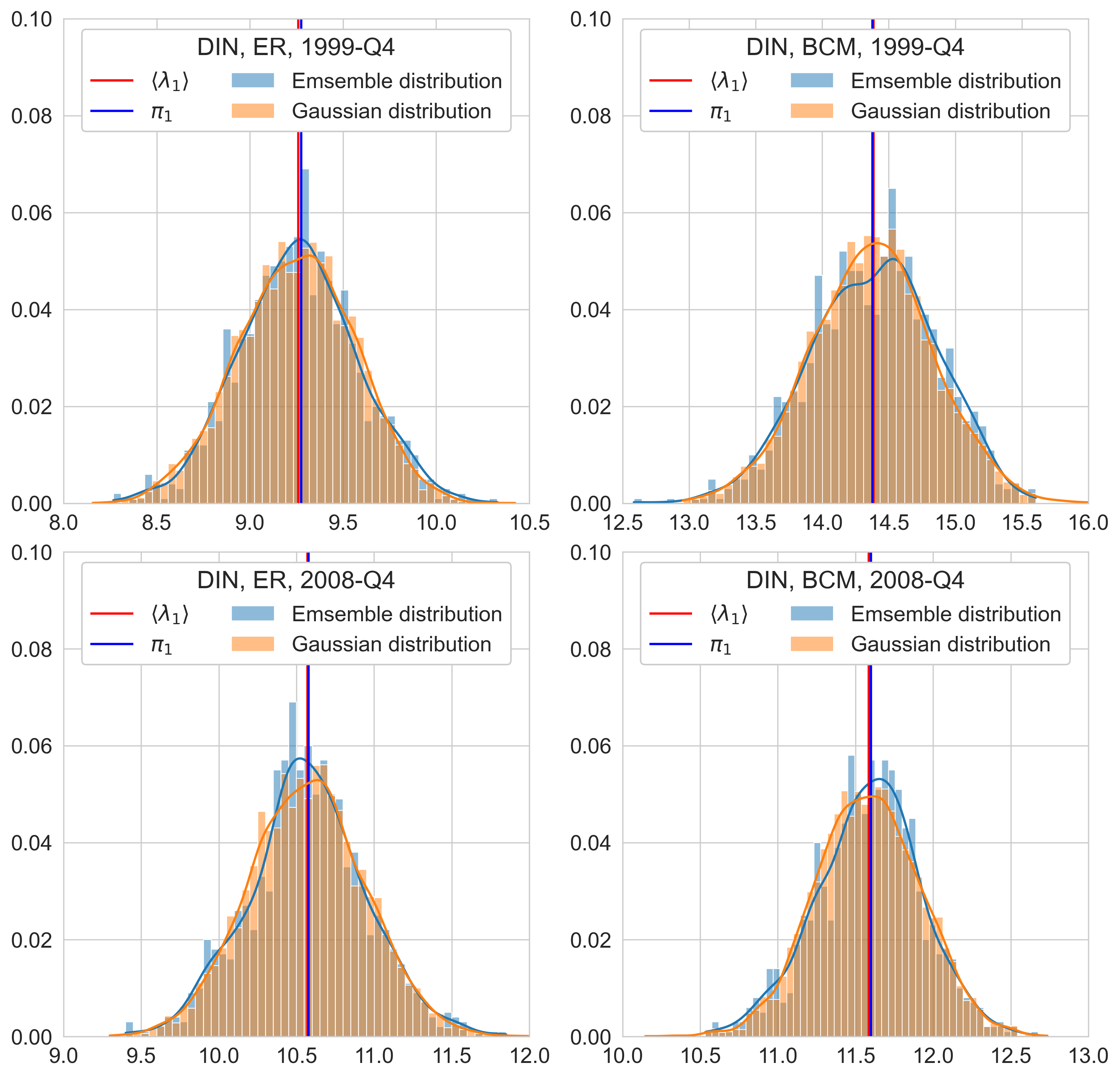}
\caption{Distribution of the spectral radius on the ensemble induced by the Erd\"os-R\'enyi Model (top panels) and the Binary Configuration Model (bottom panels), for the quarters 1999Q4 and 2008Q4 of the Dutch Interbank Network (DIN): the agreement with a Gaussian distribution whose expected value and variance coincide with those computed on the corresponding ensemble is, overall, very good. Similar results are obtained for the Electronic Market for Interbank Deposit (e-MID).}
\label{fig:11}
\end{figure*}

\clearpage

\hypertarget{AppE}{}
\section*{Appendix E.\\Dutch Interbank Network}

\begin{table*}[h!]
\begin{tabular}{l||c|c|c||c|c|c|c}
\hline
\hline
Erd\"os-R\'enyi Model & $N$ & $c$ & $r$ & $e^{\lambda_1}/\text{Tr}\left[e^{\mathbf{A}}\right]$ & $e^{\pi_1}/\text{Tr}\left[e^{\mathbf{P}}\right]$ & $\text{Tr}\left[e^{\mathbf{P}}\right]/\langle\text{Tr}\left[e^{\mathbf{A}}\right]\rangle$ & $\langle\lambda_1\rangle/\pi_1$ \\
\hline
\hline
DIN 1998-Q1 & 100 & 0.115 & 0.270 & 1.000 & 0.999 & 0.974 & 0.998 \\
DIN 1999-Q1 &  95 & 0.102 & 0.268 & 1.000 & 0.994 & 0.951 & 1.000 \\
DIN 2000-Q1 &  98 & 0.087 & 0.262 & 1.000 & 0.981 & 0.982 & 0.997 \\
DIN 2001-Q1 & 100 & 0.073 & 0.264 & 0.999 & 0.939 & 0.975 & 0.999 \\
DIN 2002-Q1 &  98 & 0.067 & 0.264 & 0.998 & 0.880 & 0.977 & 0.998 \\
DIN 2003-Q1 &  98 & 0.054 & 0.266 & 0.985 & 0.669 & 0.981 & 0.998 \\
DIN 2004-Q1 &  98 & 0.083 & 0.261 & 1.000 & 0.973 & 0.976 & 0.998 \\
DIN 2005-Q1 &  96 & 0.072 & 0.257 & 0.986 & 0.911 & 0.971 & 0.999 \\
DIN 2006-Q1 &  94 & 0.081 & 0.236 & 0.987 & 0.954 & 0.966 & 0.999 \\
DIN 2007-Q1 &  93 & 0.090 & 0.252 & 0.936 & 0.979 & 0.970 & 0.999 \\
DIN 2008-Q1 &  75 & 0.120 & 0.138 & 0.957 & 0.991 & 0.976 & 0.997 \\
\hline
\hline
Binary Configuration Model & $N$ & $c$ & $r$ & $e^{\lambda_1}/\text{Tr}\left[e^{\mathbf{A}}\right]$ & $e^{\pi_1}/\text{Tr}\left[e^{\mathbf{P}}\right]$ & $\text{Tr}\left[e^{\mathbf{P}}\right]/\langle\text{Tr}\left[e^{\mathbf{A}}\right]\rangle$ & $\langle\lambda_1\rangle/\pi_1$ \\
\hline
\hline
DIN 1998-Q1 & 100 & 0.115 & 0.270 & 1.000 & 1.000 & 0.890 & 1.000 \\
DIN 1999-Q1 &  95 & 0.102 & 0.268 & 1.000 & 1.000 & 0.923 & 0.998 \\
DIN 2000-Q1 &  98 & 0.087 & 0.262 & 1.000 & 1.000 & 0.930 & 0.998 \\
DIN 2001-Q1 & 100 & 0.073 & 0.264 & 0.999 & 1.000 & 0.908 & 1.000 \\
DIN 2002-Q1 &  98 & 0.067 & 0.264 & 0.998 & 0.999 & 0.949 & 0.996 \\
DIN 2003-Q1 &  98 & 0.054 & 0.266 & 0.985 & 0.995 & 0.927 & 0.998 \\
DIN 2004-Q1 &  98 & 0.083 & 0.261 & 1.000 & 1.000 & 0.924 & 0.999 \\
DIN 2005-Q1 &  96 & 0.072 & 0.257 & 0.986 & 1.000 & 0.926 & 0.999 \\
DIN 2006-Q1 &  94 & 0.081 & 0.236 & 0.987 & 1.000 & 0.937 & 0.997 \\
DIN 2007-Q1 &  93 & 0.090 & 0.252 & 0.936 & 1.000 & 0.939 & 0.999 \\
DIN 2008-Q1 &  75 & 0.120 & 0.138 & 0.957 & 0.998 & 0.936 & 1.001 \\
\hline
\hline
Global Reciprocity Model & $N$ & $c$ & $r$ & $e^{\lambda_1}/\text{Tr}\left[e^{\mathbf{A}}\right]$ & $e^{\pi_1}/\text{Tr}\left[e^{\mathbf{P}}\right]$ & $\text{Tr}\left[e^{\mathbf{P}}\right]/\langle\text{Tr}\left[e^{\mathbf{A}}\right]\rangle$ & $\langle\lambda_1\rangle/\pi_1$ \\
\hline
\hline
DIN 1998-Q1 & 100 & 0.115 & 0.270 & 1.000 & 1.000 & 0.879 & 1.000 \\
DIN 1999-Q1 &  95 & 0.102 & 0.268 & 1.000 & 1.000 & 0.873 & 1.002 \\
DIN 2000-Q1 &  98 & 0.087 & 0.262 & 1.000 & 1.000 & 0.889 & 1.001 \\
DIN 2001-Q1 & 100 & 0.073 & 0.264 & 0.999 & 1.000 & 0.872 & 1.003 \\
DIN 2002-Q1 &  98 & 0.067 & 0.264 & 0.998 & 0.999 & 0.860 & 1.005 \\
DIN 2003-Q1 &  98 & 0.054 & 0.266 & 0.985 & 0.996 & 0.858 & 1.007 \\
DIN 2004-Q1 &  98 & 0.083 & 0.261 & 1.000 & 1.000 & 0.863 & 1.004 \\
DIN 2005-Q1 &  96 & 0.072 & 0.257 & 0.986 & 1.000 & 0.920 & 0.999 \\
DIN 2006-Q1 &  94 & 0.081 & 0.236 & 0.987 & 1.000 & 0.925 & 0.998 \\
DIN 2007-Q1 &  93 & 0.090 & 0.252 & 0.936 & 1.000 & 0.958 & 0.997 \\
DIN 2008-Q1 &  75 & 0.120 & 0.138 & 0.957 & 0.996 & 1.066 & 0.988 \\
\hline
\hline
Reciprocal Configuration Model & $N$ & $c$ & $r$ & $e^{\lambda_1}/\text{Tr}\left[e^{\mathbf{A}}\right]$ & $e^{\pi_1}/\text{Tr}\left[e^{\mathbf{P}}\right]$ & $\text{Tr}\left[e^{\mathbf{P}}\right]/\langle\text{Tr}\left[e^{\mathbf{A}}\right]\rangle$ & $\langle\lambda_1\rangle/\pi_1$ \\
\hline
\hline
DIN 1998-Q1 & 100 & 0.115 & 0.270 & 1.000 & 1.000 & 0.846 & 1.003 \\
DIN 1999-Q1 &  95 & 0.102 & 0.268 & 1.000 & 1.000 & 0.878 & 1.003 \\
DIN 2000-Q1 &  98 & 0.087 & 0.262 & 1.000 & 1.000 & 0.881 & 1.004 \\
DIN 2001-Q1 & 100 & 0.073 & 0.264 & 0.999 & 1.000 & 0.824 & 1.009 \\
DIN 2002-Q1 &  98 & 0.067 & 0.264 & 0.998 & 0.999 & 0.799 & 1.014 \\
DIN 2003-Q1 &  98 & 0.054 & 0.266 & 0.985 & 0.996 & 0.821 & 1.013 \\
DIN 2004-Q1 &  98 & 0.083 & 0.261 & 1.000 & 1.000 & 0.787 & 1.012 \\
DIN 2005-Q1 &  96 & 0.072 & 0.257 & 0.986 & 0.999 & 0.875 & 1.003 \\
DIN 2006-Q1 &  94 & 0.081 & 0.236 & 0.987 & 0.999 & 0.852 & 1.004 \\
DIN 2007-Q1 &  93 & 0.090 & 0.252 & 0.936 & 1.000 & 0.903 & 1.002 \\
DIN 2008-Q1 &  75 & 0.120 & 0.138 & 0.957 & 0.994 & 1.010 & 0.992 \\
\hline
\hline
\end{tabular}
\caption{Check of the approximations that lead to the result $\langle\lambda_1\rangle\simeq\pi_1$, for the Dutch Interbank Network (DIN).
\label{tab:DIN}}
\end{table*}

\clearpage

\section*{Electronic Market for Interbank Deposit}

\begin{table*}[h!]
\begin{tabular}{l||c|c|c||c|c|c|c}
\hline
\hline
Erd\"os-R\'enyi Model & $N$ & $c$ & $r$ & $e^{\lambda_1}/\text{Tr}\left[e^{\mathbf{A}}\right]$ & $e^{\pi_1}/\text{Tr}\left[e^{\mathbf{P}}\right]$ & $\text{Tr}\left[e^{\mathbf{P}}\right]/\langle\text{Tr}\left[e^{\mathbf{A}}\right]\rangle$ & $\langle\lambda_1\rangle/\pi_1$ \\
\hline
\hline
e-MID 1999-Q1 & 205 & 0.181 & 0.258 & 1.000 & 1.000 & 0.940 & 1.000 \\
e-MID 2001-Q1 & 154 & 0.213 & 0.256 & 1.000 & 1.000 & 0.927 & 1.000 \\
e-MID 2003-Q1 & 124 & 0.218 & 0.199 & 1.000 & 1.000 & 0.930 & 0.999 \\
e-MID 2005-Q1 & 113 & 0.209 & 0.232 & 1.000 & 1.000 & 0.906 & 1.000 \\
e-MID 2007-Q1 & 101 & 0.237 & 0.225 & 1.000 & 1.000 & 0.908 & 1.000 \\
e-MID 2009-Q1 &  95 & 0.169 & 0.130 & 0.999 & 1.000 & 0.958 & 0.998 \\
e-MID 2011-Q1 &  90 & 0.189 & 0.149 & 1.000 & 1.000 & 0.937 & 0.999 \\
e-MID 2013-Q1 &  73 & 0.151 & 0.116 & 0.985 & 0.999 & 0.934 & 1.000 \\
\hline
\hline
Binary Configuration Model & $N$ & $c$ & $r$ & $e^{\lambda_1}/\text{Tr}\left[e^{\mathbf{A}}\right]$ & $e^{\pi_1}/\text{Tr}\left[e^{\mathbf{P}}\right]$ & $\text{Tr}\left[e^{\mathbf{P}}\right]/\langle\text{Tr}\left[e^{\mathbf{A}}\right]\rangle$ & $\langle\lambda_1\rangle/\pi_1$ \\
\hline
\hline
e-MID 1999-Q1 & 205 & 0.181 & 0.258 & 1.000 & 1.000 & 0.886 & 1.000 \\
e-MID 2001-Q1 & 154 & 0.213 & 0.256 & 1.000 & 1.000 & 0.906 & 0.999 \\
e-MID 2003-Q1 & 124 & 0.218 & 0.199 & 1.000 & 1.000 & 0.889 & 0.999 \\
e-MID 2005-Q1 & 113 & 0.209 & 0.232 & 1.000 & 1.000 & 0.871 & 1.000 \\
e-MID 2007-Q1 & 101 & 0.237 & 0.225 & 1.000 & 1.000 & 0.872 & 1.000 \\
e-MID 2009-Q1 &  95 & 0.169 & 0.130 & 0.999 & 0.999 & 0.902 & 0.997 \\
e-MID 2011-Q1 &  90 & 0.189 & 0.149 & 1.000 & 1.000 & 0.890 & 0.998 \\
e-MID 2013-Q1 &  73 & 0.151 & 0.116 & 0.985 & 0.993 & 0.867 & 0.999 \\
\hline
\hline
Global Reciprocity Model & $N$ & $c$ & $r$ & $e^{\lambda_1}/\text{Tr}\left[e^{\mathbf{A}}\right]$ & $e^{\pi_1}/\text{Tr}\left[e^{\mathbf{P}}\right]$ & $\text{Tr}\left[e^{\mathbf{P}}\right]/\langle\text{Tr}\left[e^{\mathbf{A}}\right]\rangle$ & $\langle\lambda_1\rangle/\pi_1$ \\
\hline
\hline
e-MID 1999-Q1 & 205 & 0.181 & 0.258 & 1.000 & 1.000 & 0.896 & 1.000 \\
e-MID 2001-Q1 & 154 & 0.213 & 0.256 & 1.000 & 1.000 & 0.857 & 1.001 \\
e-MID 2003-Q1 & 124 & 0.218 & 0.199 & 1.000 & 1.000 & 0.914 & 0.998 \\
e-MID 2005-Q1 & 113 & 0.209 & 0.232 & 1.000 & 1.000 & 0.856 & 1.001 \\
e-MID 2007-Q1 & 101 & 0.237 & 0.225 & 1.000 & 1.000 & 0.852 & 1.001 \\
e-MID 2009-Q1 &  95 & 0.169 & 0.130 & 0.999 & 1.000 & 0.825 & 1.004 \\
e-MID 2011-Q1 &  90 & 0.189 & 0.149 & 1.000 & 1.000 & 0.876 & 1.000 \\
e-MID 2013-Q1 &  73 & 0.151 & 0.116 & 0.985 & 0.993 & 0.901 & 0.995 \\
\hline
\hline
Reciprocal Configuration Model & $N$ & $c$ & $r$ & $e^{\lambda_1}/\text{Tr}\left[e^{\mathbf{A}}\right]$ & $e^{\pi_1}/\text{Tr}\left[e^{\mathbf{P}}\right]$ & $\text{Tr}\left[e^{\mathbf{P}}\right]/\langle\text{Tr}\left[e^{\mathbf{A}}\right]\rangle$ & $\langle\lambda_1\rangle/\pi_1$ \\
\hline
\hline
e-MID 1999-Q1 & 205 & 0.181 & 0.258 & 1.000 & 1.000 & 0.876 & 1.000 \\
e-MID 2001-Q1 & 154 & 0.213 & 0.256 & 1.000 & 1.000 & 0.844 & 1.001 \\
e-MID 2003-Q1 & 124 & 0.218 & 0.199 & 1.000 & 1.000 & 0.891 & 0.999 \\
e-MID 2005-Q1 & 113 & 0.209 & 0.232 & 1.000 & 1.000 & 0.840 & 1.001 \\
e-MID 2007-Q1 & 101 & 0.237 & 0.225 & 1.000 & 1.000 & 0.824 & 1.002 \\
e-MID 2009-Q1 &  95 & 0.169 & 0.130 & 0.999 & 1.000 & 0.821 & 1.005 \\
e-MID 2011-Q1 &  90 & 0.189 & 0.149 & 1.000 & 1.000 & 0.862 & 1.001 \\
e-MID 2013-Q1 &  73 & 0.151 & 0.116 & 0.985 & 0.992 & 0.885 & 0.997 \\
\hline
\hline
Density-Corrected Gravity Model & $N$ & $c$ & $r$ & $e^{\omega_1}/\text{Tr}\left[e^{\mathbf{W}}\right]$ & $e^{\phi_1}/\text{Tr}\left[e^{\mathbf{Q}}\right]$ & $\text{Tr}\left[e^{\mathbf{Q}}\right]/\langle\text{Tr}\left[e^{\mathbf{W}}\right]\rangle$ & $\langle\omega_1\rangle/\phi_1$ \\
\hline
\hline
e-MID 1999 & 212 & 0.279 & 0.440 & 1.000 & 1.000 & 0.987 & 1.000 \\
e-MID 2001 & 163 & 0.312 & 0.466 & 0.992 & 1.000 & 0.997 & 1.000 \\
e-MID 2003 & 128 & 0.320 & 0.433 & 1.000 & 1.000 & 0.976 & 1.000 \\
e-MID 2005 & 113 & 0.328 & 0.458 & 1.000 & 1.000 & 0.971 & 1.001 \\
e-MID 2007 & 106 & 0.369 & 0.481 & 0.990 & 0.997 & 0.993 & 1.000 \\
e-MID 2009 &  99 & 0.266 & 0.285 & 0.479 & 0.651 & 0.991 & 1.000 \\
e-MID 2011 &  92 & 0.283 & 0.336 & 0.981 & 0.961 & 0.986 & 0.999 \\
e-MID 2013 &  78 & 0.230 & 0.289 & 1.000 & 0.241 & 0.996 & 1.000 \\
\hline
\hline
\end{tabular}
\caption{Check of the approximations that lead to the results $\langle\lambda_1\rangle\simeq\pi_1$ and $\langle\omega_1\rangle\simeq\phi_1$, for the Electronic Market for Interbank Deposit (e-MID). Notice that the density-corrected Gravity Model has been solved on the yearly e-MID to prevent numerical problems related to the value of the spectral gap, i.e. $\lambda_1-\lambda_2$: let us, in fact, remind that our derivation holds in case $\lambda_1-\lambda_2$ is (much) larger than zero.
\label{tab:eMID}}
\end{table*}

\clearpage

\section*{International Trade Network}

\begin{table*}[h!]
\begin{tabular}{l||c|c|c||c|c|c|c}
\hline
\hline
Erd\"os-R\'enyi Model & $N$ & $c$ & $r$ & $e^{\lambda_1}/\text{Tr}\left[e^{\mathbf{A}}\right]$ & $e^{\pi_1}/\text{Tr}\left[e^{\mathbf{P}}\right]$ & $\text{Tr}\left[e^{\mathbf{P}}\right]/\langle\text{Tr}\left[e^{\mathbf{A}}\right]\rangle$ & $\langle\lambda_1\rangle/\pi_1$ \\
\hline
\hline
ITN 2000 & 112 & 0.753 & 0.887 & 1.000 & 1.000 & 0.899 & 1.000 \\
ITN 2003 & 112 & 0.769 & 0.884 & 1.000 & 1.000 & 0.900 & 1.000 \\
ITN 2006 & 112 & 0.790 & 0.890 & 1.000 & 1.000 & 0.921 & 1.000 \\
ITN 2009 & 112 & 0.807 & 0.903 & 1.000 & 1.000 & 0.939 & 1.000 \\
ITN 2012 & 112 & 0.828 & 0.914 & 1.000 & 1.000 & 0.910 & 1.000 \\
ITN 2015 & 112 & 0.826 & 0.912 & 1.000 & 1.000 & 0.926 & 1.000 \\
ITN 2018 & 112 & 0.838 & 0.917 & 1.000 & 1.000 & 0.919 & 1.000 \\
\hline
\hline
Binary Configuration Model & $N$ & $c$ & $r$ & $e^{\lambda_1}/\text{Tr}\left[e^{\mathbf{A}}\right]$ & $e^{\pi_1}/\text{Tr}\left[e^{\mathbf{P}}\right]$ & $\text{Tr}\left[e^{\mathbf{P}}\right]/\langle\text{Tr}\left[e^{\mathbf{A}}\right]\rangle$ & $\langle\lambda_1\rangle/\pi_1$ \\
\hline
\hline
ITN 2000 & 112 & 0.753 & 0.887 & 1.000 & 1.000 & 0.970 & 1.000 \\
ITN 2003 & 112 & 0.769 & 0.884 & 1.000 & 1.000 & 0.975 & 1.000 \\
ITN 2006 & 112 & 0.790 & 0.890 & 1.000 & 1.000 & 0.978 & 1.000 \\
ITN 2009 & 112 & 0.807 & 0.903 & 1.000 & 1.000 & 0.984 & 1.000 \\
ITN 2012 & 112 & 0.828 & 0.914 & 1.000 & 1.000 & 0.984 & 1.000 \\
ITN 2015 & 112 & 0.826 & 0.912 & 1.000 & 1.000 & 0.972 & 1.000 \\
ITN 2018 & 112 & 0.838 & 0.917 & 1.000 & 1.000 & 0.974 & 1.000 \\
\hline
\hline
Global Reciprocity Model & $N$ & $c$ & $r$ & $e^{\lambda_1}/\text{Tr}\left[e^{\mathbf{A}}\right]$ & $e^{\pi_1}/\text{Tr}\left[e^{\mathbf{P}}\right]$ & $\text{Tr}\left[e^{\mathbf{P}}\right]/\langle\text{Tr}\left[e^{\mathbf{A}}\right]\rangle$ & $\langle\lambda_1\rangle/\pi_1$ \\
\hline
\hline
ITN 2000 & 112 & 0.753 & 0.887 & 1.000 & 1.000 & 0.945 & 1.000 \\
ITN 2003 & 112 & 0.769 & 0.884 & 1.000 & 1.000 & 0.964 & 1.000 \\
ITN 2006 & 112 & 0.790 & 0.890 & 1.000 & 1.000 & 0.967 & 1.000 \\
ITN 2009 & 112 & 0.807 & 0.903 & 1.000 & 1.000 & 0.961 & 1.000 \\
ITN 2012 & 112 & 0.828 & 0.914 & 1.000 & 1.000 & 0.965 & 1.000 \\
ITN 2015 & 112 & 0.826 & 0.912 & 1.000 & 1.000 & 0.962 & 1.000 \\
ITN 2018 & 112 & 0.838 & 0.917 & 1.000 & 1.000 & 0.979 & 1.000 \\
\hline
\hline
Reciprocal Configuration Model & $N$ & $c$ & $r$ & $e^{\lambda_1}/\text{Tr}\left[e^{\mathbf{A}}\right]$ & $e^{\pi_1}/\text{Tr}\left[e^{\mathbf{P}}\right]$ & $\text{Tr}\left[e^{\mathbf{P}}\right]/\langle\text{Tr}\left[e^{\mathbf{A}}\right]\rangle$ & $\langle\lambda_1\rangle/\pi_1$ \\
\hline
\hline
ITN 2000 & 112 & 0.753 & 0.887 & 1.000 & 1.000 & 0.943 & 1.000 \\
ITN 2003 & 112 & 0.769 & 0.884 & 1.000 & 1.000 & 0.960 & 1.000 \\
ITN 2006 & 112 & 0.790 & 0.890 & 1.000 & 1.000 & 0.966 & 1.000 \\
ITN 2009 & 112 & 0.807 & 0.903 & 1.000 & 1.000 & 0.966 & 1.000 \\
ITN 2012 & 112 & 0.828 & 0.914 & 1.000 & 1.000 & 0.968 & 1.000 \\
ITN 2015 & 112 & 0.826 & 0.912 & 1.000 & 1.000 & 0.959 & 1.000 \\
ITN 2018 & 112 & 0.838 & 0.917 & 1.000 & 1.000 & 0.978 & 1.000 \\
\hline
\hline
Density-Corrected Gravity Model & $N$ & $c$ & $r$ & $e^{\omega_1}/\text{Tr}\left[e^{\mathbf{W}}\right]$ & $e^{\phi_1}/\text{Tr}\left[e^{\mathbf{Q}}\right]$ & $\text{Tr}\left[e^{\mathbf{Q}}\right]/\langle\text{Tr}\left[e^{\mathbf{W}}\right]\rangle$ & $\langle\omega_1\rangle/\phi_1$ \\
\hline
\hline
ITN 2000 & 112 & 0.753 & 0.887 & 1.000 & 1.000 & 1.000 & 1.000 \\
ITN 2003 & 112 & 0.769 & 0.884 & 1.000 & 1.000 & 1.000 & 1.000 \\
ITN 2006 & 112 & 0.790 & 0.890 & 1.000 & 1.000 & 1.000 & 1.000 \\
ITN 2009 & 112 & 0.807 & 0.903 & 1.000 & 1.000 & 1.000 & 1.000 \\
ITN 2012 & 112 & 0.828 & 0.914 & 1.000 & 1.000 & 1.000 & 1.000 \\
ITN 2015 & 112 & 0.826 & 0.912 & 1.000 & 1.000 & 1.000 & 1.000 \\
ITN 2018 & 112 & 0.838 & 0.917 & 1.000 & 1.000 & 1.000 & 1.000 \\
\hline
\hline
\end{tabular}
\caption{Check of the approximations that lead to the results $\langle\lambda_1\rangle\simeq\pi_1$ and $\langle\omega_1\rangle\simeq\phi_1$, for the International Trade Network (ITN).
\label{tab:ITN}}
\end{table*}

\clearpage

\clearpage

\hypertarget{AppF}{}
\section*{Appendix F.\\Inspecting the accuracy of the Chung-Lu approximation}

\begin{figure*}[h!]
\centering
\includegraphics[width=0.86\linewidth]{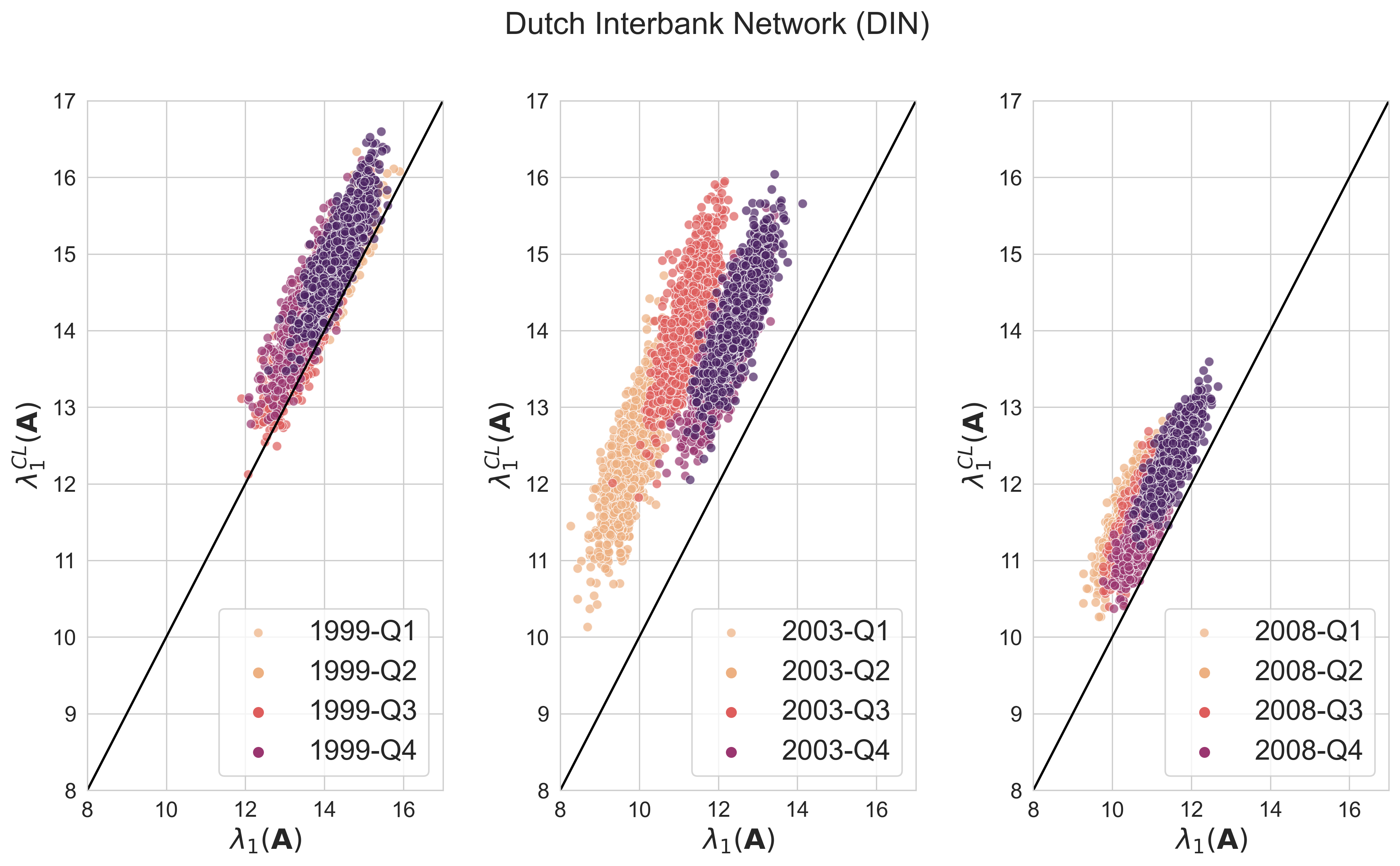}
\includegraphics[width=0.86\linewidth]{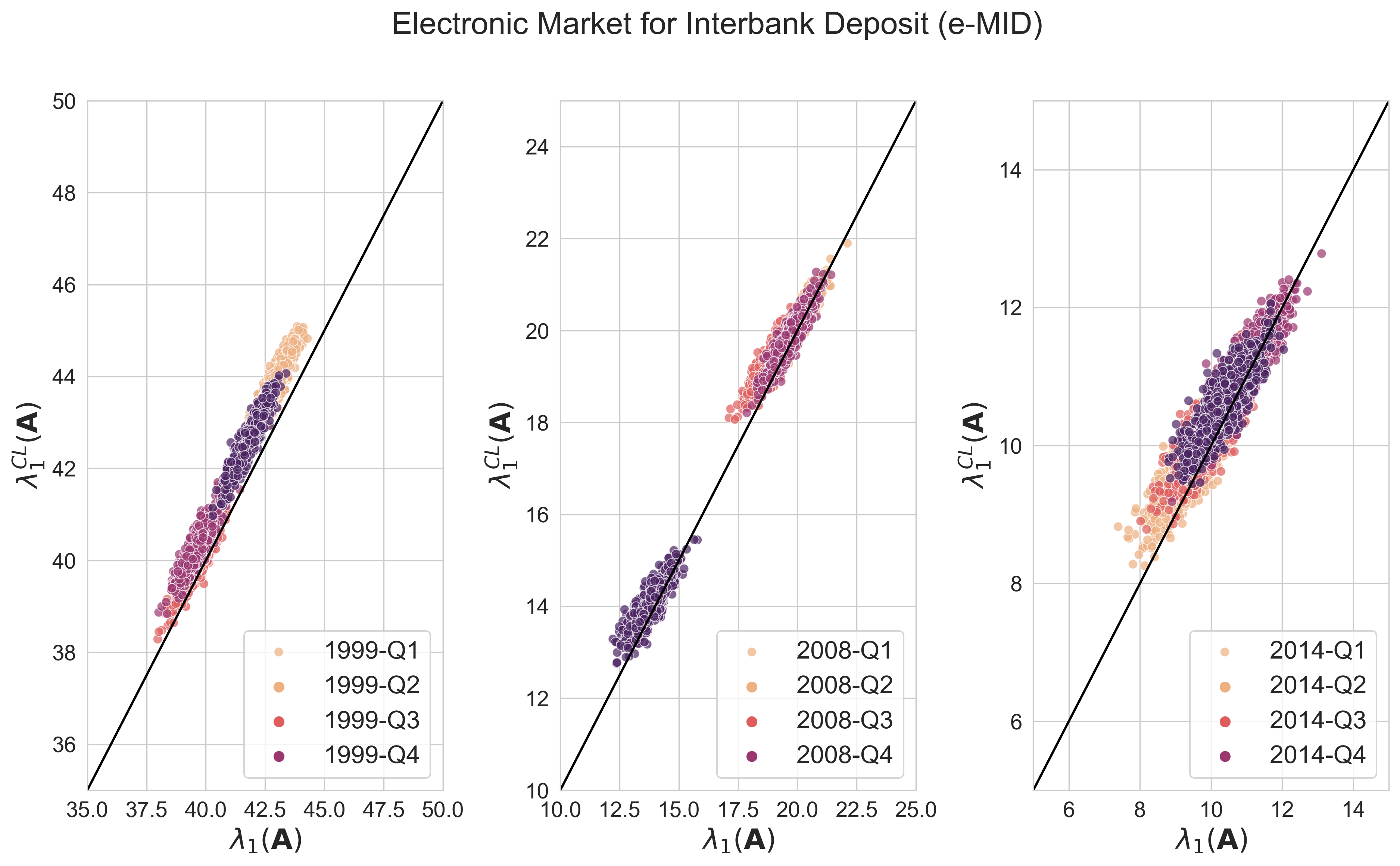}
\caption{Scattering the $10^3$ values of the BCM-induced variants of the spectral radius versus the corresponding Chung-Lu approximations may help explain the discrepancies observed in fig.~\ref{fig:2}. For instance, the evidence that $\lambda_1^\text{CL}(\mathbf{A})>\lambda_1(\mathbf{A})$ for all quarters of the Dutch Interbank Network (DIN) in 1999, 2003 and 2008 explains the overestimations provided by $\lambda_1^\text{CL}$ and $\text{Var}[\lambda_1^\text{CL}]$ and depicted in the top central and top right panels of fig.~\ref{fig:2}.}
\label{fig:13}
\end{figure*}

\end{document}